\documentclass[aps,superscriptaddress,floats,twocolumn,epsf,prr]{revtex4-2}
\usepackage{graphicx}
\usepackage{epstopdf}
\usepackage{color}
\usepackage{colortbl}
\usepackage{setspace}
\usepackage[utf8]{inputenc}
\usepackage{amsmath}
\usepackage{amssymb}
\usepackage{placeins}
\usepackage{mathtools}
\usepackage{changes}
\usepackage{tikz-feynman}
\usepackage{comment}
\usepackage[export]{adjustbox}
\usepackage{physics}
\usepackage{dsfont}
\usepackage[normalem]{ulem}
\usepackage{etoolbox}
\usepackage{enumitem}
\usepackage{pifont}
\AtEndEnvironment{subequations}{\par\nobreak\noindent}
\usepackage[colorlinks=true,urlcolor=blue,citecolor=blue,linkcolor=blue,breaklinks=true]{hyperref}
\usetikzlibrary{arrows.meta, shapes.geometric, positioning}

\tikzset{
        linePlain/.style={draw=black, thick},
        lineWithArrowCenter/.style={draw=black, thick, postaction={decorate},decoration={markings,mark=at position .6 with {\arrow[scale=1.2]{latex}}}},
        lineArrow/.style={draw=black, thick, postaction={decorate},decoration={markings,mark=at position .7 with {\arrow[scale=1.2]{latex}}}},
        lineArrowBis/.style={draw=black, thick, postaction={decorate},decoration={markings,mark=at position .4 with {\arrow[scale=1.2]{latex}}}},
        lineWithArrowCenterEnd/.style={draw=black, thick, postaction={decorate},decoration={markings,mark=at position .85 with {\arrow[scale=1.2]{latex}}}},
        lineWithArrowCenterStart/.style={draw=black, thick, postaction={decorate},decoration={markings,mark=at position .35 with {\arrow[scale=1.2]{latex}}}},
        bosonLine/.style={draw=black, thick, decorate, decoration={snake, segment length=2mm, amplitude=0.6mm}},
}

\newcommand{\tikzm}[2]{
        \tikz[baseline=-0.65ex]{#2}
}

\newcommand{\fullvertexwithlegs}[4]{
        \def\shift{0.3};  
        \def\shiftbox{0.3*#4};
        \coordinate (center) at (#2,#3);
        \coordinate (bottomleft)  at ($(center) + (-\shiftbox,-\shiftbox)$);
        \coordinate (topleft)     at ($(center) + (-\shiftbox,+\shiftbox)$);
        \coordinate (bottomright) at ($(center) + (+\shiftbox,-\shiftbox)$);
        \coordinate (topright)    at ($(center) + (+\shiftbox,+\shiftbox)$);
        \draw[linePlain] (bottomleft) rectangle (topright);
        \node at (center) {#1};
        \draw[lineWithArrowCenterEnd] (bottomleft) -- ($(bottomleft) + (-\shift,-\shift)$); 
         \draw[lineWithArrowCenterEnd] ($(topleft)     + (-\shift,+\shift)$) -- (topleft);
        \draw[lineWithArrowCenterEnd] ($(bottomright) + (+\shift,-\shift)$) -- (bottomright); 
        \draw[lineWithArrowCenterEnd] (topright) -- ($(topright) + (+\shift,+\shift)$);
}

\newcommand{\arrowslefthalf}[2]{
   \def\shift{0.3};
   \coordinate (center) at (#1,#2);
   \draw[lineWithArrowCenterEnd] (center)  -- ($(center) + (-\shift,-\shift)$);
   \draw[lineWithArrowCenterEnd] ($(center)     + (-\shift,+\shift)$) -- (center);
}

\newcommand{\arrowsrighthalf}[2]{
   \def\shift{0.3};
   \coordinate (center) at (#1,#2);
   \draw[lineWithArrowCenterEnd] ($(center) + (+\shift,-\shift)$) -- (center);
   \draw[lineWithArrowCenterEnd] (center)    -- ($(center)   + (+\shift,+\shift)$);
}

\newcommand{\arrowslefthalffull}[3]{
   \def\shift{0.3};
   \def\shiftbox{0.3*#3};
   \coordinate (center) at (#1,#2);
   \coordinate (bottomleft)  at ($(center) + (-\shiftbox,-\shiftbox)$);
   \coordinate (topleft)     at ($(center) + (-\shiftbox,+\shiftbox)$);
   \draw[lineWithArrowCenterEnd] (bottomleft)  -- ($(bottomleft) + (-\shift,-\shift)$);
   \draw[lineWithArrowCenterEnd] ($(topleft)     + (-\shift,+\shift)$) -- (topleft);
}

\newcommand{\arrowsrighthalffull}[3]{
   \def\shift{0.3};
   \def\shiftbox{0.3*#3};
   \coordinate (center) at (#1,#2);
   \coordinate (bottomright) at ($(center) + (+\shiftbox,-\shiftbox)$);
   \coordinate (topright)    at ($(center) + (+\shiftbox,+\shiftbox)$);
   \draw[lineWithArrowCenterEnd] ($(bottomright) + (+\shift,-\shift)$) -- (bottomright);
   \draw[lineWithArrowCenterEnd] (topright)    -- ($(topright)   + (+\shift,+\shift)$);
}

\newcommand{\arrowslowerhalffull}[3]{
   \def\shift{0.3};
   \def\shiftbox{0.3*#3};
   \coordinate (center) at (#1,#2);
   \coordinate (bottomleft)  at ($(center) + (-\shiftbox,-\shiftbox)$);
   \coordinate (bottomright) at ($(center) + (+\shiftbox,-\shiftbox)$);
   \draw[lineWithArrowCenterEnd] (bottomleft)  -- ($(bottomleft) + (-\shift,-\shift)$);
   \draw[lineWithArrowCenterEnd] ($(bottomright) + (+\shift,-\shift)$) -- (bottomright);
}

\newcommand{\arrowsupperhalffull}[3]{
   \def\shift{0.3};
   \def\shiftbox{0.3*#3};
   \coordinate (center) at (#1,#2);
   \coordinate (topleft)     at ($(center) + (-\shiftbox,+\shiftbox)$);
   \coordinate (topright)    at ($(center) + (+\shiftbox,+\shiftbox)$);
   \draw[lineWithArrowCenterEnd] ($(topleft)     + (-\shift,+\shift)$) -- (topleft);
   \draw[lineWithArrowCenterEnd] (topright)    -- ($(topright)   + (+\shift,+\shift)$);
}

\newcommand{\barevertexwithlegs}[2]{
   \fill (#1,#2) circle (2pt) coordinate (center);
   \arrowslefthalf{#1}{#2}
   \arrowsrighthalf{#1}{#2}
}

\newcommand{\barevertex}[2]{
   \fill (#1,#2) circle (2pt);
}

\newcommand{\arrowsallfull}[3]{
   \arrowslefthalffull{#1}{#2}{#3}
   \arrowsrighthalffull{#1}{#2}{#3}
}

\newcommand{\phxbubble}[3]{
   \draw[lineWithArrowCenter] (#1,#2) to [out=45, in=135] (#1+1.2*#3,#2);
   \draw[lineWithArrowCenter] (#1+1.2*#3,#2) to [out=225, in=315] (#1,#2);
}

\newcommand{\phbubble}[3]{
   \draw[lineWithArrowCenter] (#1,#2) to [out=225, in=135] (#1,#2-1.2*#3);
   \draw[lineWithArrowCenter] (#1,#2-1.2*#3) to [out=45, in=315] (#1,#2);
}

\newcommand{\ppbubble}[3]{
   \draw[lineWithArrowCenter] (#1+1.2*#3,#2) to [out=45, in=135] (#1,#2);
   \draw[lineWithArrowCenter] (#1+1.2*#3,#2) to [out=225, in=315] (#1,#2);
}

\newcommand{\arrowslowerhalf}[2]{
   \def\shift{0.3};
   \coordinate (center) at (#1,#2);
   \draw[lineWithArrowCenterEnd] (center)  -- ($(center) + (-\shift,-\shift)$);
   \draw[lineWithArrowCenterEnd] ($(center) + (+\shift,-\shift)$) -- (center);
}

\newcommand{\arrowsupperhalf}[2]{
   \def\shift{0.3};
   \coordinate (center) at (#1,#2);
   \draw[lineWithArrowCenterEnd] ($(center)     + (-\shift,+\shift)$) -- (center);
   \draw[lineWithArrowCenterEnd] (center)    -- ($(center)   + (+\shift,+\shift)$);
}

\newcommand{\arrowslefthalfp}[3]{
   \def\shift{0.3};
   \coordinate (center) at (#1,#2);
   \draw[lineWithArrowCenterEnd] (center) -- ($(center) + (-\shift,-\shift)$);
   \draw[lineWithArrowCenterEnd] (center) to [out=45, in=180] ($(center) + (1.2*#3+\shift,2*\shift)$);
}

\newcommand{\arrowsrighthalfp}[3]{
   \def\shift{0.3};
   \coordinate (center) at (#1,#2);
   \draw[lineWithArrowCenterEnd] ($(center) + (+\shift,-\shift)$) -- (center);
   \draw[lineWithArrowCenterStart] ($(center) + (-1.2*#3-\shift,2*\shift)$) to [out=0, in=135] (center);
}

\newcommand{\threepointvertexleft}[4]{
   \draw[linePlain] (#2,#3+0.4*#4) -- (#2+0.6*#4,#3) -- (#2,#3-0.4*#4) -- (#2,#3+0.4*#4);
   \node at (#2+0.2*#4,#3) {#1};
}

\newcommand{\threepointvertexleftarrows}[4]{
   \threepointvertexleft{#1}{#2}{#3}{#4}
   \arrowslefthalffull{#2+0.4*#4}{#3}{4./3.*#4}
}

\newcommand{\threepointvertexright}[4]{
   \draw[linePlain] (#2,#3) -- (#2+0.6*#4,#3+0.4*#4) -- (#2+0.6*#4,#3-0.4*#4) -- (#2,#3);
   \node at (#2+0.4*#4,#3) {#1};
}

\newcommand{\threepointvertexrightarrows}[4]{
   \threepointvertexright{#1}{#2}{#3}{#4}
   \arrowsrighthalffull{#2+0.2*#4}{#3}{4./3.*#4}
}

\newcommand{\threepointvertexupper}[4]{
   \draw[linePlain] (#2-0.4*#4,#3+0.3*#4) -- (#2+0.4*#4,#3+0.3*#4) -- (#2,#3-0.3*#4) -- (#2-0.4*#4,#3+0.3*#4);
   \node at (#2,#3+0.1*#4) {#1};
}

\newcommand{\threepointvertexupperarrows}[4]{
   \threepointvertexupper{#1}{#2}{#3}{#4}
   \arrowsupperhalffull{#2}{#3-0.1*#4}{4./3.*#4}
}

\newcommand{\threepointvertexlower}[4]{
   \draw[linePlain] (#2-0.4*#4,#3-0.3*#4) -- (#2+0.4*#4,#3-0.3*#4) -- (#2,#3+0.3*#4) -- (#2-0.4*#4,#3-0.3*#4);
   \node at (#2,#3-0.1*#4) {#1};
}

\newcommand{\threepointvertexlowerarrows}[4]{
   \threepointvertexlower{#1}{#2}{#3}{#4}
   \arrowslowerhalffull{#2}{#3+0.1*#4}{4./3.*#4}
}

\newcommand{\bosonfull}[4]{
   \draw[bosonLine, very thick] (#1,#2) -- (#3,#4);
}

\newcommand{\fcirc}{\mathbin{\vcenter{\hbox{\scalebox{0.65}{$\bullet$}}}}}

\newcounter{subfigure}[figure]

\definecolor{darkgreen}{rgb}{0.0, 0.7, 0.0}

\usepackage{hyperref}
\hypersetup{colorlinks=true,breaklinks,linkcolor=blue,urlcolor=blue,citecolor=blue}


\begin{document}

\newcommand{\bfk}{\mathbf{k}}
\newcommand{\bfQ}{\mathbf{Q}}
\newcommand{\bfR}{\mathbf{R}}
\newcommand{\bfe}{\mathbf{e}}
\newcommand{\bfu}{\mathbf{u}}

\newcommand{\bfX}{\mathbf{X}}
\newcommand{\X}{\mathrm{X}}
\newcommand{\SC}{\mathrm{SC}}
\newcommand{\dSC}{\mathrm{dSC}}
\newcommand{\dM}{\mathrm{dM}}
\newcommand{\dD}{\mathrm{dD}}
\newcommand{\dX}{\mathrm{dX}}
\newcommand{\D}{\mathrm{D}}
\newcommand{\M}{\mathrm{M}}
\newcommand{\pp}{\mathrm{pp}}
\newcommand{\ph}{\mathrm{ph}}
\newcommand{\xph}{{\overline{\mathrm{ph}}}}

\newcommand{\dwave}{\mathrm{dw}}
\newcommand{\swave}{\mathrm{sw}}

\title{Multiloop functional renormalization group from single bosons}

\author{Kilian Fraboulet}
\email{k.fraboulet@fkf.mpg.de}
\affiliation{Max-Planck-Institut f{\"u}r Festk{\"o}rperforschung, Heisenbergstra{\ss}e 1, 70569 Stuttgart, Germany}
\affiliation{Institute of Information Systems Engineering, TU Wien, 1040 Vienna, Austria}
\affiliation{Institute for Theoretical Physics and Center for Quantum Science, Universit\"at T\"ubingen, Auf der Morgenstelle 14, 72076 T\"ubingen, Germany}

\author{Aiman Al-Eryani}
\affiliation{Institute for Theoretical Physics III, Ruhr-Universit\"at Bochum, 44801 Bochum, Germany}

\author{Sarah Heinzelmann}
\affiliation{Institute for Theoretical Physics and Center for Quantum Science, Universit\"at T\"ubingen, Auf der Morgenstelle 14, 72076 T\"ubingen, Germany}

\author{Anna Kauch}
\affiliation{Institute of Solid State Physics, TU Wien, 1040 Vienna, Austria}

\author{Sabine Andergassen}
\affiliation{Institute of Information Systems Engineering, TU Wien, 1040 Vienna, Austria}
\affiliation{Institute of Solid State Physics, TU Wien, 1040 Vienna, Austria}

\renewcommand{\i}{{\mathrm{i}}}
\date{ \today }

\begin{abstract}
The functional renormalization group (fRG) is an established tool in the treatment of correlated electron systems, notably for the description of competing instabilities. In recent years, methodological advancements led to the multiloop extension of the fRG, which systematically includes loop corrections beyond the conventional one-loop truncation and yields a quantitatively accurate description of two-dimensional lattice systems. At the same time, the single-boson exchange (SBE) decomposition of the two-particle vertex has been shown to offer both computational and interpretative advantages, paving the way to more affordable approximation schemes. We here apply their combination coined as multiloop SBE fRG to the two-dimensional Hubbard model at weak coupling. After providing a detailed account of the underlying formalism in physical channels, we analyze the results for the frequency- and momentum-dependent vertex functions. We find that the SBE approximation, i.e., neglecting the flow of the multiboson exchange contributions, accurately reproduces the parquet approximation at loop convergence. The presented algorithmic improvement opens the route for the treatment of more challenging parameter regimes and more realistic models.
\end{abstract}

\maketitle

\section{Introduction}
\label{sec:Introduction}

Quantum many-body approaches based on the diagrammatic resummation of the one- and two-particle vertices such as parquet solvers~\cite{Eckhardt2020,Krien2020,Astretsov2020,Astleitner2020,Rohringer2020,Krien2022} and functional renormalization group (fRG)~\cite{Wetterich1993,Ellwanger1994,Morris1994,Metzner2012,Dupuis2021} methods are widespread tools to investigate the interplay between competing instabilities in many-electron systems. In particular, the accurate treatment of two-particle vertices constitutes one of the main bottlenecks in their numerical implementations. Considerable progress can be obtained by using the recently introduced single-boson exchange (SBE) decomposition~\cite{Krien2019}. It relies on a diagrammatic classification based on the reducibility with respect to bare interaction vertices, instead of fermionic propagator lines as used in the well-known parquet decomposition~\cite{Dedominicis1964,Dedominicis1964A,BickersSelfConsistent2004,Yang2009,Tam2013,Rohringer2012,Valli2015,Kauch2019,Li2019}. This alternative decomposition allows one to circumvent the divergences of two-particle irreducible vertices that hinder the theoretical description of many-electron systems at intermediate-to-strong coupling, notably in parquet-based approaches~\cite{Schaefer2013,Janis2014,Gunnarsson2016,Schaefer2016,Ribic2016,Gunnarsson2017,Vucicevic2018,Chalupa2018,Springer2020,Chalupa2021}. The SBE decomposition can also be viewed as a bosonization technique, providing a more insightful formulation of quantum many-body approaches in terms of physically relevant degrees of freedom.

As a result, this SBE formulation provides a fruitful framework for efficient and insightful approximations, motivated from both diagrammatic arguments and physical understanding. This has led to a number of studies with various quantum many-body approaches: the dynamical mean-field theory (DMFT)~\cite{Metzner1989,Georges1996} in the original work formulating the SBE decomposition~\cite{Krien2019}, parquet solvers for the Anderson impurity model~\cite{Krien2019b} as well as lattice models~\cite{Krien2021b,Krien2022}, and combinations~\cite{Krien2020,Krien2020a} with the dual fermion approach~\cite{Rubtsov2008}. The interpretative advantage of the SBE decomposition has also been used in Ref.~\cite{Adler2024} to identify the origin of the breakdown of the self-consistent many-electron perturbation theory~\cite{Schaefer2013,Kozik2015,Gunnarsson2017}, considering DMFT results for the Hubbard model on the Bethe lattice and exact solutions of the Hubbard atom and the Anderson impurity model. Most recently, the SBE approach has been used to formulate a method termed embedded multiboson exchange (eMBEX), based on an embedded cluster and used to capture correlations beyond DMFT~\cite{Kiese2024}.

Aside from the SBE decomposition, several other bosonization approaches have been proposed and successfully applied to the study of many-electron systems in recent years. Notable examples include the fluctuating local field (FLF) technique~\cite{Rubtsov2018,Rubtsov2020,Lyakhova2022}, the dual boson approach~\cite{Rubtsov2012,Loon2014,Stepanov2016,Stepanov2016b,Vandelli2020,Krien2020a}, the TRILEX method~\cite{Ayral2015,Ayral2016,Vucicevik2017,Ayral2017}, and its combination with the dual boson framework, known as D-TRILEX~\cite{Stepanov2019,Harkov2021b,Vandelli2022}. These methods provide efficient and even nonperturbative treatments of many-electron models, but differ fundamentally from the SBE framework in their underlying construction. In particular, the dual boson approach and, consequently, the D-TRILEX method rely on vertices and correlation functions of an auxiliary dual theory, whereas the SBE approach is formulated directly in terms of physical vertices (i.e., the vertices of the original theory), thus making the physical interpretation more transparent. In practical applications, the D-TRILEX approach further relies on local fermion-boson vertices, whereas the SBE approach retains momentum dependence in all of its vertices. Furthermore, the FLF technique requires identifying \emph{a priori} a dominant ordering tendency in order to provide a particularly efficient and nonperturbative description of the corresponding collective fluctuations, whereas the SBE approach offers an unbiased treatment of competing orders by treating them on an equal footing. Owing to their different underlying philosophies, all these bosonization techniques offer distinct and complementary approximations worth studying.

For the fRG, several bosonization procedures have also already been proposed~\cite{Baier2004,Krahl2007,Diehl2007,Diehl2007_II,Strack2008,Bartosch2009,Friederich2010,Friederich2011,Obert2013,Denz2020}, building on the partial bosonization of the vertex function in the channel decomposition~\cite{Karrasch2008,Husemann2009,Wang2012,Vilardi2017,TagliaviniHille2019,Vilardi2019,Stepanov2019,Hille2020,Harkov2021} (as well as its use in parquet solvers). Here, we extend the one-loop ($1\ell$) fRG in the SBE formulation presented in Refs.~\cite{Bonetti2022,Fraboulet2022} to the multiloop SBE fRG and apply it to the two-dimensional (2D) Hubbard model~\footnote{See Ref.~\cite{Qin2021} for a recent overview of computational results for the 2D Hubbard model.} at weak coupling. In particular, we assess the quality of the SBE approximation that allows one to considerably reduce the numerical effort with respect to the conventional fermionic implementation based on the parquet decomposition and high-frequency asymptotics~\cite{Hille2020,Wentzell2016} and therefore access more challenging parameter regimes. For weak to intermediate couplings, we find quantitatively accurate results that depend only slightly on the choice of the cutoff function at loop convergence.

The paper is organized as follows: In Sec.~\ref{sec:Method}, we introduce the multiloop SBE fRG formalism, providing a brief review of the notational conventions, the SBE decomposition, and the flow equations both in diagrammatic and physical channels. In Sec.~\ref{sec:NumericalImplementation}, we present the application to the 2D Hubbard model together with its algorithmic implementation. We then present the results for the susceptibilities and the Yukawa couplings in Sec.~\ref{sec:Results}, both at half filling and finite doping, and provide additional results for the self-energy in an Appendix. In particular, we (1) discuss the multiloop convergence and the cutoff dependence in the SBE approximation, (2) assess the accuracy of this numerically very advantageous approximation by comparing to results from the parquet approximation, and (3) investigate the temperature dependence in the vicinity of the Van Hove singularity. We finally conclude with a summary and an outlook in Sec.~\ref{sec:Conclusion}.

\section{Multiloop SBE fRG formalism}
\label{sec:Method}

\subsection{Classical action and notational conventions}

\begin{figure}[t!]
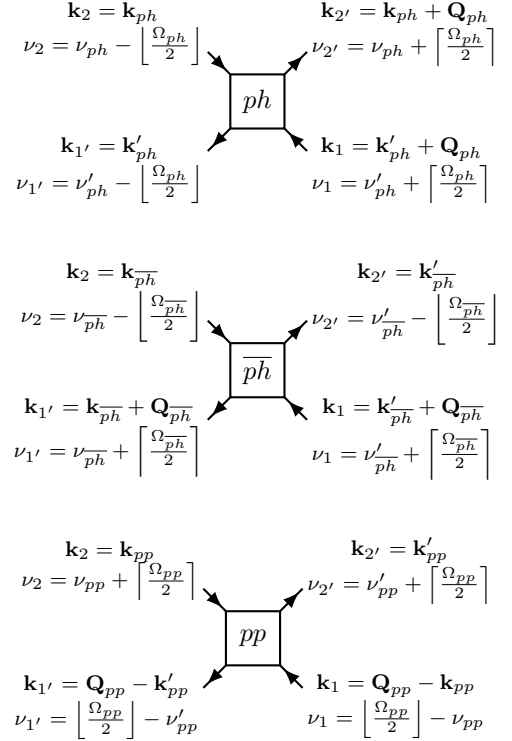

\centering
        \tikzm{A}{
                \fullvertexwithlegs{$ph$}{0}{0}{1.2}
                \node[left, align=center] (nu1) at (-0.6,0.9) {\footnotesize $\textbf{k}_2=\textbf{k}_{ph}$ \\
                        \footnotesize $\nu_2=\nu_{ph} - \left\lfloor \frac{\Omega_{ph}}{2} \right\rfloor$};
                \node[right, align=center] (nup1) at (0.6,0.9) {\footnotesize $\textbf{k}_{2^\prime}=\textbf{k}_{ph} + \textbf{Q}_{ph}$ \\
                        \footnotesize $\nu_{2^\prime}=\nu_{ph} + \left\lceil \frac{\Omega_{ph}}{2} \right\rceil$};
                \node[left, align=center] (nu2) at (-0.6,-0.9) 
                {\footnotesize $\textbf{k}_{1^\prime}=\textbf{k}^{\prime}_{ph}$ \\
                \footnotesize $\nu_{1^\prime}=\nu^{\prime}_{ph} - \left\lfloor \frac{\Omega_{ph}}{2} \right\rfloor$};
                \node[right, align=center] (nup2) at (0.6,-0.9) {\footnotesize$\textbf{k}_1=\textbf{k}^{\prime}_{ph} + \textbf{Q}_{ph}$ \\
    \footnotesize$\nu_1=\nu^{\prime}_{ph} + \left\lceil \frac{\Omega_{ph}}{2} \right\rceil$};
        }
        \\
        \vspace{0.5cm}
        \tikzm{B}{
                \fullvertexwithlegs{$\overline{ph}$}{0}{0}{1.2}
                \node[left, align=center] at (-0.6,0.9) {\footnotesize $\textbf{k}_2=\textbf{k}_{\overline{ph}}$ \\
        \footnotesize$\nu_2=\nu_{\overline{ph}} - \left\lfloor \frac{\Omega_{\overline{ph}}}{2 } \right\rfloor$};
                \node[right, align=center] at (0.6,0.9) {\footnotesize $\textbf{k}_{2^\prime}=\textbf{k}^{\prime}_{\overline{ph}}$ \\
                        \footnotesize$\nu_{2^\prime}=\nu^{\prime}_{\overline{ph}} - \left\lfloor \frac{\Omega_{\overline{ph}}}{2} \right\rfloor$};
                \node[left, align=center] at (-0.6,-0.9) {
                \footnotesize $\textbf{k}_{1^\prime}=\textbf{k}_{\overline{ph}} + \textbf{Q}_{\overline{ph}}$ \\
                \footnotesize $\nu_{1^\prime}=\nu_{\overline{ph}} + \left\lceil
                        \frac{\Omega_{\overline{ph}}}{2} \right\rceil$};
                \node[right, align=center] at (0.6,-0.9) {
                 \footnotesize $\textbf{k}_1=\textbf{k}^{\prime}_{\overline{ph}} + \textbf{Q}_{\overline{ph}}$ \\
                \footnotesize $\nu_1=\nu^{\prime}_{\overline{ph}} + \left\lceil
                        \frac{\Omega_{\overline{ph}}}{2} \right\rceil$ };
        }\\
        \vspace{0.5cm}
        \tikzm{C}{
                \fullvertexwithlegs{$pp$}{0}{0}{1.2}
                \node[left, align=center] at (-0.6,0.9) {\footnotesize $\textbf{k}_2=\textbf{k}_{pp}$ \\
                         \footnotesize $\nu_2=\nu_{pp} + \left\lceil \frac{\Omega_{pp}}{2} \right\rceil$};
                \node[right, align=center] at (0.6,0.9) {\footnotesize $\textbf{k}_{2^\prime}=\textbf{k}^{\prime}_{pp}$ \\
                         \footnotesize $\nu_{2^\prime}=\nu^{\prime}_{pp} + \left\lceil \frac{\Omega_{pp}}{2} \right\rceil$};
                \node[left, align=center] at (-0.6,-0.9) {
                \footnotesize $\textbf{k}_{1^\prime}=\textbf{Q}_{pp} - \textbf{k}^{\prime}_{pp}$ \\
                \footnotesize $\nu_{1^\prime}=\left\lfloor \frac{\Omega_{pp}}{2}
                        \right\rfloor - \nu^{\prime}_{pp}$};
                \node[right, align=center] at (0.6,-0.9) {\footnotesize $\textbf{k}_1=\textbf{Q}_{pp} - \textbf{k}_{pp}$ \\
                \footnotesize $\nu_{1}=\left\lfloor \frac{\Omega_{pp}}{2}
                        \right\rfloor - \nu_{pp}$};
        }
    \caption{Frequency and momentum conventions for the two-particle vertex in the different channel notations, where $\lceil ... \rceil$ ($\lfloor ... \rfloor$) rounds its argument up (down) to the nearest bosonic Matsubara frequency.}
    \label{fig:FrequencyMomentumParametrization}
\end{figure}

The flow equations underlying the multiloop SBE fRG approach have been derived recently in diagrammatic channels for generic fermionic models~\cite{Gievers2022}. We will explain in this section how to determine the multiloop SBE fRG equations in physical channels for translationally invariant fermionic systems relying on energy conservation and SU(2) spin symmetry, which includes, in particular, the 2D Hubbard model that we will focus on in our subsequent numerical study. Similarly to Ref.~\cite{Gievers2022}, we consider a fermionic system with the classical action
\begin{equation}
S[\overline{c},c] = - \overline{c}_{1^\prime} G_{0;1^\prime|1}^{-1} c_{1} -  \frac{1}{4} U_{1^\prime 2^\prime|12} \overline{c}_{1^\prime} \overline{c}_{2^\prime} c_{2} c_{1},
\label{eq:ClassicalActionS}
\end{equation}
where $G_0$ and $U$ are, respectively, the bare propagator and the bare interaction, with $U$ satisfying crossing symmetry (i.e., $U_{1^\prime 2^\prime|12} = -U_{2^\prime 1^\prime|12} = -U_{1^\prime 2^\prime|21} = U_{2^\prime 1^\prime|21}$). The general index $i$ encompasses all indices labeling the Grassmann field $c_i$ (or its conjugate $\overline{c}_i$). We consider here that it includes a spatial momentum $\mathbf{k}$, a Matsubara frequency $\nu$, and a spin index $\sigma$, i.e., $i = (\mathbf{k}_i,\nu_i,\sigma_i)$ or $i=(k_i,\sigma_i)$, with $k_i=(\mathbf{k}_i,\nu_i)$. Repeated indices are assumed to be integrated or summed over. Energy conservation and translational invariance impose frequency and momentum conservation laws, which translate for $G_0$ and $U$ into the relations
\begin{subequations}
\begin{align}
    G_{0;1^\prime|1} & = G_{0;\sigma_{1^\prime}|\sigma_1}(k_{1^\prime}|k_1) \nonumber \\
    & = G_{0;\sigma_{1^\prime}|\sigma_1}(k_1) \delta_{k_{1^\prime},k_1} , \label{eq:FreqMomConservationLawsG0} \\
    U_{1^\prime 2^\prime|12} & = U_{\sigma_{1^\prime}\sigma_{2^\prime}|\sigma_1\sigma_2}(k_{1^\prime},k_{2^\prime}|k_1, k_2) \nonumber \\
    & = U_{\sigma_{1^\prime}\sigma_{2^\prime}|\sigma_1\sigma_2}(Q_r,k_r,k^\prime_{r}) \delta_{k_{1^\prime}+k_{2^\prime},k_1+k_2}. \label{eq:FreqMomConservationLawsU}
\end{align}
\label{eq:FreqMomConservationLawsG0U}
\end{subequations}
The full propagator $G$ and the two-particle vertex $V$ satisfy equations identical to Eqs.~\eqref{eq:FreqMomConservationLawsG0} and~\eqref{eq:FreqMomConservationLawsU}, respectively. We parametrize the frequency and momentum dependencies of two-particle objects, like the bare interaction $U$ or the two-particle vertex $V$, in terms of bosonic momenta $Q_r=(\textbf{Q}_r,\Omega_r)$ and corresponding fermionic momenta $k^{(\prime)}_r=\Big(\textbf{k}^{(\prime)}_r,\nu^{(\prime)}_r\Big)$ that are defined differently depending on the diagrammatic channel $r$ under consideration, i.e., the particle-hole ($ph$), the particle-hole crossed ($\overline{ph}$), or the particle-particle ($pp$) channel. Our definitions of $Q_r=(\textbf{Q}_r,\Omega_r)$ and $k^{(\prime)}_r=\Big(\textbf{k}^{(\prime)}_r,\nu^{(\prime)}_r\Big)$ are given in Fig.~\ref{fig:FrequencyMomentumParametrization} for $r=ph,\overline{ph},pp$.

\begin{figure*}[t!]
\centering
\refstepcounter{subfigure}\label{fig:ParquetVsSBEdecompositions_a}
\textbf{(\alph{subfigure})}

\vspace{-0.4cm}

	\begin{gather*}
		\phi_{ph} = \ \tikzm{Diag_phi_ph_1}{
				\arrowsupperhalf{0}{0.45}
				\arrowslowerhalf{0}{-0.45}
				\phbubble{0}{0.45}{0.75}
				\barevertex{0}{0.45}
				\barevertex{0}{-0.45}
				\draw[blue,very thick,densely dashed] (-0.5,0.0) -- (0.5,0.0);
			} \ + \ \tikzm{Diag_phi_ph_2}{
				\arrowsupperhalffull{0}{0}{1.5}
				\phxbubble{-0.45}{0.45}{0.75}
				\arrowslowerhalf{0}{-0.45}
				\draw[lineWithArrowCenter] (-0.45,0.45) to [out=225, in=135] (0,-0.45);
				\draw[lineWithArrowCenter] (0,-0.45) to [out=45, in=315] (0.45,0.45);
				\barevertex{-0.45}{0.45}
				\barevertex{0.45}{0.45}
				\barevertex{0}{-0.45}
				\draw[blue,very thick,densely dashed] (-0.8,-0.05) -- (0.8,-0.05);
			} \ + \ \tikzm{Diag_phi_ph_3}{
			\arrowsupperhalffull{0}{0}{1.5}
			\arrowslowerhalffull{0}{0}{1.5}
			\phxbubble{-0.45}{0.45}{0.75}
			\phxbubble{-0.45}{-0.45}{0.75}
			\draw[lineWithArrowCenter] (-0.45,0.45) to [out=225, in=135] (-0.45,-0.45);
			\draw[lineWithArrowCenter] (0.45,-0.45) to [out=45, in=315] (0.45,0.45);
			\barevertex{0.45}{0.45}
			\barevertex{0.45}{-0.45}
			\barevertex{-0.45}{0.45}
			\barevertex{-0.45}{-0.45}
			\draw[blue,very thick,densely dashed] (-0.9,-0.) -- (0.9,0.);
		} \ + \dots \hspace{0.7cm} \phi_{\overline{ph}} = \ \tikzm{Diag_phi_phx_1}{
				\arrowslefthalf{0}{0}
				\phxbubble{0}{0}{0.75}
				\arrowsrighthalf{0.9}{0}
				\barevertex{0}{0}
				\barevertex{0.9}{0}
				\draw[blue,very thick,densely dashed] (0.45,-0.5) -- (0.45,0.5);
			} \ + \ \tikzm{Diag_phi_phx_2}{
				\arrowslefthalffull{0.45}{0}{1.5}
				\phbubble{0}{0.45}{0.75}
				\draw[lineWithArrowCenter] (0,0.45) to [out=45, in=135] (0.9,0);
				\draw[lineWithArrowCenter] (0.9,0) to [out=225, in=315] (0,-0.45);
				\arrowsrighthalf{0.9}{0};
				\barevertex{0}{0.45}
				\barevertex{0}{-0.45}
				\barevertex{0.9}{0}
				\draw[blue,very thick,densely dashed] (0.5,-0.8) -- (0.5,0.8);
			} \ + \ \tikzm{Diag_phi_phx_3}{
			\phbubble{-0.45}{0.45}{0.75}
			\phbubble{0.45}{0.45}{0.75}
			\draw[lineWithArrowCenter] (-0.45,0.45) to [out=45, in=135] (0.45,0.45);
			\draw[lineWithArrowCenter] (0.45,-0.45) to [out=225, in=315] (-0.45,-0.45);
			\arrowsallfull{0}{0}{1.5}
			\barevertex{0.45}{0.45}
			\barevertex{0.45}{-0.45}
			\barevertex{-0.45}{0.45}
			\barevertex{-0.45}{-0.45}
			\draw[blue,very thick,densely dashed] (-0.,-0.9) -- (0.,0.9);
		} \ + \dots \\
		\phi_{pp} = \ \tikzm{Diag_phi_pp_1}{
				\arrowslefthalfp{0}{0}{0.75}
				\arrowsrighthalfp{0.9}{0}{0.75}
				\node at (0,0.6) {};
				\ppbubble{0}{0}{0.75}
				\node at (0.9,0.6) {};
				\barevertex{0}{0}
				\barevertex{0.9}{0}
				\draw[blue,very thick,densely dashed] (0.47,-0.5) -- (0.47,0.5);
			} \ + \ \tikzm{Diag_phi_pp_2}{
				\draw[lineWithArrowCenterEnd] (1.2,-0.3) -- (0.9,0);
				\draw[lineWithArrowCenterEnd] (0,-0.45) -- (-0.3,-0.75);
				\draw[lineWithArrowCenterEnd] (0,0.45) to [out=60, in=180] (1.2,0.85);
				\draw[lineWithArrowCenterStart] (-0.3,0.85) to [out=0, in=135] (0.9,0);
				\phbubble{0}{0.45}{0.75}
				\draw[lineWithArrowCenter] (0.9,0) .. controls ++(45:0.7) and ++(135:0.7) .. (0,0.45);
				\draw[lineWithArrowCenter] (0.9,0) to [out=225, in=315] (0,-0.45);
				\barevertex{0}{0.45}
				\barevertex{0}{-0.45}
				\barevertex{0.9}{0}
				\draw[blue,very thick,densely dashed] (0.5,-0.8) -- (0.5,0.8);
			} \ + \ \tikzm{Diag_phi_pp_3}{
			\arrowslowerhalffull{0}{0}{1.5}
			\draw[lineWithArrowCenterEnd] (-0.45,0.45) to [out=60, in=180] (0.65,1.05);
			\node at (0.65,1.05) {};
			\draw[lineWithArrowCenterStart] (-0.65,1.05) to [out=0, in=120] (0.45,0.45);
			\node at (-0.65,1.05) {};
			\phbubble{-0.45}{0.45}{0.75}
			\phbubble{0.45}{0.45}{0.75}
			\draw[lineWithArrowCenter] (0.45,0.45) .. controls ++(45:0.5) and ++(135:0.5) .. (-0.45,0.45);
			\draw[lineWithArrowCenter] (0.45,-0.45) to [out=225, in=315] (-0.45,-0.45);
			\barevertex{0.45}{0.45}
			\barevertex{0.45}{-0.45}
			\barevertex{-0.45}{0.45}
			\barevertex{-0.45}{-0.45}
			\draw[blue,very thick,densely dashed] (-0.,-0.9) -- (-0.,0.9);
		} \ + \dots \\[8pt]
            I^{\text{2PI}} = \ \tikzm{I2PIbarevertex}{
			\barevertexwithlegs{0}{0}
		} \ + \ \tikzm{I2PInontrivialdiagram}{
			\draw[lineArrowBis] (-0.4,0.4) -- (0.4,-0.4);
			\draw[lineArrowBis] (-0.4,-0.4) -- (0.4,0.4);
			\draw[lineArrow] (0.4,0.4) -- (-0.4,0.4);
			\draw[lineArrow] (0.4,-0.4) -- (-0.4,-0.4);
			\draw[lineArrow] (0.4,-0.4) -- (0.4,0.4);
			\draw[lineArrow] (-0.4,0.4) -- (-0.4,-0.4);
			\barevertex{-0.4}{0.4}
			\barevertex{-0.4}{-0.4}
			\barevertex{0.4}{0.4}
			\barevertex{0.4}{-0.4}
			\arrowsallfull{0}{0}{4./3.}
		} \ + \dots
	\end{gather*}
    
\vspace{0.5cm}

\refstepcounter{subfigure}\label{fig:ParquetVsSBEdecompositions_b}
  \textbf{(\alph{subfigure})}
	\begin{gather*}
		\nabla_{ph} = \ \tikzm{nablaphbarevertex}{
			\barevertexwithlegs{0}{0}
                \draw[red,very thick,densely dotted] (-0.5,0) -- (0.5,0);
		} \ + \ \tikzm{Diag_nabla_ph_1}{
				\arrowsupperhalf{0}{0.45}
				\arrowslowerhalf{0}{-0.45}
				\phbubble{0}{0.45}{0.75}
				\barevertex{0}{0.45}
				\barevertex{0}{-0.45}
                    \draw[red,very thick,densely dotted] (-0.5,0.45) -- (0.5,0.45);
				\draw[red,very thick,densely dotted] (-0.5,-0.45) -- (0.5,-0.45);
			} \ + \ \tikzm{Diag_nabla_ph_2}{
				\arrowsupperhalffull{0}{0}{1.5}
				\phxbubble{-0.45}{0.45}{0.75}
				\arrowslowerhalf{0}{-0.45}
				\draw[lineWithArrowCenter] (-0.45,0.45) to [out=225, in=135] (0,-0.45);
				\draw[lineWithArrowCenter] (0,-0.45) to [out=45, in=315] (0.45,0.45);
				\barevertex{-0.45}{0.45}
				\barevertex{0.45}{0.45}
				\barevertex{0}{-0.45}
				\draw[red,very thick,densely dotted] (-0.5,-0.45) -- (0.5,-0.45);
			} \ + \dots \hspace{0.7cm} \nabla_{\overline{ph}} = \ \tikzm{nablaphxbarevertex}{
			\barevertexwithlegs{0}{0}
                \draw[red,very thick,densely dotted] (0,-0.5) -- (0,0.5);
		} \ + \ \tikzm{Diag_nabla_phx_1}{
				\arrowslefthalf{0}{0}
				\phxbubble{0}{0}{0.75}
				\arrowsrighthalf{0.9}{0}
				\barevertex{0}{0}
				\barevertex{0.9}{0}
                    \draw[red,very thick,densely dotted] (0,-0.5) -- (0,0.5);
				\draw[red,very thick,densely dotted] (0.9,-0.5) -- (0.9,0.5);
			} \ + \ \tikzm{Diag_nabla_phx_2}{
				\arrowslefthalffull{0.45}{0}{1.5}
				\phbubble{0}{0.45}{0.75}
				\draw[lineWithArrowCenter] (0,0.45) to [out=45, in=135] (0.9,0);
				\draw[lineWithArrowCenter] (0.9,0) to [out=225, in=315] (0,-0.45);
				\arrowsrighthalf{0.9}{0};
				\barevertex{0}{0.45}
				\barevertex{0}{-0.45}
				\barevertex{0.9}{0}
				\draw[red,very thick,densely dotted] (0.9,-0.5) -- (0.9,0.5);
			} \ + \dots \\
		\nabla_{pp} = \ \tikzm{nablappbarevertex}{
			\barevertexwithlegs{0}{0}
                \draw[red,very thick,densely dotted] (0,-0.5) -- (0,0.5);
		} \ + \ \tikzm{Diag_nabla_pp_1}{
				\arrowslefthalfp{0}{0}{0.75}
				\arrowsrighthalfp{0.9}{0}{0.75}
				\node at (0,0.6) {};
				\ppbubble{0}{0}{0.75}
				\node at (0.9,0.6) {};
				\barevertex{0}{0}
				\barevertex{0.9}{0}
                    \draw[red,very thick,densely dotted] (0,-0.5) -- (0,0.5);
				\draw[red,very thick,densely dotted] (0.9,-0.5) -- (0.9,0.5);
			} \ + \ \tikzm{Diag_nabla_pp_2}{
				\draw[lineWithArrowCenterEnd] (1.2,-0.3) -- (0.9,0);
				\draw[lineWithArrowCenterEnd] (0,-0.45) -- (-0.3,-0.75);
				\draw[lineWithArrowCenterEnd] (0,0.45) to [out=60, in=180] (1.2,0.85);
				\draw[lineWithArrowCenterStart] (-0.3,0.85) to [out=0, in=135] (0.9,0);
				\phbubble{0}{0.45}{0.75}
				\draw[lineWithArrowCenter] (0.9,0) .. controls ++(45:0.7) and ++(135:0.7) .. (0,0.45);
				\draw[lineWithArrowCenter] (0.9,0) to [out=225, in=315] (0,-0.45);
				\barevertex{0}{0.45}
				\barevertex{0}{-0.45}
				\barevertex{0.9}{0}
				\draw[red,very thick,densely dotted] (0.9,-0.5) -- (0.9,0.5);
			} \ + \dots \\
            M_{ph} = \ \tikzm{Diag_M_ph}{
			\arrowsupperhalffull{0}{0}{1.5}
			\arrowslowerhalffull{0}{0}{1.5}
			\phxbubble{-0.45}{0.45}{0.75}
			\phxbubble{-0.45}{-0.45}{0.75}
			\draw[lineWithArrowCenter] (-0.45,0.45) to [out=225, in=135] (-0.45,-0.45);
			\draw[lineWithArrowCenter] (0.45,-0.45) to [out=45, in=315] (0.45,0.45);
			\barevertex{0.45}{0.45}
			\barevertex{0.45}{-0.45}
			\barevertex{-0.45}{0.45}
			\barevertex{-0.45}{-0.45}
		} \ + \dots \hspace{0.7cm} M_{\overline{ph}} = \ \tikzm{Diag_M_phx}{
			\phbubble{-0.45}{0.45}{0.75}
			\phbubble{0.45}{0.45}{0.75}
			\draw[lineWithArrowCenter] (-0.45,0.45) to [out=45, in=135] (0.45,0.45);
			\draw[lineWithArrowCenter] (0.45,-0.45) to [out=225, in=315] (-0.45,-0.45);
			\arrowsallfull{0}{0}{1.5}
			\barevertex{0.45}{0.45}
			\barevertex{0.45}{-0.45}
			\barevertex{-0.45}{0.45}
			\barevertex{-0.45}{-0.45}
		} \ + \dots \hspace{0.7cm} M_{pp} = \ \tikzm{Diag_nabla_pp}{
			\arrowslowerhalffull{0}{0}{1.5}
			\draw[lineWithArrowCenterEnd] (-0.45,0.45) to [out=60, in=180] (0.65,1.05);
			\node at (0.65,1.05) {};
			\draw[lineWithArrowCenterStart] (-0.65,1.05) to [out=0, in=120] (0.45,0.45);
			\node at (-0.65,1.05) {};
			\phbubble{-0.45}{0.45}{0.75}
			\phbubble{0.45}{0.45}{0.75}
			\draw[lineWithArrowCenter] (0.45,0.45) .. controls ++(45:0.5) and ++(135:0.5) .. (-0.45,0.45);
			\draw[lineWithArrowCenter] (0.45,-0.45) to [out=225, in=315] (-0.45,-0.45);
			\barevertex{0.45}{0.45}
			\barevertex{0.45}{-0.45}
			\barevertex{-0.45}{0.45}
			\barevertex{-0.45}{-0.45}
		} \ + \dots
	\end{gather*}
    \caption{Illustration of the diagrammatic criteria underlying the parquet and the SBE decompositions of the two-particle vertex $V$. {(a)} Different contributions to the parquet decomposition~\eqref{eq:parquetDecomposition}, which relies on two-particle reducibility. The vertices $\phi_r$ include all diagrams that are 2PR in channel $r$, whereas the diagrams that are 2PI (i.e., not 2PR in any channel) are contained in $I^{\text{2PI}}$. Hence, by definition, the diagrams contributing to $\phi_r$ can be split into two disconnected parts by cutting a $\Pi_r$ bubble, as indicated by the dashed blue lines. The $\Pi_{ph}$, $\Pi_{\overline{ph}}$, and $\Pi_{pp}$ bubbles include two transverse antiparallel, two antiparallel, and two parallel lines, respectively. {(b)} Different contributions to the SBE decomposition~\eqref{eq:SBEDecomposition}, which relies on $U$ reducibility. In the SBE formalism, the 2PR vertices $\phi_r$ are split into $U$-reducible and $U$-irreducible objects according to the relation $\phi_r = \nabla_r + M_r -U$. The vertices $\nabla_r$ contain all diagrams that are 2PR and $U$ reducible in channel $r$, i.e., all diagrams that can be split into two disconnected parts by cutting a bare interaction $U$ connected to a $\Pi_r$ bubble, as marked by the red dotted lines. The 2PR diagrams in channel $r$ that are not $U$ reducible are assigned to the SBE rest function $M_r$.
    }
    \label{fig:ParquetVsSBEdecompositions}
\end{figure*}

\begin{figure}[t!]
    \centering
    \adjustbox{max width=0.5\textwidth, scale=0.75}
    {\includegraphics{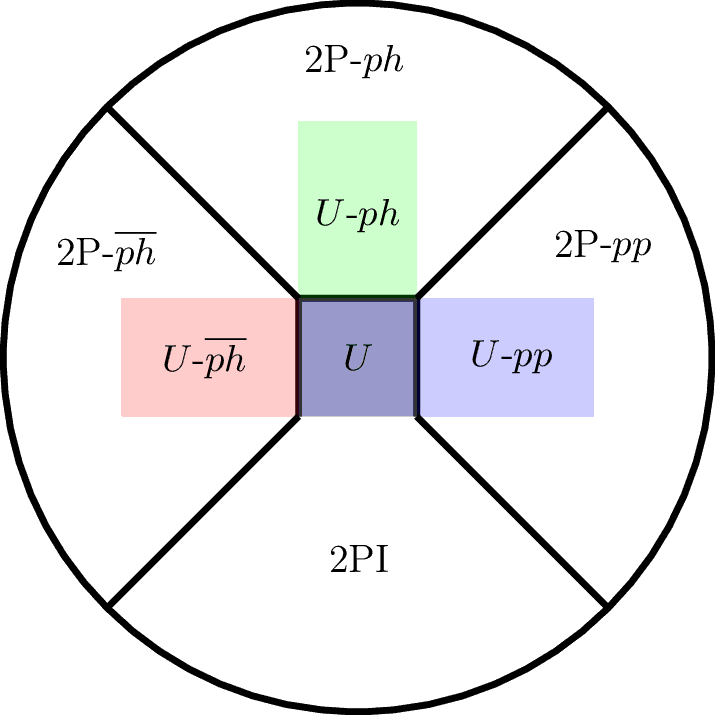}}
    \caption{Venn diagram illustrating the connection between 2PR and $U$-reducible diagrams. All $U$-reducible diagrams in a given channel $r$ (i.e., all $U$-$r$ diagrams) are also 2PR in the same channel $r$ (i.e., are also 2P-$r$), with the exception of the diagram of the bare interaction $U$, which is both 2PI and $U$ reducible in all channels. Figure adapted from Ref.~\cite{Krien2019}.}
    \label{fig:VennDiagram}
\end{figure}

\begin{figure*}[t!]
\begin{align*}
\tikzm{SBEdecomposition_V}{
			\fullvertexwithlegs{$V$}{0}{0}{1}
			\node[above, align=center] at (-0.6,0.6) {\footnotesize $\textbf{k}_2$ \\
                        \footnotesize $\nu_2$};
			\node[above, align=center] at (0.6,0.6) {\footnotesize $\textbf{k}^\prime_2$ \\
                        \footnotesize $\nu^\prime_2$};
			\node[below, align=center] at (-0.6,-0.6) {\footnotesize $\textbf{k}^\prime_1$ \\
                        \footnotesize $\nu^\prime_1$};
			\node[below, align=center] at (0.6,-0.6) {\footnotesize $\textbf{k}_1$ \\
                        \footnotesize $\nu_1$};
		}
		& = \ \tikzm{SBEdecomposition_nablaph}{
				\threepointvertexupperarrows{$\bar\lambda_{ph}$}{0}{0.82}{1.4}
				\bosonfull{0}{0.4}{0}{-0.4}
				\node at (0.4,0) {$w_{ph}$};
				\threepointvertexlowerarrows{$\lambda_{ph}$}{0}{-0.82}{1.4}
				\node[left, align=center] at (-0.3,2.1) {\footnotesize $\textbf{k}_2=\textbf{k}_{ph}$ \\
                        \footnotesize $\nu_2=\nu_{ph} - \left\lfloor \frac{\Omega_{ph}}{2} \right\rfloor$};
				\node[right, align=center] at (0.3,2.1) {\footnotesize $\textbf{k}_{2^\prime}=\textbf{k}_{ph} + \textbf{Q}_{ph}$ \\
                        \footnotesize $\nu_{2^\prime}=\nu_{ph} + \left\lceil \frac{\Omega_{ph}}{2} \right\rceil$};
				\node[left, align=center] at (-0.3,-2.1) {\footnotesize $\textbf{k}_{1^\prime}=\textbf{k}^{\prime}_{ph}$ \\
                \footnotesize $\nu_{1^\prime}=\nu^{\prime}_{ph} - \left\lfloor \frac{\Omega_{ph}}{2} \right\rfloor$};
				\node[right, align=center] at (0.3,-2.1) {\footnotesize$\textbf{k}_1=\textbf{k}^{\prime}_{ph} + \textbf{Q}_{ph}$ \\
    \footnotesize$\nu_1=\nu^{\prime}_{ph} + \left\lceil \frac{\Omega_{ph}}{2} \right\rceil$};
			}
		\ + \ \tikzm{SBEdecomposition_nablaphx}{\threepointvertexleftarrows{$\bar\lambda_{\overline{ph}}$}{-0.18}{0}{1.4}
				\bosonfull{0.66}{0}{1.46}{0}
				\node at (1.06,0.3) {$w_{\overline{ph}}$};
				\threepointvertexrightarrows{$\lambda_{\overline{ph}}$}{1.46}{0}{1.4}
				\node[above, align=center] at (-0.58,0.9) {\footnotesize $\textbf{k}_2=\textbf{k}_{\overline{ph}}$ \\
        \footnotesize$\nu_2=\nu_{\overline{ph}} - \left\lfloor \frac{\Omega_{\overline{ph}}}{2 } \right\rfloor$};
				\node[above, align=center] at (2.7,0.9) {\footnotesize $\textbf{k}_{2^\prime}=\textbf{k}^{\prime}_{\overline{ph}}$ \\
                        \footnotesize$\nu_{2^\prime}=\nu^{\prime}_{\overline{ph}} - \left\lfloor \frac{\Omega_{\overline{ph}}}{2} \right\rfloor$};
				\node[below, align=center] at (-0.58,-0.9) {
                \footnotesize $\textbf{k}_{1^\prime}=\textbf{k}_{\overline{ph}} + \textbf{Q}_{\overline{ph}}$ \\
                \footnotesize $\nu_{1^\prime}=\nu_{\overline{ph}} + \left\lceil
                        \frac{\Omega_{\overline{ph}}}{2} \right\rceil$};
				\node[below, align=center] at (2.7,-0.9) {
                 \footnotesize $\textbf{k}_1=\textbf{k}^{\prime}_{\overline{ph}} + \textbf{Q}_{\overline{ph}}$ \\
                \footnotesize $\nu_1=\nu^{\prime}_{\overline{ph}} + \left\lceil
                        \frac{\Omega_{\overline{ph}}}{2} \right\rceil$ };
			} \\
            & \\
            & \phantom{=} + \ \tikzm{SBEdecomposition_nablapp}{
				\threepointvertexleft{$\bar\lambda_{pp}$}{-0.18}{0}{1.4}
				\bosonfull{0.66}{0}{1.46}{0}
				\node at (1.06,0.3) {$w_{pp}$};
				\node at (1.06,0.74) {};
				\threepointvertexright{$\lambda_{pp}$}{1.46}{0}{1.4}
				\draw[lineWithArrowCenterEnd] (-0.18,-0.56) -- (-0.48,-0.86);
				\draw[lineWithArrowCenterEnd] (2.6,-0.86) -- (2.3,-0.56);
				\draw[lineWithArrowCenterEnd] (-0.18,0.56) to [out=20, in=180] (2.6,0.86);
				\draw[lineWithArrowCenterStart] (-0.48,0.86) to [out=0, in=160] (2.3,0.56);
				\node[above, align=center] at (-0.58,0.9) {\footnotesize $\textbf{k}_2=\textbf{k}_{pp}$ \\
                         \footnotesize $\nu_2=\nu_{pp} + \left\lceil \frac{\Omega_{pp}}{2} \right\rceil$};
				\node[above, align=center] at (2.7,0.9) {\footnotesize $\textbf{k}_{2^\prime}=\textbf{k}^{\prime}_{pp}$ \\
                         \footnotesize $\nu_{2^\prime}=\nu^{\prime}_{pp} + \left\lceil \frac{\Omega_{pp}}{2} \right\rceil$};
				\node[below, align=center] at (-0.58,-0.9) {
                \footnotesize $\textbf{k}_{1^\prime}=\textbf{Q}_{pp} - \textbf{k}^{\prime}_{pp}$ \\
                \footnotesize $\nu_{1^\prime}=\left\lfloor \frac{\Omega_{pp}}{2}
                        \right\rfloor - \nu^{\prime}_{pp}$};
				\node[below, align=center] at (2.7,-0.9) {\footnotesize $\textbf{k}_1=\textbf{Q}_{pp} - \textbf{k}_{pp}$ \\
                \footnotesize $\nu_{1}=\left\lfloor \frac{\Omega_{pp}}{2}
                        \right\rfloor - \nu_{pp}$};
			}
		\ + \ \tikzm{SBEdecomposition_IUirr}{
			\fullvertexwithlegs{$\mathcal{I}^{U\text{irr}}$}{0}{0}{1.35}
			\node[above, align=center] at (-0.7,0.7) {\footnotesize $\textbf{k}_2$ \\
                        \footnotesize $\nu_2$};
			\node[above, align=center] at (0.7,0.7) {\footnotesize $\textbf{k}^\prime_2$ \\
                        \footnotesize $\nu^\prime_2$};
			\node[below, align=center] at (-0.7,-0.7) {\footnotesize $\textbf{k}^\prime_1$ \\
                        \footnotesize $\nu^\prime_1$};
			\node[below, align=center] at (0.7,-0.7){\footnotesize $\textbf{k}_1$ \\
                        \footnotesize $\nu_1$};
		}
            \ -2 \ 
            \tikzm{SBE_decomposition_Gamma0}{
			\barevertexwithlegs{0}{0}
                \node[above, align=center] at (-0.35,0.3) {\footnotesize $\textbf{k}_2$ \\
                        \footnotesize $\nu_2$};
			\node[above, align=center] at (0.35,0.3) {\footnotesize $\textbf{k}^\prime_2$ \\
                        \footnotesize $\nu^\prime_2$};
			\node[below, align=center] at (-0.35,-0.3) {\footnotesize $\textbf{k}^\prime_1$ \\
                        \footnotesize $\nu^\prime_1$};
			\node[below, align=center] at (0.35,-0.3){\footnotesize $\textbf{k}_1$ \\
                        \footnotesize $\nu_1$};
		} \\
            & \\
            & = \left[\nabla_{ph}(Q_{ph},k_{ph},k^\prime_{ph})+\nabla_{\overline{ph}}(Q_{\overline{ph}},k_{\overline{ph}},k^\prime_{\overline{ph}})+\nabla_{pp}(Q_{pp},k_{pp},k^\prime_{pp})\right]\delta_{k_{1^\prime}+k_{2^\prime},k_1+k_2} \\
            & \phantom{=} + \mathcal{I}^{U\text{irr}}(k_{1^\prime},k_{2^\prime}|k_1, k_2) -2 U(k_{1^\prime},k_{2^\prime}|k_1, k_2)
        \end{align*}
\caption{Diagrammatic representation of the SBE decomposition expressed by Eqs.~\eqref{eq:SBEDecomposition2} and~\eqref{eq:nablaSBE}, using the frequency and momentum notations defined in Fig.~\ref{fig:FrequencyMomentumParametrization}.
}
\label{fig:SBEdecomposition}
\end{figure*}

\subsection{Parquet decomposition}

The diagrammatic channels are at the heart of the parquet decomposition of the two-particle vertex $V$, which is widely used in quantum many-body approaches, notably in extensions of DMFT~\cite{BickersSelfConsistent2004,Rohringer2018}. The parquet decomposition can be written as follows:
\begin{equation}
    V = \phi_{ph} + \phi_{\overline{ph}} + \phi_{pp} + I^{\text{2PI}},
    \label{eq:parquetDecomposition}
\end{equation}
where the vertices $\phi_r$ contain all \emph{two-particle reducible} (2PR) diagrams contributing to $V$ in each diagrammatic channel $r$, whereas $I^{\text{2PI}}$ comprises the \emph{two-particle irreducible} (2PI) ones. By definition, 2PR diagrams can be split into two disconnected parts by cutting two propagator lines, which are referred to as bubble [see Eqs.~\eqref{eq:Bubbles}]: A 2PR diagram is then assigned to a given vertex $\phi_r$ depending on the nature of that bubble [see Fig.~\ref{fig:ParquetVsSBEdecompositions}(\ref{fig:ParquetVsSBEdecompositions_a}) for more details and concrete examples]. Finally, we note that, by construction, $I^{\text{2PI}}$ is the sum of all diagrams of $V$ that are not 2PR, i.e., 2PI.

It can be shown that the vertices $\phi_r$ satisfy the Bethe-Salpeter equations~\cite{Bickers1991,BickersSelfConsistent2004,KuglerPhDthesis}:
\begin{equation}
\phi_r = \left(V - \phi_r\right) \circ \Pi_r \circ V = V \circ \Pi_r \circ \left(V - \phi_r\right),
\label{eq:BetheSalpeterEquations}
\end{equation}
where the bubbles $\Pi_r$ are defined as
\begin{subequations}
\begin{align}
	\Pi_{ph;1^\prime 2^\prime|12} &= - G_{1^\prime |2}G_{2^\prime |1} , \\
	\Pi_{\overline{ph};1^\prime 2^\prime|12} &= G_{1^\prime |1}G_{2^\prime |2} , \\
	\Pi_{pp;1^\prime 2^\prime|12} &= \frac{1}{2} G_{1^\prime |1}G_{2^\prime |2} .
\end{align}
\label{eq:Bubbles}
\end{subequations}
The $\circ$ product is also defined differently depending on the channel $r$, i.e.,
\begin{subequations}
    \begin{align}
       ph~:\quad [A\circ B]_{1^\prime 2^\prime|12} &= A_{42^\prime |32}B_{1^\prime 3|14}, \\
       \overline{ph}~:\quad [A\circ B]_{1^\prime 2^\prime|12} &= A_{1^\prime 4|32}B_{32^\prime |14}, \\
        pp~:\quad [A\circ B]_{1^\prime 2^\prime|12} &= A_{1^\prime 2^\prime|34}B_{34|12},
    \end{align}
\label{eq:Definitioncircproduct}
\end{subequations}
where $A$ and $B$ are arbitrary four-point functions, such as $\Pi_r$, $\phi_r$, or $V$~\cite{Gievers2022}.

\subsection{SBE decomposition}

The SBE decomposition of the two-particle vertex $V$ relies on an additional diagrammatic criterion, called $U$ reducibility, used to split the set of 2PR diagrams in $\phi_r$ into two subclasses of diagrams~\cite{Krien2019}. The propagator lines that make a diagram 2PR (i.e., that can be cut in such a way that the diagram becomes disconnected) are all attached to a bare interaction $U$ at both ends. If cutting any of those bare interactions also splits the diagram into two disconnected parts, then the diagram is considered \emph{$U$ reducible}, and \emph{$U$ irreducible} otherwise [see Fig.~\ref{fig:ParquetVsSBEdecompositions}(\ref{fig:ParquetVsSBEdecompositions_b}) for more details and concrete examples]. Furthermore, the $U$-reducible diagrams contributing to $\phi_r$ are said to be $U$ reducible in channel $r$. Hence, all $U$-reducible diagrams are also 2PR in the same channel $r$, with the exception of the bare interaction $U=\tikzm{barevertexText}{\barevertexwithlegs{0}{0}}$, which is 2PI but $U$ reducible in all channels. This classification based on two-particle and $U$ reducibility is summarized by Fig.~\ref{fig:VennDiagram}. For a more exhaustive discussion on these diagrammatic properties, see Refs.~\cite{Gievers2022,GieversPhDthesis}.

Therefore, using the criterion of $U$ reducibility, the 2PR vertex $\phi_r$ can be rewritten as
\begin{equation}
    \phi_r = \nabla_r + M_r -U,
\label{eq:phinablaM}
\end{equation}
where $\nabla_r$ contains all $U$-reducible diagrams in channel $r$, which includes the diagram for the bare interaction $\tikzm{barevertexText2}{\barevertexwithlegs{0}{0}}$. As mentioned previously, the latter is 2PI and should therefore not be contained in $\phi_r$, hence the double-counting correction $-U$ in Eq.~\eqref{eq:phinablaM}. By construction, $M_r$ thus corresponds to the set of 2PR diagrams in channel $r$ that are not $U$ reducible, namely, $U$ irreducible. The SBE decomposition is then directly obtained from the parquet decomposition~\eqref{eq:parquetDecomposition} combined with Eq.~\eqref{eq:phinablaM}, which yields~\cite{Krien2019}
\begin{align}
    V & = \nabla_{ph} + \nabla_{\overline{ph}} + \nabla_{pp} + M_{ph} + M_{\overline{ph}} + M_{pp}  - 3 U + I^{\text{2PI}},
    \label{eq:SBEDecomposition}
\end{align}
or equivalently, in a more compact form
\begin{equation}
    V = \sum_r \nabla_r + \mathcal{I}^{U\text{irr}} - 2U,
    \label{eq:SBEDecomposition2}
\end{equation}
with the $U$-irreducible part of the two-particle vertex $V$ defined as
\begin{equation}
    \mathcal{I}^{U\text{irr}} = \sum_r M_r + I^{\text{2PI}} - U.
\end{equation}
The heart of the SBE decomposition is the parametrization of the vertices $\nabla_r$ in terms of bosonic propagators $w_r$ and Yukawa couplings~\cite{Sadovskii1979,Schmalian1999} (or Hedin vertices~\cite{Hedin1965}) $\lambda_r$. Since $\nabla_r$ only contains $U$-reducible diagrams in channel $r$, it can be rewritten exactly as (see also Fig.~\ref{fig:SBEdecomposition})
\begin{equation}
    \nabla_r (Q_r,k_r,k^{\prime}_r) = \overline{\lambda}_r(Q_r,k_r) \fcirc w_r(Q_r) \fcirc \lambda_r(Q_r,k^{\prime}_r),
    \label{eq:nablaSBE}
\end{equation}
where the $\fcirc$ product is defined in the same way as $\circ$ through Eqs.~\eqref{eq:Definitioncircproduct}, but involves only the summation over spin indices (i.e., without summation over frequencies or momenta). In that way, the two-particle vertex $V(Q_r,k_r,k^{\prime}_r)$ with its complex frequency and momentum dependencies is replaced by simpler objects retaining the frequency and momentum conservation laws, namely, $\lambda_r(Q_r,k_r)$ and $w_r(Q_r)$, which depend on fewer frequencies and momenta. The remarkably simple frequency and momentum parametrization of the $\nabla_r$ functions thus makes the SBE framework particularly amenable to efficient and computationally tractable approximations. At the same time, it also offers a deeper physical insight into the many-fermion system under consideration. Indeed, competing orders in many-fermion systems are naturally described by introducing bosonic degrees of freedom, and this is precisely what the SBE approach does: The parametrization~\eqref{eq:nablaSBE} implicitly~\footnote{By ``implicitly'', we mean here that no bosons were introduced at the level of the path integral by means of a Hubbard-Stratonovich transformation; i.e., we are in principle still dealing with a purely fermionic theory.} introduces bosons in our framework through the bosonic propagators $w_r$ and the associated Yukawa couplings $\lambda_r$, also called fermion-boson couplings. More concretely, this parametrization amounts to interpreting all $U$-reducible diagrams contributing to $\nabla_r$ as the exchange of a single boson with momentum $Q_r$, hence the name of the SBE decomposition. Moreover, the 2PR and $U$-irreducible diagrams contributing to the SBE rest functions $M_r$ can be viewed as multiboson exchange processes. Finally, it is worth noting that similar parametrizations of four-point fermionic correlation functions in terms of SBE processes are also used in the context of quantum chromodynamics~\cite{Fukushima2022}.

There is also another important point concerning the SBE parametrization of Eq.~\eqref{eq:nablaSBE}. If the bare interaction of the studied model is nonlocal, which means that $U_{\sigma_{1^\prime}\sigma_{2^\prime}|\sigma_1\sigma_2}(Q_r,k_r,k^\prime_{r})$ does depend on $Q_r$, $k_r$, and $k^\prime_{r}$, then the bosonic propagators and the Yukawa couplings would in principle also depend on those three momenta, i.e., $w_r=w_r(Q_r,k_r,k^\prime_{r})$ and $\lambda_r=\lambda_r(Q_r,k_r,k^\prime_{r})$. This can directly be seen for the bosonic propagators $w_r$ since $U(Q_r,k_r,k^\prime_{r})$ is, by construction, the leading order of their perturbative expansion, i.e., $w_r(Q_r,k_r,k^\prime_{r})=U(Q_r,k_r,k^\prime_{r})+\mathcal{O}(U^2)$. With these momentum dependencies of $w_r$ and $\lambda_r$, the computational advantage of the SBE approach is spoiled. Hence, the parametrization~\eqref{eq:nablaSBE} in terms of $w_r(Q_r)$ and $\lambda_r(Q_r,k_r)$ is in principle only valid for local interactions if one uses the SBE formalism based on $U$ reducibility, as originally developed in Ref.~\cite{Krien2019}. However, the SBE approach can still be efficiently generalized to nonlocal interactions by splitting the bare interaction $U(Q_r,k_r,k^\prime_{r})$ into a bosonic part $\mathcal{B}_r$ and a fermionic part $\mathcal{F}_r$, where $\mathcal{F}_r$ encompasses the whole fermionic momentum dependence of $U(Q_r,k_r,k^\prime_{r})$~\cite{AlEryani2024,AlEryani2025}. This translates into the following relation:
\begin{equation}
    U(Q_r,k_r,k^\prime_{r}) = \mathcal{B}_r(Q_r) + \mathcal{F}_r(Q_r,k_r,k^\prime_{r}).
    \label{eq:UrBrFr}
\end{equation}
A parametrization of the form of Eq.~\eqref{eq:nablaSBE} is then possible for the vertex $\nabla^{(\mathcal{B})}_r$ containing all diagrams that are reducible with respect to the bosonic part $\mathcal{B}_r$ instead of the whole bare interaction $U$, which introduces the notion of \emph{$\mathcal{B}$ reducibility} as a new diagrammatic criterion. Namely, we have the equality
\begin{equation}
    \nabla^{(\mathcal{B})}_r(Q_r,k_r,k^\prime_{r}) = \overline{\lambda}^{(\mathcal{B})}_r(Q_r,k_r) \fcirc w^{(\mathcal{B})}_r(Q_r) \fcirc \lambda^{(\mathcal{B})}_r(Q_r,k^{\prime}_r),
    \label{eq:nablaBreducibility}
\end{equation}
which still allows us to work with objects whose frequency and momentum dependencies are much simpler than those of the full vertex $V$, i.e., $w^{(\mathcal{B})}_r(Q_r)$ and $\lambda^{(\mathcal{B})}_r(Q_r,k_r)$. Relations connecting $w^{(\mathcal{B})}_r(Q_r)$ and $\lambda^{(\mathcal{B})}_r(Q_r,k_r)$ to their counterparts $w_r(Q_r,k_r,k^\prime_{r})$ and $\lambda_r(Q_r,k_r,k^\prime_{r})$ defined from $U$ reducibility can also be worked out. In the following, we will employ the SBE formalism based on $U$ reducibility but the relations derived in the rest of Sec.~\ref{sec:Method} are in principle still valid for nonlocal interactions after reintroducing the ``$(\mathcal{B})$'' superscript and making only a few additional modifications. For more details on the latter, we refer to the recent studies~\cite{AlEryani2024,AlEryani2025} that treat the extended Hubbard model~\cite{Zhang1989,Terletska2017,Terletska2018,Paki2019} and the Hubbard-Holstein model~\cite{Holstein1959,Freericks1995,Capone2004} using the $1\ell$ fRG formulated within this generalization of the SBE approach based on $\mathcal{B}$ reducibility~\footnote{As an example, we can mention that the bosonic propagators $w_r^{(\mathcal{B})}$ must be set to $\mathcal{B}_r$, instead of $U$ for $w_r$, for the initial conditions of conventional implementations of the fRG flow~\cite{Husemann2009}.}.

\subsection{Multiloop SBE fRG equations in diagrammatic channels}
\label{sec:mSBEfRGequationsDiagChannels}

As a next step, we outline the derivation of the multiloop fRG equations in the SBE framework that we have just defined. The multiloop fRG was originally developed for generic fermionic models based on the classical action~\eqref{eq:ClassicalActionS}~\cite{Kugler2018b,KuglerPhDthesis}, with a first application to the X-ray-edge singularity~\cite{Kugler2018a}. There are also closely related theoretical developments in the context of a bosonic $\varphi^4$ theory in Refs.~\cite{Blaizot2011,Blaizot2021}. The general idea underlying this approach is to derive flow equations from the parquet equations~\cite{kugler2018c}, which include the Bethe-Salpeter equations~\eqref{eq:BetheSalpeterEquations}, instead of following the more common procedure relying on the Wetterich equation~\cite{Wetterich1993,Ellwanger1994,Morris1994} from which the conventional $1\ell$ fRG equations were originally derived~\cite{Morris1994,Metzner2012}. In that way, one can show that solving the resulting multiloop fRG equations amounts to adding loop corrections to the full vertex $V$ on top of the conventional $1\ell$ approximation, until one reaches the well-known parquet approximation at convergence~\cite{Kugler2018a}. The latter is obtained by iteratively solving the Bethe-Salpeter equations together with the Schwinger-Dyson equation, with the 2PI vertex set to the bare interaction, i.e., $I^{\text{2PI}}=U$. Such a connection is particularly appealing since the parquet equations have been used to design conserving and quantitative approximation schemes~\cite{Bickers1991,Chen1992,BickersSelfConsistent2004}, with numerous successful applications and insightful studies over the years~\cite{Tam2013,Valli2015,Li2016,Janis2017,Li2019,Kauch2019,Pudleiner2019,Astretsov2020,Eckhardt2020,Kauch2020}. Hence, the multiloop fRG has recently also been investigated in the framework of various models: quantum spin systems~\cite{Thoenniss2020,Kiese2022,Ritter2022}, the Anderson impurity model~\cite{Chalupa2020,Ge2024,Ritz2024}, and the 2D Hubbard model~\cite{TagliaviniHille2019,Hille2020a,Hille2020,Heinzelmann2023,HillePhDThesis}. However, none of the multiloop fRG applications have been carried out within the SBE scheme. Due to the advantages of this approach, the multiloop fRG method could be made substantially more powerful if reformulated within this bosonized framework. We will now show how this reformulation can be achieved. For completeness, we first review the derivation of the generic multiloop SBE fRG equations in diagrammatic channels~\cite{Gievers2022,GieversPhDthesis}, before using physical channels to treat the SU(2)-spin-symmetric case.

We first determine the self-consistent equations for the bosonic propagators $w_r$, the Yukawa couplings $\lambda_r$, and the SBE rest functions $M_r$. For this, we introduce the following identity vertices:
\begin{subequations}
	\begin{align}
		\mathds{1}_{ph;1^{\prime}2^{\prime}|12} &= \delta_{1^{\prime},2} \phantom{,} \delta_{2^{\prime},1},\\
		\mathds{1}_{\overline{ph};1^{\prime}2^{\prime}|12} &= \delta_{1^{\prime},1} \phantom{,} \delta_{2^{\prime},2},\\
		\mathds{1}_{pp;1^{\prime}2^{\prime}|12} &= \delta_{1^{\prime},1} \phantom{,} \delta_{2^{\prime},2},
	\end{align}
    \label{eq:Identitycirc}
\end{subequations}
with $\delta_{i,j} = \delta_{k_i,k_j} \phantom{,} \delta_{\sigma_i,\sigma_j}$. This enables us to define an inverse $A^{-1}$ for any four-point object such as the full vertex $V_{\sigma_{1^\prime}\sigma_{2^\prime}|\sigma_1\sigma_2}(k_{1^\prime},k_{2^\prime}|k_1, k_2)$ as follows:
\begin{equation}
A^{-1} \circ A = A \circ A^{-1} = \mathds{1}_{r},
\end{equation}
where we recall that the definition of the $\circ$ product depends on the channel $r$, as stated by Eqs.~\eqref{eq:Definitioncircproduct}. In the SBE parametrization~\eqref{eq:nablaSBE}, we have also introduced the $\fcirc$ product, defined by Eqs.~\eqref{eq:Definitioncircproduct} like the $\circ$ product but \emph{only} for spin indices. In the same way, we introduce the identity vertices $\mathbf{1}_r$ associated with the $\fcirc$ product. Thus, we have
\begin{equation}
B^{-1} \fcirc B = B \fcirc B^{-1} = \mathbf{1}_r,
\label{eq:InverseSpinIndices}
\end{equation}
where $B$ must be a four-point object with respect to spin indices. The corresponding identity vertices $\mathbf{1}_{r}$ satisfy
\begin{subequations}
	\begin{align}
		\mathbf{1}_{ph;\sigma_{1^{\prime}}\sigma_{2^{\prime}}|\sigma_1 \sigma_2} &= \delta_{\sigma_{1^{\prime}},\sigma_2} \phantom{,} \delta_{\sigma_{2^{\prime}},\sigma_1},\\
		\mathbf{1}_{\overline{ph};\sigma_{1^{\prime}}\sigma_{2^{\prime}}|\sigma_1 \sigma_2} &= \delta_{\sigma_{1^{\prime}},\sigma_1} \phantom{,} \delta_{\sigma_{2^{\prime}},\sigma_2},\\
	\mathbf{1}_{pp;\sigma_{1^{\prime}} \sigma_{2^{\prime}}|\sigma_1 \sigma_2} &= \delta_{\sigma_{1^{\prime}},\sigma_1} \phantom{,} \delta_{\sigma_{2^{\prime}},\sigma_2}.
	\end{align}
    \label{eq:Identityfcirc}
\end{subequations}
The identity vertices $\mathds{1}_r$ and $\mathbf{1}_r$, with their definitions given by Eqs.~\eqref{eq:Identitycirc} and~\eqref{eq:Identityfcirc}, are constructed from the relations $V = \mathds{1}_r \circ V = V \circ \mathds{1}_r$ and $V = \mathbf{1}_r \fcirc V = V \fcirc \mathbf{1}_r$, respectively~\footnote{More details on the construction of the identity vertices $\mathds{1}_r$ and $\mathbf{1}_r$ can be found in Appendix C.2 of Ref.~\cite{Gievers2022}.}. Note also that the bosonic propagators $w_r=w_{r;\sigma_{1^{\prime}}\sigma_{2^{\prime}}|\sigma_1 \sigma_2}(Q_r)$ and the Yukawa couplings $\lambda_r=\lambda_{r;\sigma_{1^{\prime}}\sigma_{2^{\prime}}|\sigma_1 \sigma_2}(Q_r,k_r)$ introduced by the SBE parametrization~\eqref{eq:nablaSBE} are both four-point objects with respect to spin indices, but not with respect to frequencies or momenta. For our forthcoming derivations, we also need to introduce the irreducible vertices:
\begin{subequations}
	\begin{align}
		I_r &= V - \phi_r, \label{eq:Ir} \\
		\mathcal{I}_r &= V - \nabla_r. \label{eq:mathcalIr}
	\end{align}
\label{eq:IrreducibleVertices}
\end{subequations}
Hence, $I_r$ and $\mathcal{I}_r$ collect all diagrams that are, respectively, 2PI and $U$ irreducible in channel $r$. With those new definitions, the Bethe-Salpeter equations~\eqref{eq:BetheSalpeterEquations} can be rewritten as
\begin{align}
\phi_r & = I_r \circ \Pi_r \circ (\nabla_r + \mathcal{I}_r) \nonumber \\
& = I_r \circ \Pi_r \circ \nabla_r + U \circ \Pi_r \circ \mathcal{I}_r + (I_r - U) \circ \Pi_r \circ \mathcal{I}_r. \label{eq:phirIrreducibleVertices}
\end{align}
In the last line, we have disentangled the $U$-reducible and $U$-irreducible terms with respect to channel $r$, i.e., the $U$-$r$-reducible and $U$-$r$-irreducible terms. The first two contributions $I_r \circ \Pi_r \circ \nabla_r$ and $U \circ \Pi_r \circ \mathcal{I}_r$ are both $U$-$r$ reducible due to the presence of $\nabla_r$ and $U$, whereas $(I_r - U) \circ \Pi_r \circ \mathcal{I}_r$ is $U$-$r$ irreducible. Comparing to Eq.~\eqref{eq:phinablaM}, we thus identify
\begin{subequations}
	\begin{align}
		\nabla_r - U &= I_r \circ \Pi_r \circ \nabla_r + U \circ \Pi_r \circ \mathcal{I}_r, \label{eq:nablaMinusU} \\
		M_r &= (I_r - U) \circ \Pi_r \circ \mathcal{I}_r. \label{eq:SelfConsistentEqMr}
	\end{align}
\end{subequations}
Focusing on Eq.~\eqref{eq:nablaMinusU}, we can isolate $\nabla_r$ on the left-hand side and obtain
\begin{equation}
    (\mathds{1}_r - I_r \circ \Pi_r) \circ \nabla_r = U + U \circ \Pi_r \circ \mathcal{I}_r,
\end{equation}
which is equivalent to
\begin{equation}
    \nabla_r = (\mathds{1}_r - I_r \circ \Pi_r)^{-1} \circ U \circ (\mathds{1}_r + \Pi_r \circ \mathcal{I}_r).
    \label{eq:nablaInverse}
\end{equation}
In order to rewrite the inverse $(\mathds{1}_r - I_r \circ \Pi_r)^{-1}$, we can also use the definitions~\eqref{eq:IrreducibleVertices} to rearrange the Bethe-Salpeter equations~\eqref{eq:BetheSalpeterEquations} as
\begin{equation}
    V = I_r + I_r \circ \Pi_r \circ V.
    \label{eq:ExtendedBSEfirstStep}
\end{equation}
Isolating $V$ on the left-hand side leads to the relation
\begin{equation}
    V = (\mathds{1}_r - I_r \circ \Pi_r)^{-1} \circ I_r ,
\end{equation}
which can directly be used to obtain
\begin{align}
    \mathds{1}_r + V \circ \Pi_r & = \mathds{1}_r + (\mathds{1}_r - I_r \circ \Pi_r)^{-1} \circ I_r \circ \Pi_r \nonumber \\
    & = \mathds{1}_r + (\mathds{1}_r - I_r \circ \Pi_r)^{-1} \nonumber \\
    & \phantom{=} \circ (- (\mathds{1}_r - I_r \circ \Pi_r ) + \mathds{1}_r) \nonumber \\
    & = (\mathds{1}_r - I_r \circ \Pi_r)^{-1}.
    \label{eq:ExtendedBSElastStep}
\end{align}
Using the last equality, which is referred to as extended Bethe-Salpeter equation, Eq.~\eqref{eq:nablaInverse} reduces to
\begin{equation}
    \nabla_r = (\mathds{1}_r + V \circ \Pi_r) \circ U \circ (\mathds{1}_r + \Pi_r \circ \mathcal{I}_r).
\end{equation}
As a next step, we use relations such as $\mathds{1}_r \circ U = U = \mathbf{1}_r \fcirc U$ and $U \circ \mathds{1}_r = U = U \fcirc \mathbf{1}_r$ to introduce the identity vertices for spin indices [see Eq.~\eqref{eq:InverseSpinIndices}], leading to
\begin{equation}
    \nabla_r = (\mathbf{1}_r + V \circ \Pi_r \circ \mathbf{1}_r) \fcirc U \fcirc (\mathbf{1}_r + \mathbf{1}_r \circ \Pi_r \circ \mathcal{I}_r).
    \label{eq:nablarmathbf1}
\end{equation}
We then consider the following expressions of the Yukawa couplings and their conjugates~\cite{Krien2021b,Gievers2022}:
\begin{subequations}
    \begin{align}
        \lambda_r & = \mathbf{1}_r + \mathbf{1}_r \circ \Pi_r \circ \mathcal{I}_r, \label{eq:SelfConsistentEqlambdar} \\
        \overline{\lambda}_r & = \mathbf{1}_r + \mathcal{I}_r \circ \Pi_r \circ \mathbf{1}_r. \label{eq:SelfConsistentEqlambdabarr}
    \end{align}
    \label{eq:SelfConsistentEqlambdarlambdabarr}
\end{subequations}
Combining Eqs.~\eqref{eq:nablarmathbf1} and~\eqref{eq:SelfConsistentEqlambdar} directly yields
\begin{equation}
    \nabla_r = (\mathbf{1}_r + V \circ \Pi_r \circ \mathbf{1}_r) \fcirc U \fcirc \lambda_r.
\end{equation}
With $V=\mathcal{I}_r + \nabla_r$, this becomes
\begin{align}
    \nabla_r & = (\mathbf{1}_r + \mathcal{I}_r \circ \Pi_r \circ \mathbf{1}_r + \nabla_r \circ \Pi_r \circ \mathbf{1}_r) \fcirc U \fcirc \lambda_r \nonumber \\
    & = \overline{\lambda}_r \fcirc (U + w_r \fcirc \lambda_r \circ \Pi_r \circ U) \fcirc \lambda_r , \label{eq:nablarlambdabarrlambdar}
\end{align}
where the last line was obtained using Eq.~\eqref{eq:SelfConsistentEqlambdabarr} and the SBE parametrization~\eqref{eq:nablaSBE}. From the latter, we know that $\nabla_r = \overline{\lambda}_r \fcirc w_r \fcirc \lambda_r$ and the term between $\overline{\lambda}_r$ and $\lambda_r$ in our last result can therefore be identified as $w_r$, i.e.,
\begin{equation}
    w_r = U + w_r \fcirc \lambda_r \circ \Pi_r \circ U.
\end{equation}
This result corresponds to a Dyson equation for the bosonic propagator. By isolating $w_r$ on the left-hand side, it can also be rewritten as
\begin{equation}
  w_r = U \fcirc (\mathbf{1}_r - \lambda_r \circ \Pi_r \circ U)^{-1}.
  \label{eq:SelfConsistentEqwr}
\end{equation}
We note that starting from $\phi_r = (\nabla_r + \mathcal{I}_r) \circ \Pi_r \circ I_r$ instead of $\phi_r = I_r \circ \Pi_r \circ (\nabla_r + \mathcal{I}_r)$ in Eq.~\eqref{eq:phirIrreducibleVertices} and following the same reasoning to obtain Eq.~\eqref{eq:SelfConsistentEqwr} yields the following equivalent expression of $w_r$:
\begin{equation}
    w_r = (\mathbf{1}_r - U \circ \Pi_r \circ \overline{\lambda}_r)^{-1} \fcirc U.
  \label{eq:SelfConsistentEqwrbis}
\end{equation}
We have now derived all self-consistent equations needed for our derivation of the multiloop SBE fRG equations, i.e., Eq.~\eqref{eq:SelfConsistentEqMr} for $M_r$, Eqs.~\eqref{eq:SelfConsistentEqlambdarlambdabarr} for $\lambda_r$ and $\overline{\lambda}_r$, and Eq.~\eqref{eq:SelfConsistentEqwr} [or~\eqref{eq:SelfConsistentEqwrbis}] for $w_r$. Note that these self-consistent equations are also called \emph{SBE equations} in Refs.~\cite{Gievers2022,GieversPhDthesis}.

As a next step, we introduce a scale or flow parameter $\Lambda$ in the bare propagator via $G_0\rightarrow G_{0}^\Lambda$. The scale $\Lambda$ can have different meanings and corresponds, e.g., to a frequency or a temperature, which will be specified in Sec.~\ref{sec:NumericalImplementation}, where we define our cutoff schemes. As a consequence, all objects that depend on the bare propagator $G_0$ also depend on the scale $\Lambda$, which is the case of most entities introduced so far (including the bosonic propagators $w_r$, the Yukawa couplings $\lambda_r$, and the SBE rest functions $M_r$) with the notable exception of the bare interaction $U$ ($\partial_\Lambda U = 0$). Hence, by differentiating the $\Lambda$-dependent self-consistent equations above, we obtain the flow equations.

The \emph{flow equation for $M_r$} is determined by differentiating Eq.~\eqref{eq:SelfConsistentEqMr}:
    \begin{align}
        \dot{M}_r & = \dot{I}_r \circ \Pi_r \circ \mathcal{I}_r + (I_r - U) \circ \dot{\Pi}_r \circ \mathcal{I}_r \nonumber \\
        & \phantom{=} + (I_r - U) \circ \Pi_r \circ \dot{\mathcal{I}}_r, \label{eq:MrdotFirstExpression}
    \end{align}
    where we have used the shorthand notation $\dot{X} \equiv \partial_\Lambda X$ for any vertex $X$. We can exploit the relation $\mathcal{I}_r = M_r + I_r - U$, or rather its differentiated version $\dot{\mathcal{I}}_r = \dot{M}_r + \dot{I}_r$ to introduce $\dot{M}_r$ in the last term of Eq.~\eqref{eq:MrdotFirstExpression}. Isolating $\dot{M}_r$ on the left-hand side then yields
    \begin{align}
        \dot{M}_r & = (\mathds{1}_r - (I_r - U) \circ \Pi_r )^{-1} \nonumber \\
        & \phantom{=} \circ \Big[\dot{I}_r \circ \Pi_r \circ \mathcal{I}_r + (I_r - U) \circ \dot{\Pi}_r \circ \mathcal{I}_r \nonumber \\
        & \phantom{= \circ \Big[} + (I_r - U) \circ \Pi_r \circ \dot{I}_r \Big]. \label{eq:IntermediaryExpressionMrdot}
    \end{align}
    The inverse $(\mathds{1}_r - (I_r - U) \circ \Pi_r )^{-1}$ can also be expressed via the extended Bethe-Salpeter equation
    \begin{equation}
        \mathds{1}_r + \mathcal{I}_r \circ \Pi_r = (\mathds{1}_r - (I_r - U) \circ \Pi_r)^{-1}.
        \label{eq:extendedBSEmathcalI}
    \end{equation}
    This equality can be derived following the same steps as in Eqs.~\eqref{eq:ExtendedBSEfirstStep}--\eqref{eq:ExtendedBSElastStep}, starting instead from $\mathcal{I}_r = (I_r-U) + (I_r-U) \circ \Pi_r \circ \mathcal{I}_r$, which is directly obtained from $M_r = \mathcal{I}_r - (I_r - U)$ and Eq.~\eqref{eq:SelfConsistentEqMr}. Rewriting the result~\eqref{eq:IntermediaryExpressionMrdot} by using Eq.~\eqref{eq:extendedBSEmathcalI}, one can show that
    \begin{align}
        \dot{M}_r & = \mathcal{I}_r \circ \dot{\Pi}_r \circ \mathcal{I}_r + \dot{I}_r \circ \Pi_r \circ \mathcal{I}_r + \mathcal{I}_r \circ \Pi_r \circ \dot{I}_r \nonumber \\
        & \phantom{=} + \mathcal{I}_r \circ \Pi_r \circ \dot{I}_r \circ \Pi_r \circ \mathcal{I}_r. \label{eq:FlowEquationFullMrdot}
    \end{align}

The \emph{flow equations for $\lambda_r$ and $\overline{\lambda}_r$} can be derived by differentiating Eqs.~\eqref{eq:SelfConsistentEqlambdarlambdabarr}:
    \begin{subequations}
    \begin{align}
        \dot{\lambda}_r & = \mathbf{1}_r \circ \dot{\Pi}_r \circ \mathcal{I}_r + \mathbf{1}_r \circ \Pi_r \circ \dot{\mathcal{I}}_r \nonumber \\
        & = \mathbf{1}_r \circ \dot{\Pi}_r \circ \mathcal{I}_r + \mathbf{1}_r \circ \Pi_r \circ (\dot{M}_r + \dot{I}_r), \\
        \dot{\overline{\lambda}}_r & = \dot{\mathcal{I}}_r \circ \Pi_r \circ \mathbf{1}_r + \mathcal{I}_r \circ \dot{\Pi}_r \circ \mathbf{1}_r \nonumber \\
        & = (\dot{M}_r + \dot{I}_r) \circ \Pi_r \circ \mathbf{1}_r + \mathcal{I}_r \circ \dot{\Pi}_r \circ \mathbf{1}_r ,
    \end{align}
\end{subequations}
    where the relation $\dot{\mathcal{I}}_r = \dot{M}_r + \dot{I}_r$ has been used. We then replace $\dot{M}_r$ in the expressions of $\dot{\lambda}_r$ and $\dot{\overline{\lambda}}_r$ by the right-hand side of the flow equation~\eqref{eq:FlowEquationFullMrdot} and simplify the resulting terms using Eqs.~\eqref{eq:SelfConsistentEqlambdarlambdabarr}, which leads to
    \begin{subequations}
    \begin{align}
        \dot{\lambda}_r & = \lambda_r \circ \dot{\Pi}_r \circ \mathcal{I}_r + \lambda_r \circ \Pi_r \circ \dot{I}_r \nonumber \\
        & \phantom{=} + \lambda_r \circ \Pi_r \circ \dot{I}_r \circ \Pi_r \circ \mathcal{I}_r, \label{eq:FlowEquationsFulllambdardot} \\
        \dot{\overline{\lambda}}_r & = \mathcal{I}_r \circ \dot{\Pi}_r \circ \overline{\lambda}_r + \dot{I}_r \circ \Pi_r \circ \overline{\lambda}_r \nonumber \\
        & \phantom{=} + \mathcal{I}_r \circ \Pi_r \circ \dot{I}_r \circ \Pi_r \circ \overline{\lambda}_r. \label{eq:FlowEquationsFulllambdabarrdot}
    \end{align}
    \label{eq:FlowEquationsFulllambdarlambdabarrdot}
\end{subequations}

Finally, the \emph{flow equation for $w_r$} is obtained by differentiating Eq.~\eqref{eq:SelfConsistentEqwr}:
    \begin{align}
        \dot{w}_r & = U \fcirc (\mathbf{1}_r - \lambda_r \circ \Pi_r \fcirc U)^{-1} \fcirc (\dot{\lambda}_r \circ \Pi_r + \lambda_r \circ \dot{\Pi}_r) \nonumber \\
        & \phantom{=} \circ U \fcirc (\mathbf{1}_r - \lambda_r \circ \Pi_r \fcirc U)^{-1} \nonumber \\
        & = w_r \fcirc (\dot{\lambda}_r \circ \Pi_r + \lambda_r \circ \dot{\Pi}_r) \circ w_r,
    \end{align}
    where the last line also follows from Eq.~\eqref{eq:SelfConsistentEqwr}. Replacing $\dot{\lambda}_r$ using the flow equation~\eqref{eq:FlowEquationsFulllambdardot} and then introducing $\overline{\lambda}_r$ via its self-consistent equation~\eqref{eq:SelfConsistentEqlambdabarr} yields
    \begin{align}
        \dot{w}_r & = w_r \fcirc \lambda_r \circ \dot{\Pi}_r \circ \overline{\lambda}_r \fcirc w_r \nonumber \\
        & \phantom{=} + w_r \fcirc \lambda_r \circ \Pi_r \circ \dot{I}_r \circ \Pi_r \circ \overline{\lambda}_r \fcirc w_r. \label{eq:FlowEquationFullwrdot}
    \end{align}
    We note that a similar reasoning starting from Eq.~\eqref{eq:SelfConsistentEqwrbis} instead of Eq.~\eqref{eq:SelfConsistentEqwr} would have also led to the flow equation~\eqref{eq:FlowEquationFullwrdot}.

Hence, Eqs.~\eqref{eq:FlowEquationFullMrdot},~\eqref{eq:FlowEquationsFulllambdarlambdabarrdot}, and~\eqref{eq:FlowEquationFullwrdot} are the flow equations for all objects introduced via the SBE decomposition, i.e., $w_r$, $\lambda_r$, $\overline{\lambda}_r$, and $M_r$, respectively. Their right-hand side still involves $\dot{w}_r$, $\dot{\lambda}_r$, $\dot{\overline{\lambda}}_r$, and $\dot{M}_r$ through $\dot{I}_r$, reflecting the self-consistent character of the Bethe-Salpeter equations that underpins our derivation. In practice, one reshuffles this structure by performing a loop expansion, implemented by expanding the left-hand sides as
\begin{subequations}
    \begin{align}
        \dot{w}_r & = \sum_{\ell=1}^\infty \dot{w}_r^{(\ell)}, \\
        \dot{\lambda}_r & = \sum_{\ell=1}^\infty \dot{\lambda}_r^{(\ell)}, \\
        \dot{\overline{\lambda}}_r & = \sum_{\ell=1}^\infty \dot{\overline{\lambda}}_r^{(\ell)}, \\
        \dot{M}_r & = \sum_{\ell=1}^\infty \dot{M}_r^{(\ell)},
    \end{align}
    \label{eq:LoopExpansionSBEobjects}
\end{subequations}
whereas $\dot{I}_r$ is expanded on the right-hand side of each of those equations according to
\begin{equation}
    \dot{I}_r = \sum_{r^\prime \neq r} \dot{\phi}_{r^\prime} + \dot{I}^{\text{2PI}} = \sum_{r^\prime \neq r} \sum_{\ell=1}^\infty \dot{\phi}_{r^\prime}^{(\ell)} + \dot{I}^{\text{2PI}}.
    \label{eq:dotIrwithI2PI}
\end{equation}
Imposing the condition
\begin{equation}
    \dot{I}^{\text{2PI}} = 0
    \label{eq:I2PIapprox}
\end{equation}
closes our system of equations. Specifically, in the conventional weak-coupling fRG relying on the parquet approximation, we have ${I}^{\text{2PI}}=U$, while in the strong-coupling extension based on the combination with the DFMT as starting point of the fRG flow within the so-called DMF$^2$RG \cite{Taranto2014,Vilardi2019}, ${I}^{\text{2PI}}={I}^{\text{2PI}}_\text{DMFT}$ is used. 
We stress that Eq.~\eqref{eq:I2PIapprox} 
is the only approximation imposed so far: Eqs.~\eqref{eq:FlowEquationFullMrdot},~\eqref{eq:FlowEquationsFulllambdarlambdabarrdot}, and~\eqref{eq:FlowEquationFullwrdot} are all exact. Using Eq.~\eqref{eq:I2PIapprox}, we rewrite our expression~\eqref{eq:dotIrwithI2PI} for $\dot{I}_r$ as
\begin{equation}
    \dot{I}_r = \sum_{\ell=1}^\infty \dot{I}_r^{(\ell)},
    \label{eq:dotIrwithoutI2PI}
\end{equation}
with
\begin{equation}
    \dot{I}_r^{(\ell)} = \sum_{r^\prime \neq r} \dot{\phi}_{r^\prime}^{(\ell)} ,
\end{equation}
where $\dot{\phi}_r^{(\ell)}$ is related to the different loop corrections introduced in Eqs.~\eqref{eq:LoopExpansionSBEobjects} according to
\begin{equation}
    \dot{\phi}_{r}^{(\ell)} = \dot{\overline{\lambda}}_r^{(\ell)} \fcirc w_r \fcirc \lambda_r + \overline{\lambda}_r \fcirc \dot{w}_r^{(\ell)} \fcirc \lambda_r + \overline{\lambda}_r \fcirc w_r \fcirc \dot{\lambda}_r^{(\ell)} + \dot{M}_r^{(\ell)}.
\end{equation}
Expanding the flow equations and identifying the different loop orders on both sides of those equations, we obtain the multiloop SBE fRG equations expressed as
\begin{subequations}
    \begin{align}
    \dot{w}_r^{(1)} & = w_r \fcirc \lambda_r \circ \dot{\Pi}_r \circ \overline{\lambda}_r \fcirc w_r, \label{eq:SBEmfRGEq_1l_DiagChannels_wr} \\
    \dot{\overline{\lambda}}_r^{(1)} & = \mathcal{I}_r \circ \dot{\Pi}_r \circ \overline{\lambda}_r, \label{eq:SBEmfRGEq_1l_DiagChannels_lambdabarr} \\
    \dot{\lambda}_r^{(1)} & = \lambda_r \circ \dot{\Pi}_r \circ \mathcal{I}_r, \label{eq:SBEmfRGEq_1l_DiagChannels_lambdar} \\
    \dot{M}_r^{(1)} & = \mathcal{I}_r \circ \dot{\Pi}_r \circ \mathcal{I}_r, \label{eq:SBEmfRGEq_1l_DiagChannels_Mr}
    \end{align}
\label{eq:SBEmfRGEq_1l_DiagChannels}
\end{subequations}
for $1\ell$, 
\begin{subequations}
    \begin{align}
    \dot{w}_r^{(2)} & = 0, \label{eq:SBEmfRGEq_2l_DiagChannels_wr} \\
    \dot{\overline{\lambda}}_r^{(2)} & = \dot{I}_{r}^{(1)} \circ \Pi_r \circ \overline{\lambda}_r , \label{eq:SBEmfRGEq_2l_DiagChannels_lambdabarr} \\
    \dot{\lambda}_r^{(2)} & = \lambda_r \circ \Pi_r \circ \dot{I}_{r}^{(1)}, \label{eq:SBEmfRGEq_2l_DiagChannels_lambdar} \\
    \dot{M}_r^{(2)} & = \dot{I}_{r}^{(1)} \circ \Pi_r \circ \mathcal{I}_r + \mathcal{I}_r \circ \Pi_r \circ \dot{I}_{r}^{(1)} , \label{eq:SBEmfRGEq_2l_DiagChannels_Mr}
    \end{align}
\label{eq:SBEmfRGEq_2l_DiagChannels}
\end{subequations}
for $2\ell$, and
\begin{subequations}
    \begin{align}
    \dot{w}_r^{(\ell)} & = w_r \fcirc \lambda_r \circ \Pi_r \circ \dot{I}_{r}^{(\ell-2)} \circ \Pi_r \circ \overline{\lambda}_r \fcirc w_r , \label{eq:SBEmfRGEq_nl_DiagChannels_wr} \\
    \dot{\overline{\lambda}}_r^{(\ell)} & = \dot{I}_{r}^{(\ell-1)} \circ \Pi_r \circ \overline{\lambda}_r  + \mathcal{I}_r \circ \Pi_r \circ \dot{I}_{r}^{(\ell-2)} \circ \Pi_r \circ \overline{\lambda}_r , \label{eq:SBEmfRGEq_nl_DiagChannels_lambdabarr} \\
    \dot{\lambda}_r^{(\ell)} & = \lambda_r \circ \Pi_r \circ \dot{I}_{r}^{(\ell-1)} + \lambda_r \circ \Pi_r \circ \dot{I}_{r}^{(\ell-2)} \circ \Pi_r \circ \mathcal{I}_r , \label{eq:SBEmfRGEq_nl_DiagChannels_lambdar} \\
    \dot{M}_r^{(\ell)} & = \dot{I}_{r}^{(\ell-1)} \circ \Pi_r \circ \mathcal{I}_r + \mathcal{I}_r \circ \Pi_r \circ \dot{I}_{r}^{(\ell-1)} \nonumber \\
    & \phantom{=} + \mathcal{I}_r \circ \Pi_r \circ \dot{I}_{r}^{(\ell-2)} \circ \Pi_r \circ \mathcal{I}_r , \label{eq:SBEmfRGEq_nl_DiagChannels_Mr}
    \end{align}
\label{eq:SBEmfRGEq_nl_DiagChannels}
\end{subequations}
for $\ell \geq 3$. The flow equations~\eqref{eq:SBEmfRGEq_1l_DiagChannels},~\eqref{eq:SBEmfRGEq_2l_DiagChannels}, and~\eqref{eq:SBEmfRGEq_nl_DiagChannels} correspond to those reported and derived in Ref.~\cite{Gievers2022}.

\subsection{Multiloop SBE fRG equations in physical channels}
\label{sec:mSBEfRGequationsPhysChannels}

As a next step, we assume spin rotation invariance, i.e., SU(2) spin symmetry, which imposes constraints on the spin components of the objects underlying our formalism. For the full propagator and the two-particle vertex, these constraints translate into the relations~\cite{Salmhofer2001}
\begin{subequations}
\begin{align}
    G_{1^\prime|1}  & = G_{\sigma_{1^\prime}|\sigma_1}(k_{1^\prime}|k_1) \nonumber \\
    & = G(k_{1^\prime}|k_1) \delta_{\sigma_{1^\prime},\sigma_1} \label{eq:FullPropagatorSU2symmetry} \\
    V_{1^\prime 2^\prime|12} & = V_{\sigma_{1^\prime}\sigma_{2^\prime}|\sigma_1\sigma_2}(k_{1^\prime},k_{2^\prime}|k_1, k_2) \nonumber \\
    & = V^{\uparrow\downarrow}(k_{1^\prime},k_{2^\prime}|k_1, k_2) \delta_{\sigma_{1^\prime},\sigma_1} \delta_{\sigma_{2^\prime},\sigma_2} \nonumber \\
    & \phantom{=} + V^{\widehat{\uparrow\downarrow}}(k_{1^\prime},k_{2^\prime}|k_1, k_2) \delta_{\sigma_{1^\prime},\sigma_2} \delta_{\sigma_{2^\prime},\sigma_1}. \label{eq:FullVertexSU2symmetry}
\end{align}
\label{eq:GVSU2symmetry}
\end{subequations}
Note that Eq.~\eqref{eq:FullVertexSU2symmetry} relies on some of the following shorthand notations that will be conveniently used throughout the rest of this section:
\begin{subequations}
    \begin{align}
       A_{\sigma\sigma|\sigma\sigma} & = A^{\sigma\sigma}, \\
       A_{\sigma\overline{\sigma}|\sigma\overline{\sigma}} & = A^{\sigma\overline{\sigma}}, \\
       A_{\sigma\overline{\sigma}|\overline{\sigma}\sigma} & = A^{\widehat{\sigma\overline{\sigma}}},
    \end{align}
\end{subequations}
with $\mathord{\overline{\uparrow}} = \mathord{\downarrow}$, $\mathord{\overline{\downarrow}} = \mathord{\uparrow}$, and $A$ an arbitrary four-point object with respect to spin indices. Furthermore, the relation $V_{1^\prime 2^\prime|12} = -V_{2^\prime 1^\prime|12} = -V_{1^\prime 2^\prime|21}$, which follows from the crossing symmetry of $V$, imposes that $V^{\widehat{\uparrow\downarrow}}(k_{1^\prime},k_{2^\prime}|k_1, k_2) = -V^{\uparrow\downarrow}(k_{2^\prime},k_{1^\prime}|k_1, k_2) = -V^{\uparrow\downarrow}(k_{1^\prime},k_{2^\prime}|k_2, k_1)$. Therefore, Eq.~\eqref{eq:FullVertexSU2symmetry} can be rewritten as
\begin{align}
    V_{1^\prime 2^\prime|12} & = V^{\uparrow\downarrow}(k_{1^\prime},k_{2^\prime}|k_1, k_2) \delta_{\sigma_{1^\prime},\sigma_1} \delta_{\sigma_{2^\prime},\sigma_2} \nonumber \\
    & \phantom{=} - V^{\uparrow\downarrow}(k_{2^\prime},k_{1^\prime}|k_1, k_2) \delta_{\sigma_{1^\prime},\sigma_2} \delta_{\sigma_{2^\prime},\sigma_1}.
\label{eq:FullVertexSU2Crossingsymmetry}
\end{align}
Hence, the two-particle vertex $V$ has in principle $2^4=16$ spin components. However, in the presence of spin rotation invariance, only six of those components are finite and satisfy
\begin{subequations}
    \begin{align}
       V^{\uparrow\uparrow} & = V^{\downarrow\downarrow}, \\
       V^{\uparrow\downarrow} & = V^{\downarrow\uparrow}, \\
       V^{\widehat{\uparrow\downarrow}} & = V^{\widehat{\downarrow\uparrow}}.
    \end{align}
\end{subequations}
Note that this property also applies to the objects introduced in the SBE formalism, i.e., $w_r$, $\lambda_r$, and $M_r$, which will be used repeatedly in the forthcoming derivations. Moreover, according to Eq.~\eqref{eq:FullVertexSU2Crossingsymmetry}, the vertex $V$ is fully determined by its $\uparrow\downarrow$ spin component. In this case, one can thus only consider the SBE decomposition parametrizing $V^{\uparrow\downarrow}$ without losing information. 

We thus define the physical channels in terms of the spin components of $w_r$, $\lambda_r$, $\overline{\lambda}_r$, and $M_r$. Introduced to diagonalize the Bethe-Salpeter equations with respect to their spin indices for SU(2) spin-symmetric systems~\cite{BickersSelfConsistent2004}, the three physical channels, i.e., the magnetic (M), the density (D), and the superconducting (SC) channels, were also used in the original formulation of the SBE decomposition by Krien \textit{et al.}~\cite{Krien2019}. In the following, we will refer to physical channels with the letter $\text{X}$ (as opposed to $r$ for diagrammatic channels)~\footnote{Formulating Eq.~\eqref{eq:FullVertexSU2symmetry} for $A_{ph}$ yields $A_{ph}^{\uparrow\uparrow}=A_{ph}^{\uparrow\downarrow}+A_{ph}^{\widehat{\uparrow\downarrow}}$, while the crossing symmetry of $A_{ph}$ imposes that $A_{ph}^{\widehat{\uparrow\downarrow}}=-A_{\overline{ph}}^{\uparrow\downarrow}$. By combining these two equalities, we obtain the relation $A^{\uparrow\uparrow}_{ph} - A^{\uparrow\downarrow}_{ph} = - A^{\uparrow\downarrow}_{\overline{ph}}$ used in Eq.~\eqref{eq:DefinitionMchannel}.}:
\begin{subequations}
    \begin{align}
       A_{\text{M}} & = A^{\uparrow\uparrow}_{ph} - A^{\uparrow\downarrow}_{ph} = - A^{\uparrow\downarrow}_{\overline{ph}}, \label{eq:DefinitionMchannel} \\
       A_{\text{D}} & = A^{\uparrow\uparrow}_{ph} + A^{\uparrow\downarrow}_{ph}, \label{eq:DefinitionDchannel} \\
       A_{\text{SC}} & = A^{\uparrow\downarrow}_{pp}, \label{eq:DefinitionSCchannel}
    \end{align}
    \label{eq:DefinitionsPhysicalChannels}
\end{subequations}
where the vertices $A_r$ on the right-hand sides can be any four-point object with respect to spin indices (e.g., $w_r$, $M_r$, $\nabla_r$, or $\phi_r$) except the Yukawa couplings $\lambda_r$ [and the bubbles $\Pi_r$, whose physical channels are defined later in this section by Eqs.~\eqref{eq:BubblesPiX}]. We use different definitions for the physical channels of the Yukawa couplings in order to be able to express the $U$-reducible vertices $\nabla_{\text{X}}$ as
\begin{equation}
    \nabla_{\text{X}} (Q,k,k^{\prime}) = \overline{\lambda}_{\text{X}}(Q,k) w_{\text{X}}(Q) \lambda_{\text{X}}(Q,k^{\prime}),
    \label{eq:nablaSBEphysicalChannel}
\end{equation}
similarly to Eq.~\eqref{eq:nablaSBE} in diagrammatic channels. More precisely, we determine the definitions of the physical channels for the Yukawa couplings starting from Eqs.~\eqref{eq:DefinitionsPhysicalChannels} for $\nabla_{\text{X}}$ and then use the definitions of the $\fcirc$ products [see Eqs.~\eqref{eq:Definitioncircproduct}] to rewrite the objects $\overline{\lambda}_r \fcirc w_r \fcirc \lambda_r$ in terms of the bosonic propagators and the Yukawa couplings in physical channels, namely, $w_{\text{X}}$ and $\lambda_{\text{X}}$ (as well as $\overline{\lambda}_{\text{X}}$). For example, for the $\text{SC}$ channel, this gives
\begin{align}
   \nabla_{\text{SC}} & = \nabla_{pp}^{\uparrow\downarrow} \nonumber \\
   & = \left[\overline{\lambda}_{pp} \fcirc w_{pp} \fcirc \lambda_{pp}\right]^{\uparrow\downarrow} \nonumber \\
   & = \overline{\lambda}_{pp;\uparrow\downarrow|\sigma_1 \sigma_{1^\prime}} w_{pp;\sigma_1 \sigma_{1^\prime}|\sigma_2 \sigma_{2^\prime}} \lambda_{pp;\sigma_2 \sigma_{2^\prime}|\uparrow\downarrow} \nonumber \\
   & = \overline{\lambda}_{pp}^{\uparrow\downarrow} w_{pp}^{\uparrow\downarrow} \lambda_{pp}^{\uparrow\downarrow} + \overline{\lambda}_{pp}^{\widehat{\uparrow\downarrow}} w_{pp}^{\uparrow\downarrow} \lambda_{pp}^{\widehat{\uparrow\downarrow}} + \overline{\lambda}_{pp}^{\widehat{\uparrow\downarrow}} w_{pp}^{\widehat{\uparrow\downarrow}} \lambda_{pp}^{\uparrow\downarrow} + \overline{\lambda}_{pp}^{\uparrow\downarrow} w_{pp}^{\widehat{\uparrow\downarrow}} \lambda_{pp}^{\widehat{\uparrow\downarrow}} \nonumber \\
   & = \left( \overline{\lambda}_{pp}^{\uparrow\downarrow} - \overline{\lambda}_{pp}^{\widehat{\uparrow\downarrow}} \right) w^{\uparrow\downarrow}_{pp} \left( \lambda_{pp}^{\uparrow\downarrow} - \lambda_{pp}^{\widehat{\uparrow\downarrow}} \right) ,
\end{align}
where the last line was obtained using the relation $w_{pp}^{\widehat{\uparrow\downarrow}}(Q)=-w_{pp}^{\uparrow\downarrow}(Q)$, which follows directly from the crossing symmetry of $w_{pp}(Q)$. Introducing $w_{\text{SC}} = w_{pp}^{\uparrow\downarrow}$ [consistently with Eq.~\eqref{eq:DefinitionSCchannel}] as well as $\lambda_{\text{SC}} = \lambda_{pp}^{\uparrow\downarrow} - \lambda_{pp}^{\widehat{\uparrow\downarrow}}$ (and $\overline{\lambda}_{\text{SC}} = \overline{\lambda}_{pp}^{\uparrow\downarrow} - \overline{\lambda}_{pp}^{\widehat{\uparrow\downarrow}}$), we indeed recover Eq.~\eqref{eq:nablaSBEphysicalChannel} for $\text{X}=\text{SC}$. Our definitions of $\lambda^{\text{M}}$ and $\lambda^{\text{D}}$ are determined following the same reasoning. We summarize below our definitions of the Yukawa couplings in physical channels:
\begin{subequations}
    \begin{align}
       \lambda_{\text{M}} & = \lambda^{\uparrow\uparrow}_{ph} - \lambda^{\uparrow\downarrow}_{ph} = +\lambda^{\uparrow\downarrow}_{\overline{ph}}, \label{eq:DefinitionlambdaMchannel} \\
       \lambda_{\text{D}} & = \lambda^{\uparrow\uparrow}_{ph} + \lambda^{\uparrow\downarrow}_{ph}, \label{eq:DefinitionlambdaDchannel} \\
       \lambda_{\text{SC}} & = \lambda_{pp}^{\uparrow\downarrow} - \lambda_{pp}^{\widehat{\uparrow\downarrow}}. \label{eq:DefinitionlambdaSCchannel}
    \end{align}
    \label{eq:DefinitionslambdasPhysicalChannels}
\end{subequations}
Our definition~\eqref{eq:DefinitionlambdaDchannel} for $\lambda_{\text{D}}$ is therefore consistent with Eq.~\eqref{eq:DefinitionDchannel}. In the case of the magnetic channel, we have the equality $\lambda^{\uparrow\uparrow}_{ph} - \lambda^{\uparrow\downarrow}_{ph} = +\lambda^{\uparrow\downarrow}_{\overline{ph}}$ in Eq.~\eqref{eq:DefinitionlambdaMchannel}, whereas Eq.~\eqref{eq:DefinitionMchannel} exhibits an opposite sign with $A^{\uparrow\uparrow}_{ph} - A^{\uparrow\downarrow}_{ph} = -A^{\uparrow\downarrow}_{\overline{ph}}$. This sign discrepancy results from the fact that the Yukawa couplings follow different crossing symmetry relations as compared to other vertices like $w_r$ or $\phi_r$, and therefore $A_r$ in Eq.~\eqref{eq:DefinitionMchannel}~\footnote{The equality $\lambda^{\widehat{\uparrow\downarrow}}_{ph}=+\lambda^{\uparrow\downarrow}_{\overline{ph}}$ holds [see, e.g., Eq.~(4.31a) in Ref.~\cite{Gievers2022}], which is to be compared with $A^{\widehat{\uparrow\downarrow}}_{ph}=-A^{\uparrow\downarrow}_{\overline{ph}}$ for Eq.~\eqref{eq:DefinitionMchannel}. The definition $\lambda_{\text{M}}=+\lambda^{\uparrow\downarrow}_{\overline{ph}}$ of Eq.~\eqref{eq:DefinitionlambdaMchannel} directly follows from the relations $\lambda^{\widehat{\uparrow\downarrow}}_{ph}=+\lambda^{\uparrow\downarrow}_{\overline{ph}}$ and $\lambda_{ph}^{\uparrow\uparrow}=\lambda_{ph}^{\uparrow\downarrow}+\lambda_{ph}^{\widehat{\uparrow\downarrow}}$.}. Furthermore, relations identical to Eqs.~\eqref{eq:DefinitionslambdasPhysicalChannels} also apply to the conjugates $\overline{\lambda}_{\text{X}}$. However, under time-reversal and crossing symmetry, one can show that~\cite{Bonetti2022,GieversPhDthesis}
\begin{equation}
    \overline{\lambda}_{\text{X}}=\lambda_{\text{X}},
    \label{eq:lambdabarXtimereversal}
\end{equation}
for $\text{X}=\text{M},\text{D},\text{SC}$. We will use this property in what follows, thus assuming time-reversal symmetry.

As a next step, we will outline the derivation of the multiloop SBE fRG equations in physical channels, in the form that we used for the numerical implementation to obtain the results for the 2D Hubbard model presented afterward. For this, we start from the generic flow equations~\eqref{eq:SBEmfRGEq_1l_DiagChannels}--\eqref{eq:SBEmfRGEq_nl_DiagChannels} and simplify them by evaluating the sums over momenta and frequencies using the conservation laws of Eqs.~\eqref{eq:FreqMomConservationLawsG0U} and the sums over spin indices exploiting the SU(2) spin symmetry. We first explain how those sums are treated in a general fashion.
 
    \emph{Frequencies and momenta.}     
    In the generic flow equations~\eqref{eq:SBEmfRGEq_1l_DiagChannels}--\eqref{eq:SBEmfRGEq_nl_DiagChannels}, summations over frequencies and momenta are included in terms of the form
    \begin{equation}
        A \circ \Pi_r \circ B,
    \end{equation}
    with the bubbles $\Pi_r$ introduced in Eqs.~\eqref{eq:Bubbles} [more precisely, $\Pi_r$ is replaced by its derivative $\dot{\Pi}_r$ in the $1\ell$ flow equations~\eqref{eq:SBEmfRGEq_1l_DiagChannels}], and $A$ and $B$ generic four-point objects with respect to momenta and frequencies (we will leave spin indices implicit in most equations here). 
    The objects $A$ and $B$ satisfy frequency and momentum conservation laws 
    \begin{subequations}
        \begin{align}
            A(k_{1^\prime},k_{2^\prime}|k_1, k_2) & = A(Q_r,k_r,k^\prime_{r}) \delta_{k_{1^\prime}+k_{2^\prime},k_1+k_2}, \\
            B(k_{1^\prime},k_{2^\prime}|k_1, k_2) & = B(Q_r,k_r,k^\prime_{r}) \delta_{k_{1^\prime}+k_{2^\prime},k_1+k_2},
        \end{align}
        \label{eq:AandBFreqMomentum}
    \end{subequations}
    similarly to the bare interaction in Eq.~\eqref{eq:FreqMomConservationLawsU}.
    The same applies to the whole product, i.e.,
    \begin{align}
        [A \circ \Pi_r \circ B](k_{1^\prime},k_{2^\prime}|k_1, k_2) & = [A \circ \Pi_r \circ B](Q_r,k_r,k^\prime_{r}) \nonumber \\
        & \phantom{=} \times \delta_{k_{1^\prime}+k_{2^\prime},k_1+k_2}. \label{eq:ApiBFreqMomentum}
    \end{align}
    Furthermore, we know that the frequency and momentum conservation for the full propagator read
    \begin{equation}
        G(k_{1^\prime}|k_1)=G(k_1)\delta_{k_{1^\prime},k_1}.
        \label{eq:GMomentum}
    \end{equation}
    With this, the definitions~\eqref{eq:Bubbles} for the bubbles can be simplified as
    \begin{subequations}
\begin{align}
	\Pi_{ph}(k_{1^\prime},k_{2^\prime}|k_1, k_2) &= - G(k_2) G(k_1) \delta_{k_{1^\prime},k_2} \delta_{k_{2^\prime},k_1} , \\
	\Pi_{\overline{ph}}(k_{1^\prime},k_{2^\prime}|k_1, k_2) &= G(k_1) G(k_2) \delta_{k_{1^\prime},k_1} \delta_{k_{2^\prime},k_2} , \\
	\Pi_{pp}(k_{1^\prime},k_{2^\prime}|k_1, k_2) &= \frac{1}{2} G(k_1) G(k_2) \delta_{k_{1^\prime},k_1} \delta_{k_{2^\prime},k_2} .
\end{align}
\label{eq:BubblesFreqMomentum}
    \end{subequations}
    Rewriting the product $[A \circ \Pi_r \circ B](k_{1^\prime},k_{2^\prime}|k_1, k_2)$ with Eqs.~\eqref{eq:AandBFreqMomentum} and~\eqref{eq:BubblesFreqMomentum}, we can identify the expressions for $[A \circ \Pi_r \circ B](Q_r,k_r,k^\prime_{r})$ in Eq.~\eqref{eq:ApiBFreqMomentum} for each channel $r$~\footnote{The procedure to derive Eqs.~\eqref{eq:EvaluateSumsFreqMomenta} is notably discussed in detail in Appendix A of Ref.~\cite{Patricolo2025}. Regarding those equations, it can also be noted that our frequency and momentum conventions in the $pp$ channel defined in Fig.~\ref{fig:FrequencyMomentumParametrization} can be modified in such a way that the right-hand side of Eq.~\eqref{eq:EvaluateSumsFreqMomentappchannel} can be rewritten as $\sum_{k^{\prime\prime}_{pp}} A(Q_{pp},k_{pp},k^{\prime\prime}_{pp}) \fcirc \Pi_{pp}(Q_{pp},k^{\prime\prime}_{pp}) \fcirc B(Q_{pp},k^{\prime\prime}_{pp},k_{pp})$, which matches the structure of momentum and frequency arguments in the $ph$ and $\overline{ph}$ channels [in Eqs.~\eqref{eq:EvaluateSumsFreqMomentaphchannel} and~\eqref{eq:EvaluateSumsFreqMomentaphxchannel}], and would also be consistent with Eq.~(20) of Ref.~\cite{Gievers2022}. This would, however, lead to the same final flow equations derived in this paper [Eqs.~\eqref{eq:mfRGequationsPhysicalChannels1l}--\eqref{eq:mfRGequationsPhysicalChannelsbeyond2l}].}:
    \begin{widetext}
    \begin{subequations}
        \begin{align}
            [A\circ\Pi_{ph}\circ B](Q_{ph},k_{ph},k^{\prime}_{ph}) &= \sum_{k^{\prime\prime}_{ph}}A(Q_{ph},k_{ph},k^{\prime\prime}_{ph}) \fcirc \Pi_{ph}(Q_{ph},k^{\prime\prime}_{ph}) \fcirc B(Q_{ph},k^{\prime\prime}_{ph},k^{\prime}_{ph}), \label{eq:EvaluateSumsFreqMomentaphchannel} \\
            [A\circ\Pi_{\overline{ph}}\circ B](Q_{\overline{ph}},k_{\overline{ph}},k^{\prime}_{\overline{ph}}) &= \sum_{k^{\prime\prime}_{\overline{ph}}}A(Q_{\overline{ph}},k_{\overline{ph}},k^{\prime\prime}_{\overline{ph}}) \fcirc \Pi_{\overline{ph}}(Q_{\overline{ph}},k^{\prime\prime}_{\overline{ph}}) \fcirc B(Q_{\overline{ph}},k^{\prime\prime}_{\overline{ph}},k^{\prime}_{\overline{ph}}), \label{eq:EvaluateSumsFreqMomentaphxchannel} \\
            [A\circ\Pi_{pp}\circ B](Q_{pp},k_{pp},k^{\prime}_{pp}) &= \sum_{k^{\prime\prime}_{pp}}A(Q_{pp},k^{\prime\prime}_{pp},k^{\prime}_{pp}) \fcirc \Pi_{pp}(Q_{pp},k^{\prime\prime}_{pp}) \fcirc B(Q_{pp},k_{pp},k^{\prime\prime}_{pp}), \label{eq:EvaluateSumsFreqMomentappchannel}
        \end{align}
        \label{eq:EvaluateSumsFreqMomenta}
    \end{subequations}
    with
    \begin{subequations}
        \begin{align}
        \Pi_{ph;\sigma_{1^\prime} \sigma_{2^\prime}|\sigma_1 \sigma_2}(Q_{ph},k_{ph}) & = - G_{\sigma_{2^\prime}|\sigma_1}\Bigg(\mathbf{k}_{ph}+\mathbf{Q}_{ph},\nu_{ph}+\left\lceil\frac{\Omega_{ph}}{2}\right\rceil\Bigg)G_{\sigma_{1^\prime}|\sigma_2}\Bigg(\mathbf{k}_{ph},\nu_{ph}-\left\lfloor\frac{\Omega_{ph}}{2}\right\rfloor\Bigg), \\
        \Pi_{\overline{ph};\sigma_{1^\prime} \sigma_{2^\prime}|\sigma_1 \sigma_2}(Q_{\overline{ph}},k_{\overline{ph}}) & = G_{\sigma_{1^\prime}|\sigma_1}\Bigg(\mathbf{k}_{\overline{ph}}+\mathbf{Q}_{\overline{ph}},\nu_{\overline{ph}}+\left\lceil\frac{\Omega_{\overline{ph}}}{2}\right\rceil\Bigg)G_{\sigma_{2^\prime}|\sigma_2}\Bigg(\mathbf{k}_{\overline{ph}},\nu_{\overline{ph}}-\left\lfloor\frac{\Omega_{\overline{ph}}}{2}\right\rfloor\Bigg), \\
        \Pi_{pp;\sigma_{1^\prime} \sigma_{2^\prime}|\sigma_1 \sigma_2}(Q_{pp},k_{pp}) & = \frac{1}{2} G_{\sigma_{1^\prime}|\sigma_1}\Bigg(\mathbf{Q}_{pp}-\mathbf{k}_{pp},\left\lfloor\frac{\Omega_{pp}}{2}\right\rfloor-\nu_{pp}\Bigg)G_{\sigma_{2^\prime}|\sigma_2}\Bigg(\mathbf{k}_{pp},\nu_{pp}+\left\lceil\frac{\Omega_{pp}}{2}\right\rceil\Bigg).
        \end{align}
        \label{eq:BubblesEvaluateSumsFreqMomenta}
    \end{subequations}
    \end{widetext}
    Note that Eqs.~\eqref{eq:EvaluateSumsFreqMomenta} still hold if the bubbles $\Pi_r$ are replaced by their derivatives $\dot{\Pi}_r$.

   \emph{Spin indices.}     
    In the generic flow equations~\eqref{eq:SBEmfRGEq_1l_DiagChannels}--\eqref{eq:SBEmfRGEq_nl_DiagChannels}, we also have to evaluate products between four-point objects with respect to spin indices (we will leave frequency and momentum indices implicit here). They have the form
    \begin{equation}
        A \fcirc B,
    \end{equation}
    which can be written explicitly in each diagrammatic channel using the definitions of Eqs.~\eqref{eq:Definitioncircproduct}, i.e.,
    \begin{subequations}
    \begin{align}
       ph~:\quad [A\fcirc B]_{\sigma_{1^\prime} \sigma_{2^\prime}|\sigma_{1} \sigma_{2}} &= A_{\sigma_{4} \sigma_{2^\prime} |\sigma_{3} \sigma_{2}}B_{\sigma_{1^\prime} \sigma_{3}|\sigma_{1} \sigma_{4}}, \\
       \overline{ph}~:\quad [A\fcirc B]_{\sigma_{1^\prime} \sigma_{2^\prime}|\sigma_{1} \sigma_{2}} &= A_{\sigma_{1^\prime} \sigma_{4}|\sigma_{3} \sigma_{2}}B_{\sigma_{3} \sigma_{2^\prime} |\sigma_{1} \sigma_{4}}, \\
        pp~:\quad [A\fcirc B]_{\sigma_{1^\prime} \sigma_{2^\prime}|\sigma_{1} \sigma_{2}} &= A_{\sigma_{1^\prime} \sigma_{2^\prime}|\sigma_{3} \sigma_{4}}B_{\sigma_{3}\sigma_{4}|\sigma_{1}\sigma_{2}},
    \end{align}
\label{eq:Definitionfcircproduct}
\end{subequations}
where we recall that the repeated spin indices ($\sigma_3$ and $\sigma_4$ here) are summed over their two configurations $\uparrow$ and $\downarrow$. Their products 
can be reformulated efficiently in terms of standard matrix multiplications in the presence of SU(2) spin symmetry. In this case, the four-point objects $A = w_r, \lambda_r, M_r, ...$ have only six nonvanishing spin components ($A^{\uparrow\uparrow}$, $A^{\downarrow\downarrow}$, $A^{\uparrow\downarrow}$, $A^{\downarrow\uparrow}$, $A^{\widehat{\uparrow\downarrow}}$, and $A^{\widehat{\downarrow\uparrow}}$). Evaluating the sums over $\sigma_3$ and $\sigma_4$ in Eqs.~\eqref{eq:Definitionfcircproduct} then amounts to carrying out products between the matrices~\cite{Patricolo2025,GieversPhDthesis}
\begin{subequations}
    \begin{align}
       ph~:\quad {A}=\begin{bmatrix}
        A^{\widehat{\uparrow\downarrow}} & 0 & 0 & 0 \\
        0 & A^{\widehat{\downarrow\uparrow}}& 0 & 0 \\
        0 & 0 & A^{\uparrow\uparrow} & A^{\downarrow\uparrow} \\
        0 & 0 & A^{\uparrow\downarrow} & A^{\downarrow\downarrow}
    \end{bmatrix}, \\
       \overline{ph}~:\quad {A}=\begin{bmatrix}
        A^{\uparrow\downarrow} & 0 & 0 & 0 \\
        0 & A^{\downarrow\uparrow} & 0 & 0 \\
        0 & 0 & A^{\uparrow\uparrow} & A^{\widehat{\uparrow\downarrow}}\\
        0 & 0 & A^{\widehat{\downarrow\uparrow}}& A^{\downarrow\downarrow}
    \end{bmatrix}, \\
        pp~:\quad A=\begin{bmatrix}
        A^{\uparrow\uparrow}& 0 & 0 & 0 \\
        0 & A^{\downarrow\downarrow} & 0 & 0 \\
        0 & 0 & A^{\uparrow\downarrow}& A^{\widehat{\uparrow\downarrow}} \\
        0 & 0 & A^{\widehat{\downarrow\uparrow}} & A^{\downarrow\uparrow}
    \end{bmatrix},
    \end{align}
\label{eq:Matricesfcircproduct}
\end{subequations}
and analogously for $B$. In the $ph$ channel, for example, this means that we can write
\begin{align}
    &\begin{bmatrix}
        [A\fcirc B]^{\widehat{\uparrow\downarrow}} & 0 & 0 & 0 \\
        0 & [A\fcirc B]^{\widehat{\downarrow\uparrow}}& 0 & 0 \\
        0 & 0 & [A\fcirc B]^{\uparrow\uparrow} & [A\fcirc B]^{\downarrow\uparrow} \\
        0 & 0 & [A\fcirc B]^{\uparrow\downarrow} & [A\fcirc B]^{\downarrow\downarrow}
    \end{bmatrix} \nonumber \\
    & = \begin{bmatrix}
        A^{\widehat{\uparrow\downarrow}} & 0 & 0 & 0 \\
        0 & A^{\widehat{\downarrow\uparrow}}& 0 & 0 \\
        0 & 0 & A^{\uparrow\uparrow} & A^{\downarrow\uparrow} \\
        0 & 0 & A^{\uparrow\downarrow} & A^{\downarrow\downarrow}
    \end{bmatrix} \begin{bmatrix}
        B^{\widehat{\uparrow\downarrow}} & 0 & 0 & 0 \\
        0 & B^{\widehat{\downarrow\uparrow}}& 0 & 0 \\
        0 & 0 & B^{\uparrow\uparrow} & B^{\downarrow\uparrow} \\
        0 & 0 & B^{\uparrow\downarrow} & B^{\downarrow\downarrow}
    \end{bmatrix}.
\end{align}
In that way, we find for each diagrammatic channel
\begin{subequations}
    \begin{align}
       ph~:\quad [A\fcirc B]^{\widehat{\uparrow\downarrow}} &= A^{\widehat{\uparrow\downarrow}} B^{\widehat{\uparrow\downarrow}}, \nonumber \\
       [A\fcirc B]^{\uparrow\downarrow} &= A^{\uparrow\downarrow}B^{\uparrow\uparrow}+A^{\downarrow\downarrow}B^{\uparrow\downarrow}, \nonumber \\
       [A\fcirc B]^{\uparrow\uparrow} &= A^{\uparrow\uparrow}B^{\uparrow\uparrow}+A^{\downarrow\uparrow}B^{\uparrow\downarrow}, \\
       \overline{ph}~:\quad [A\fcirc B]^{\uparrow\downarrow} &= A^{\uparrow\downarrow} B^{\uparrow\downarrow}, \nonumber \\
       [A\fcirc B]^{\uparrow\uparrow} &= A^{\uparrow\uparrow}B^{\uparrow\uparrow}+A^{\widehat{\uparrow\downarrow}}B^{\widehat{\downarrow\uparrow}}, \nonumber \\
       [A\fcirc B]^{\widehat{\uparrow\downarrow}} &= A^{\uparrow\uparrow}B^{\widehat{\uparrow\downarrow}}+A^{\widehat{\uparrow\downarrow}}B^{\downarrow\downarrow}, \\
        pp~:\quad [A\fcirc B]^{\uparrow\uparrow} &= A^{\uparrow\uparrow} B^{\uparrow\uparrow}, \nonumber \\
       [A\fcirc B]^{\widehat{\uparrow\downarrow}} &= A^{\uparrow\downarrow}B^{\widehat{\uparrow\downarrow}}+A^{\widehat{\uparrow\downarrow}}B^{\downarrow\uparrow}, \nonumber \\
       [A\fcirc B]^{\uparrow\downarrow} &= A^{\uparrow\downarrow}B^{\uparrow\downarrow}+A^{\widehat{\uparrow\downarrow}}B^{\widehat{\downarrow\uparrow}}.
    \end{align}
\label{eq:ABFormulasfcircproduct}
\end{subequations}
One can straightforwardly extend this reasoning to products between more than two four-point objects. In the following, we will actually exploit the identities: 
\begin{subequations}
    \begin{align}
       ph~:\quad [A\fcirc B\fcirc C]^{\uparrow\uparrow} &= A^{\uparrow\uparrow} B^{\uparrow\uparrow} C^{\uparrow\uparrow} + A^{\uparrow\uparrow} B^{\downarrow\uparrow} C^{\uparrow\downarrow} \nonumber \\
       &\phantom{=} + A^{\downarrow\uparrow} B^{\uparrow\downarrow} C^{\uparrow\uparrow} + A^{\downarrow\uparrow} B^{\downarrow\downarrow} C^{\uparrow\downarrow}, \nonumber \\
       \quad [A\fcirc B\fcirc C]^{\uparrow\downarrow} &= A^{\uparrow\downarrow} B^{\uparrow\uparrow} C^{\uparrow\uparrow} + A^{\uparrow\downarrow} B^{\downarrow\uparrow} C^{\uparrow\downarrow} \nonumber \\
       &\phantom{=} + A^{\downarrow\downarrow} B^{\uparrow\downarrow} C^{\uparrow\uparrow} + A^{\downarrow\downarrow} B^{\downarrow\downarrow} C^{\uparrow\downarrow}, \\
       \overline{ph}~:\quad [A\fcirc B\fcirc C]^{\uparrow\downarrow} &= A^{\uparrow\downarrow} B^{\uparrow\downarrow} C^{\uparrow\downarrow}, \\
        pp~:\quad [A\fcirc B\fcirc C]^{\uparrow\downarrow} &= A^{\uparrow\downarrow} B^{\uparrow\downarrow} C^{\uparrow\downarrow} + A^{\widehat{\uparrow\downarrow}} B^{\widehat{\downarrow\uparrow}} C^{\uparrow\downarrow} \nonumber \\
        &\phantom{=} + A^{\uparrow\downarrow} B^{\widehat{\uparrow\downarrow}} C^{\widehat{\downarrow\uparrow}} + A^{\widehat{\uparrow\downarrow}} B^{\downarrow\uparrow} C^{\widehat{\downarrow\uparrow}}, \nonumber \\
       \quad [A\fcirc B\fcirc C]^{\widehat{\uparrow\downarrow}} &= A^{\uparrow\downarrow} B^{\uparrow\downarrow} C^{\widehat{\uparrow\downarrow}} + A^{\widehat{\uparrow\downarrow}} B^{\widehat{\downarrow\uparrow}} C^{\widehat{\uparrow\downarrow}} \nonumber \\
       &\phantom{=} + A^{\uparrow\downarrow} B^{\widehat{\uparrow\downarrow}} C^{\downarrow\uparrow} + A^{\widehat{\uparrow\downarrow}} B^{\downarrow\uparrow} C^{\downarrow\uparrow}.
    \end{align}
\label{eq:ABCFormulasfcircproduct}
\end{subequations}
In most cases, we will use Eqs.~\eqref{eq:ABCFormulasfcircproduct} with $B=\Pi_r$ (or $B=\dot{\Pi}_r$). However, some spin components of the bubbles $\Pi_r$ (or their derivatives $\dot{\Pi}_r$) vanish in the context of spin rotation invariance, since the $\uparrow\downarrow$ and $\downarrow\uparrow$ spin components of the full propagator $G$ vanish in that situation [see Eq.~\eqref{eq:FullPropagatorSU2symmetry}]. According to the definitions of the bubbles [see Eqs.~\eqref{eq:Bubbles}, or equivalently Eqs.~\eqref{eq:BubblesEvaluateSumsFreqMomenta}], we thus have $\Pi_{ph}^{\uparrow\downarrow}=\Pi_{ph}^{\downarrow\uparrow}=0$ and $\Pi_{pp}^{\widehat{\uparrow\downarrow}}=\Pi_{pp}^{\widehat{\downarrow\uparrow}}=0$. With this, Eqs.~\eqref{eq:ABCFormulasfcircproduct} become 
\begin{subequations}
    \begin{align}
       ph~:\quad [A\fcirc \Pi_{ph}\fcirc B]^{\uparrow\uparrow} &= A^{\uparrow\uparrow} \Pi_{ph}^{\uparrow\uparrow} B^{\uparrow\uparrow} + A^{\downarrow\uparrow} \Pi_{ph}^{\downarrow\downarrow} B^{\uparrow\downarrow}, \nonumber \\
       \quad [A\fcirc \Pi_{ph}\fcirc B]^{\uparrow\downarrow} &= A^{\uparrow\downarrow} \Pi_{ph}^{\uparrow\uparrow} B^{\uparrow\uparrow} + A^{\downarrow\downarrow} \Pi_{ph}^{\downarrow\downarrow} B^{\uparrow\downarrow}, \\
       \overline{ph}~:\quad [A\fcirc \Pi_{\overline{ph}}\fcirc B]^{\uparrow\downarrow} &= A^{\uparrow\downarrow} \Pi_{\overline{ph}}^{\uparrow\downarrow} B^{\uparrow\downarrow}, \\
        pp~:\quad [A\fcirc \Pi_{pp}\fcirc B]^{\uparrow\downarrow} &= A^{\uparrow\downarrow} \Pi_{pp}^{\uparrow\downarrow} B^{\uparrow\downarrow} + A^{\widehat{\uparrow\downarrow}} \Pi_{pp}^{\downarrow\uparrow} B^{\widehat{\downarrow\uparrow}}, \nonumber \\
       \quad [A\fcirc \Pi_{pp}\fcirc B]^{\widehat{\uparrow\downarrow}} &= A^{\uparrow\downarrow} \Pi_{pp}^{\uparrow\downarrow} B^{\widehat{\uparrow\downarrow}} + A^{\widehat{\uparrow\downarrow}} \Pi_{pp}^{\downarrow\uparrow} B^{\downarrow\uparrow},
    \end{align}
\label{eq:APiBFormulasfcircproduct}
\end{subequations}
where we relabeled $C$ as $B$ for consistency. We stress that Eqs.~\eqref{eq:APiBFormulasfcircproduct} are still valid if the bubbles $\Pi_r$ are replaced by their derivatives $\dot{\Pi}_r$.

We are now equipped to derive the multiloop SBE fRG equations in physical channels. We will illustrate how to obtain these equations by focusing on the $1\ell$ contribution of the bosonic propagators. Specifically, we will start from the flow equations in diagrammatic channels that express $w_r^{(1\ell)}$ for $r=ph,\overline{ph},pp$ and determine the corresponding expressions of $w_{\text{X}}^{(1\ell)}$ for $\text{X}=\text{M},\text{D},\text{SC}$. According to the definition of the physical channels given by Eqs.~\eqref{eq:DefinitionsPhysicalChannels}, we have
\begin{subequations}
    \begin{align}
       \dot{w}^{(1\ell)}_{\text{M}} & = - \dot{w}^{(1\ell)\uparrow\downarrow}_{\overline{ph}}, \\
       \dot{w}^{(1\ell)}_{\text{D}} & = \dot{w}^{(1\ell)\uparrow\uparrow}_{ph} + \dot{w}^{(1\ell)\uparrow\downarrow}_{ph}, \\
       \dot{w}^{(1\ell)}_{\text{SC}} & = \dot{w}^{(1\ell)\uparrow\downarrow}_{pp}.
    \end{align}
    \label{eq:DerivationwX1loopStep1}
\end{subequations}
The right-hand sides of those three equations can be rewritten with the help of the corresponding flow equations for $w_{ph}^{(1\ell)}$, $w_{\overline{ph}}^{(1\ell)}$, and $w_{pp}^{(1\ell)}$ [see Eq.~\eqref{eq:SBEmfRGEq_1l_DiagChannels_wr}]. This gives
\begin{subequations}
    \begin{align}
       \dot{w}^{(1\ell)}_{\text{M}} & = - \left[w_{\overline{ph}} \fcirc \lambda_{\overline{ph}} \circ \dot{\Pi}_{\overline{ph}} \circ \lambda_{\overline{ph}} \fcirc w_{\overline{ph}}\right]^{\uparrow\downarrow}, \\
       \dot{w}^{(1\ell)}_{\text{D}} & = \left[w_{ph} \fcirc \lambda_{ph} \circ \dot{\Pi}_{ph} \circ \lambda_{ph} \fcirc w_{ph}\right]^{\uparrow\uparrow} \nonumber \\
       & \phantom{=} + \left[w_{ph} \fcirc \lambda_{ph} \circ \dot{\Pi}_{ph} \circ \lambda_{ph} \fcirc w_{ph}\right]^{\uparrow\downarrow}, \\
       \dot{w}^{(1\ell)}_{\text{SC}} & = \left[w_{pp} \fcirc \lambda_{pp} \circ \dot{\Pi}_{pp} \circ \lambda_{pp} \fcirc w_{pp}\right]^{\uparrow\downarrow}.
    \end{align}
    \label{eq:DerivationwX1loopStep2}
\end{subequations}
We can then evaluate some of the sums over spin indices by making use of Eqs.~\eqref{eq:ABCFormulasfcircproduct}, with $A=C=w_r$ and $B=\lambda_r \circ \dot{\Pi}_r \circ \lambda_r$ for $r=ph,\overline{ph},pp$. This leads to
\begin{subequations}
    \begin{align}
       \dot{w}^{(1\ell)}_{\text{M}} & = -\left(w_{\text{M}}\right)^2\left[\lambda_{\overline{ph}} \circ \dot{\Pi}_{\overline{ph}} \circ \lambda_{\overline{ph}}\right]^{\uparrow\downarrow}, \\
       \dot{w}^{(1\ell)}_{\text{D}} & = \left(w_{\text{D}}\right)^2 \bigg(\left[\lambda_{ph} \circ \dot{\Pi}_{ph} \circ \lambda_{ph}\right]^{\uparrow\uparrow} \nonumber \\
       & \phantom{= \left(w_{\text{D}}\right)^2 \bigg(} + \left[\lambda_{ph} \circ \dot{\Pi}_{ph} \circ \lambda_{ph}\right]^{\uparrow\downarrow}\bigg), \\
       \dot{w}^{(1\ell)}_{\text{SC}} & = 2\left(w_{\text{SC}}\right)^2 \bigg(\left[\lambda_{pp} \circ \dot{\Pi}_{pp} \circ \lambda_{pp}\right]^{\uparrow\downarrow} \nonumber \\
       & \phantom{= 2\left(w_{\text{SC}}\right)^2 \bigg(} - \left[\lambda_{pp} \circ \dot{\Pi}_{pp} \circ \lambda_{pp}\right]^{\widehat{\uparrow\downarrow}}\bigg), \label{eq:DerivationwSC1loopStep3}
    \end{align}
    \label{eq:DerivationwX1loopStep3}
\end{subequations}
where $w_{\text{M}}=-w_{\overline{ph}}^{\uparrow\downarrow}$, $w_{\text{D}}=w_{ph}^{\uparrow\uparrow}+w_{ph}^{\uparrow\downarrow}$, and $w_{\text{SC}}=w^{\uparrow\downarrow}_{pp}$. Note that we also used the crossing symmetry of $w_{pp}$ that implies $w_{pp}^{\widehat{\uparrow\downarrow}}=-w_{pp}^{\uparrow\downarrow}$ to obtain Eq.~\eqref{eq:DerivationwSC1loopStep3}. We evaluate the sums over frequencies and momenta encoded in the $\circ$ product by using Eqs.~\eqref{eq:EvaluateSumsFreqMomenta}  and thus rewrite Eqs.~\eqref{eq:DerivationwX1loopStep3} as
\begin{widetext}
\begin{subequations}
    \begin{align}
       \dot{w}^{(1\ell)}_{\text{M}}(Q) & = -\left(w_{\text{M}}(Q)\right)^2 \sum_k \left[\lambda_{\overline{ph}}(Q,k) \fcirc \dot{\Pi}_{\overline{ph}}(Q,k) \fcirc \lambda_{\overline{ph}}(Q,k)\right]^{\uparrow\downarrow}, \\
       \dot{w}^{(1\ell)}_{\text{D}}(Q) & = \left(w_{\text{D}}(Q)\right)^2 \sum_k \bigg(\left[\lambda_{ph}(Q,k) \fcirc \dot{\Pi}_{ph}(Q,k) \fcirc \lambda_{ph}(Q,k)\right]^{\uparrow\uparrow} + \left[\lambda_{ph}(Q,k) \fcirc \dot{\Pi}_{ph}(Q,k) \fcirc \lambda_{ph}(Q,k)\right]^{\uparrow\downarrow}\bigg), \\
       \dot{w}^{(1\ell)}_{\text{SC}}(Q) & = 2\left(w_{\text{SC}}(Q)\right)^2 \sum_k \bigg(\left[\lambda_{pp}(Q,k) \fcirc \dot{\Pi}_{pp}(Q,k) \fcirc \lambda_{pp}(Q,k)\right]^{\uparrow\downarrow} - \left[\lambda_{pp}(Q,k) \fcirc \dot{\Pi}_{pp}(Q,k) \fcirc \lambda_{pp}(Q,k)\right]^{\widehat{\uparrow\downarrow}}\bigg).
    \end{align}
    \label{eq:DerivationwX1loopStep4}
\end{subequations}
\end{widetext}
Finally, the remaining $\fcirc$ products are directly treated with the help of Eqs.~\eqref{eq:APiBFormulasfcircproduct}, yielding
\begin{equation}
    \dot{w}_{\text{X}}^{(1)}(Q) = \left(w_{\text{X}}(Q)\right)^2 \sum_k \lambda_{\text{X}}(Q,k) \dot{\Pi}_{\text{X}}(Q,k) \lambda_{\text{X}}(Q,k)
    \label{eq:flowequationswXPhysicalChannels1l}
\end{equation}
for $\text{X}=\text{M},\text{D},\text{SC}$. 
The Yukawa couplings in physical channels $\lambda_{\text{X}}$ have been introduced in Eqs.~\eqref{eq:DefinitionslambdasPhysicalChannels}, while the bubbles $\Pi_{\text{X}}$ are defined by Eqs.~\eqref{eq:BubblesEvaluateSumsFreqMomenta} as follows:
\begin{subequations}
        \begin{align}
        \Pi_{\text{M}}(Q,k) & = \Pi_{\text{D}}(Q,k) = -\Pi_{\overline{ph}}^{\uparrow\downarrow}(Q,k) = \Pi_{ph}^{\uparrow\uparrow}(Q,k) \nonumber \\
        & = -G\bigg(\mathbf{k}+\mathbf{Q},\nu+\left\lceil\frac{\Omega}{2}\right\rceil\bigg)G\bigg(\mathbf{k},\nu-\left\lfloor\frac{\Omega}{2}\right\rfloor\bigg), \\
        \Pi_{\text{SC}}(Q,k) & = 2\Pi_{pp}^{\uparrow\downarrow}(Q,k) \nonumber \\
        & = G\bigg(\mathbf{Q}-\mathbf{k},\left\lfloor\frac{\Omega}{2}\right\rfloor-\nu\bigg)G\bigg(\mathbf{k},\nu+\left\lceil\frac{\Omega}{2}\right\rceil\bigg),
        \end{align}
        \label{eq:BubblesPiX}
    \end{subequations}
where $G(k)=G(\mathbf{k},\nu)$ is the diagonal part of the full propagator on the right-hand side of Eq.~\eqref{eq:GMomentum}. Hence, Eq.~\eqref{eq:flowequationswXPhysicalChannels1l} is the $1\ell$ flow equation for the bosonic propagators in physical channels. The corresponding multiloop SBE fRG equations for the bosonic propagators, the Yukawa couplings, and the SBE rest functions can be derived analogously. In that way, we find the following flow equations:
\begin{subequations}
    \begin{align}
        \dot{w}_{\text{X}}^{(1)}(Q) & = \left(w_{\text{X}}(Q)\right)^2 \sum_k \lambda_{\text{X}}(Q,k) \dot{\Pi}_{\text{X}}(Q,k) \lambda_{\text{X}}(Q,k), \label{eq:flowequationswXPhysicalChannels1lv2} \\
        \dot{\lambda}_{\text{X}}^{(1)}(Q,k) & = \sum_{k^{\prime}} \lambda_{\text{X}}(Q,k^{\prime}) \dot{\Pi}_{\text{X}}(Q,k^{\prime}) \mathcal{I}_{\text{X}}(Q,k^{\prime},k), \label{eq:flowequationslambdaXPhysicalChannels1lv2} \\
        \dot{M}_{\text{X}}^{(1)}(Q,k,k^{\prime}) & = \sum_{k^{\prime\prime}} \mathcal{I}_{\text{X}}(Q,k,k^{\prime\prime}) \dot{\Pi}_{\text{X}}(Q,k^{\prime\prime}) \mathcal{I}_{\text{X}}(Q,k^{\prime\prime},k^{\prime}), \label{eq:flowequationsMXPhysicalChannels1l}
    \end{align}
    \label{eq:mfRGequationsPhysicalChannels1l}
\end{subequations}
for $1\ell$, with the bubbles defined by Eqs.~\eqref{eq:BubblesPiX},
\begin{subequations}
    \begin{align}
        \dot{w}_{\text{X}}^{(2)}(Q) & = 0, \\
        \dot{\lambda}_{\text{X}}^{(2)}(Q,k) & = \sum_{k^{\prime}} \lambda_{\text{X}}(Q,k^{\prime}) \Pi_{\text{X}}(Q,k^{\prime}) \dot{I}_{\text{X}}^{(1)}(Q,k^{\prime},k), \\
        \dot{M}_{\text{X}}^{(2)}(Q,k,k^{\prime}) & = \sum_{k^{\prime\prime}} \dot{I}_{\text{X}}^{(1)}(Q,k,k^{\prime\prime}) \Pi_{\text{X}}(Q,k^{\prime\prime}) \mathcal{I}_{\text{X}}(Q,k^{\prime\prime},k^{\prime}) \nonumber\\&\phantom{=} \hspace{-0.25cm}+ \sum_{k^{\prime\prime}} \mathcal{I}_{\text{X}}(Q,k,k^{\prime\prime}) \Pi_{\text{X}}(Q,k^{\prime\prime}) \dot{I}_{\text{X}}^{(1)}(Q,k^{\prime\prime},k^{\prime}), \label{eq:flowequationsMXPhysicalChannels2l}
    \end{align}
    \label{eq:mfRGequationsPhysicalChannels2l}
\end{subequations}
for $2\ell$, and for $\ell\geq 3$,
\begin{widetext}
\begin{subequations}
    \begin{align}
        \dot{w}_{\text{X}}^{(\ell)}(Q) & = \left(w_{\text{X}}(Q)\right)^2 \sum_{k,k^{\prime}} \lambda_{\text{X}}(Q,k) \Pi_{\text{X}}(Q,k) \dot{I}^{(\ell-2)}_{\text{X}}(Q,k,k^{\prime}) \Pi_{\text{X}}(Q,k^{\prime}) \lambda_{\text{X}}(Q,k^{\prime}), \\
        \dot{\lambda}_{\text{X}}^{(\ell)}(Q,k) & = \sum_{k^{\prime}} \lambda_{\text{X}}(Q,k^{\prime}) \Pi_{\text{X}}(Q,k^{\prime}) \dot{I}_{\text{X}}^{(\ell-1)}(Q,k^{\prime},k) \nonumber \\
        & \phantom{=} + \sum_{k^{\prime},k^{\prime\prime}} \lambda_{\text{X}}(Q,k^{\prime}) \Pi_{\text{X}}(Q,k^{\prime}) \dot{I}^{(\ell-2)}_{\text{X}}(Q,k^{\prime},k^{\prime\prime}) \Pi_{\text{X}}(Q,k^{\prime\prime}) \mathcal{I}_{\text{X}}(Q,k^{\prime\prime},k), \\
        \dot{M}_{\text{X}}^{(\ell)}(Q,k,k^{\prime}) & = \sum_{k^{\prime\prime}} \dot{I}_{\text{X}}^{(\ell-1)}(Q,k,k^{\prime\prime}) \Pi_{\text{X}}(Q,k^{\prime\prime}) \mathcal{I}_{\text{X}}(Q,k^{\prime\prime},k^{\prime})  + \sum_{k^{\prime\prime}} \mathcal{I}_{\text{X}}(Q,k,k^{\prime\prime}) \Pi_{\text{X}}(Q,k^{\prime\prime}) \dot{I}_{\text{X}}^{(\ell-1)}(Q,k^{\prime\prime},k^{\prime}) \nonumber \\
        & \phantom{=} + \sum_{k^{\prime\prime},k^{\prime\prime\prime}} \mathcal{I}_{\text{X}}(Q,k,k^{\prime\prime}) \Pi_{\text{X}}(Q,k^{\prime\prime}) \dot{I}^{(\ell-2)}_{\text{X}}(Q,k^{\prime\prime},k^{\prime\prime\prime}) \Pi_{\text{X}}(Q,k^{\prime\prime\prime}) \mathcal{I}_{\text{X}}(Q,k^{\prime\prime\prime},k^{\prime}), \label{eq:flowequationsMXPhysicalChannelsbeyond2l}
    \end{align}
    \label{eq:mfRGequationsPhysicalChannelsbeyond2l}
\end{subequations}
\end{widetext}
where $\text{X}=\text{M},\text{D},\text{SC}$. The vertices $\mathcal{I}_{\text{X}}$ can be expressed directly in terms of $w_{\text{X}}$, $\lambda_{\text{X}}$, and $M_{\text{X}}$, whereas the derivatives $\dot{I}^{(\ell)}_{\text{X}}$ are also related to $\dot{w}^{(\ell)}_{\text{X}}$, $\dot{\lambda}^{(\ell)}_{\text{X}}$, and $\dot{M}^{(\ell)}_{\text{X}}$. For $\mathcal{I}_{\text{X}}$, this translates into
\begin{subequations}
\begin{align}
    \mathcal{I}_{\text{M}} & = M_{\text{M}} + \frac{1}{2} P^{ph\rightarrow \overline{ph}}\left(\phi_{\text{M}}-\phi_{\text{D}}\right) - P^{pp\rightarrow \overline{ph}}\phi_{\text{SC}}, \\
    \mathcal{I}_{\text{D}} & = M_{\text{D}} - 2 P^{\overline{ph} \rightarrow ph}\phi_{\text{M}} + 2 P^{pp\rightarrow ph}\phi_{\text{SC}} - P^{pp\rightarrow \overline{ph}}\phi_{\text{SC}} \nonumber \\
    & \phantom{=} + \frac{1}{2} P^{ph\rightarrow \overline{ph}}\left(\phi_{\text{M}}-\phi_{\text{D}}\right), \\
    \mathcal{I}_{\text{SC}} & = M_{\text{SC}} + \frac{1}{2} P^{ph\rightarrow pp}\left(\phi_{\text{D}} - \phi_{\text{M}}\right) - P^{\overline{ph}\rightarrow pp}\phi_{\text{M}},
\end{align}
\label{eq:ExpressionsMathcalIX}
\end{subequations}
where
\begin{equation}
\phi_{\text{X}}=\nabla_{\text{X}}+M_{\text{X}}-U_{\text{X}},
\label{eq:ExpressionphiX}
\end{equation}
with $\nabla_{\text{X}}(Q,k,k^{\prime})=\lambda_{\text{X}}(Q,k)w_{\text{X}}(Q)\lambda_{\text{X}}(Q,k^{\prime})$, and $U_{\text{X}}$ is defined consistently with Eqs.~\eqref{eq:DefinitionsPhysicalChannels}, i.e.,
\begin{subequations}
    \begin{align}
       U_{\text{M}} & = U^{\uparrow\uparrow} - U^{\uparrow\downarrow} = - U^{\uparrow\downarrow}, \\
       U_{\text{D}} & = U^{\uparrow\uparrow} + U^{\uparrow\downarrow}, \\
       U_{\text{SC}} & = U^{\uparrow\downarrow}.
    \end{align}
\end{subequations}
The projection matrices $P^{r\rightarrow r^{\prime}}$ translate the conventions between our different channel notations (defined by Fig.~\ref{fig:FrequencyMomentumParametrization}) for frequencies and momenta. We can also use these projection matrices to derive the following relations for $\dot{I}^{(\ell)}_{\text{X}}(Q,k,k^{\prime})$:
\begin{subequations}
\begin{align}
    \dot{I}_{\text{M}}^{(\ell)} & = \frac{1}{2} P^{ph\rightarrow \overline{ph}}\left(\dot{\phi}_{\text{M}}^{(\ell)}-\dot{\phi}_{\text{D}}^{(\ell)}\right) - P^{pp\rightarrow \overline{ph}}\dot{\phi}_{\text{SC}}^{(\ell)}, \\
    \dot{I}_{\text{D}}^{(\ell)} & = - 2 P^{\overline{ph} \rightarrow ph}\dot{\phi}_{\text{M}}^{(\ell)} + 2 P^{pp\rightarrow ph}\dot{\phi}_{\text{SC}}^{(\ell)} - P^{pp\rightarrow \overline{ph}}\dot{\phi}_{\text{SC}}^{(\ell)} \nonumber \\
    &\phantom{=} + \frac{1}{2} P^{ph\rightarrow \overline{ph}}\left(\dot{\phi}_{\text{M}}^{(\ell)}-\dot{\phi}_{\text{D}}^{(\ell)}\right), \\
    \dot{I}_{\text{SC}}^{(\ell)} & = \frac{1}{2} P^{ph\rightarrow pp}\left(\dot{\phi}_{\text{D}}^{(\ell)} - \dot{\phi}_{\text{M}}^{(\ell)}\right) - P^{\overline{ph}\rightarrow pp}\dot{\phi}_{\text{M}}^{(\ell)},
\end{align}
\label{eq:ExpressionsdotIellX}
\end{subequations}
with
\begin{equation}
\dot{\phi}^{(\ell)}_{\text{X}}=\dot{\nabla}^{(\ell)}_{\text{X}}+\dot{M}^{(\ell)}_{\text{X}},
\end{equation}
and
\begin{align}
\dot{\nabla}^{(\ell)}_{\text{X}}(Q,k,k^{\prime})&=\dot{\lambda}^{(\ell)}_{\text{X}}(Q,k)w_{\text{X}}(Q)\lambda_{\text{X}}(Q,k^{\prime}) \nonumber \\
& \phantom{=} +\lambda_{\text{X}}(Q,k)\dot{w}^{(\ell)}_{\text{X}}(Q)\lambda_{\text{X}}(Q,k^{\prime}) \nonumber \\
& \phantom{=} +\lambda_{\text{X}}(Q,k)w_{\text{X}}(Q)\dot{\lambda}^{(\ell)}_{\text{X}}(Q,k^{\prime}).
\end{align}

Equations~\eqref{eq:mfRGequationsPhysicalChannels1l}--\eqref{eq:mfRGequationsPhysicalChannelsbeyond2l}, together with Eqs.~\eqref{eq:ExpressionsMathcalIX}--\eqref{eq:ExpressionsdotIellX}, constitute the multiloop SBE fRG equations for the two-particle vertex. To summarize, the equations for a fermionic model with the classical action~\eqref{eq:ClassicalActionS} have been obtained by using the frequency and momentum notations depicted in Fig.~\ref{fig:FrequencyMomentumParametrization}, assuming translational invariance and energy conservation [see Eqs.~\eqref{eq:FreqMomConservationLawsG0U}], SU(2) spin symmetry [see Eqs.~\eqref{eq:GVSU2symmetry}], and time-reversal symmetry [see Eq.~\eqref{eq:lambdabarXtimereversal}].

The flow equations for the two-particle vertex (namely, for $w_{\text{X}}$, $\lambda_{\text{X}}$, and $M_{\text{X}}$ in the SBE framework) go hand in hand with a flow equation for the self-energy that can be derived by differentiating the Schwinger-Dyson equation~\cite{Hille2020,HillePhDThesis}, thus reflecting the procedure leading to the multiloop equations for the two-particle vertex from the Bethe-Salpeter equations. Compared to the multiloop extension of the conventional self-energy flow equation determined by Kugler and von Delft~\cite{Kugler2018b,kugler2018c}, 
this form is preferable in the truncated-unity fRG~\cite{Husemann2009,Wang2012,Lichtenstein2017}, where the fermionic momentum dependence is expanded in form factors. Since the implementations of 2D systems widely rely on the truncated-unity fRG (see, e.g., Refs.~\cite{Schober2018,Profe2022,Gneist2022,Bonetti2022,Fraboulet2022,Beyer2023,Profe2024}), we consider the flow equation obtained from the derivative of the Schwinger-Dyson equation in the numerical treatment of the Hubbard model presented in Sec.~\ref{sec:Results}. Furthermore, it was 
recently shown~\cite{Patricolo2025} that, in the SBE framework, the Schwinger-Dyson equation exhibits a particularly simple one-loop form. For the repulsive Hubbard model that we will consider in the following sections, the formulation in the dominating magnetic channel is advantageous~\cite{Patricolo2025}. It reads
\begin{equation}
    \Sigma(k) = \sum_{Q} w_{\mathrm{M}}(Q) \lambda_{\mathrm{M}}\Big(Q,\widetilde{k}\Big) G(k+Q),
    \label{eq:SigmaMEquation}
\end{equation}
in the frequency and momentum notations defined by Fig.~\ref{fig:FrequencyMomentumParametrization}, with $\widetilde{k}=\left(\mathbf{k},\nu + \left\lfloor\frac{\Omega}{2}\right\rfloor\right)$, $Q=(\mathbf{Q},\Omega)$, and $k=(\mathbf{k},\nu)$. Introducing the flow parameter $\Lambda$ in the bare propagator $G_0$ as before ($G_0\rightarrow G_0^\Lambda$), the corresponding self-energy flow equation therefore reads
\begin{align}
    \dot{\Sigma}(k) & = \sum_{Q} \dot{w}_{\mathrm{M}}(Q) \lambda_{\mathrm{M}}\Big(Q,\widetilde{k}\Big) G(k+Q) \nonumber \\
    & \phantom{=} + \sum_{Q} w_{\mathrm{M}}(Q) \dot{\lambda}_{\mathrm{M}}\Big(Q,\widetilde{k}\Big) G(k+Q) \nonumber \\
    & \phantom{=} + \sum_{Q} w_{\mathrm{M}}(Q) \lambda_{\mathrm{M}}\Big(Q,\widetilde{k}\Big) \dot{G}(k+Q). \label{eq:SigmaMdotFlowEquation}
\end{align}
Although Eqs.~\eqref{eq:SigmaMEquation} and~\eqref{eq:SigmaMdotFlowEquation} were derived under the same assumptions as the flow equations for the two-particle vertex [Eqs.~\eqref{eq:mfRGequationsPhysicalChannels1l}--\eqref{eq:mfRGequationsPhysicalChannelsbeyond2l}], we note that its extension to nonlocal interactions is less straightforward \footnote{While the flow equations for $w_{\text{X}}(Q)$, $\lambda_{\text{X}}(Q,k)$, and $M_{\text{X}}(Q,k,k^\prime)$ are also valid for nonlocal interactions, i.e., for a nonzero fermionic bare interaction $\mathcal{F}_r$ in Eq.~\eqref{eq:UrBrFr}, as long as the SBE parametrization $\nabla_r(Q,k,k^\prime)=\lambda_r(Q,k)w_r(Q)\lambda_r(Q,k^\prime)$ applies to the $\mathcal{B}$-reducible part of the full vertex $V$ (and not its $U$-reducible part), a nontrivial fermionic bare interaction $\mathcal{F}_r$ can generate extra terms to the right-hand side of Eq.~\eqref{eq:SigmaMEquation} or~\eqref{eq:SigmaMdotFlowEquation}, even after redefining $\nabla_r$ as the sum of $\mathcal{B}$-reducible diagrams (see Appendix~B of Ref.~\cite{Patricolo2025} for more details).}. This point is, however, not relevant for the present study of the Hubbard model since $\mathcal{F}_r=0$ for all $r$.

Importantly, the self-energy flow equation~\eqref{eq:SigmaMdotFlowEquation} is \emph{exact}. Therefore, the only \emph{approximation} used within the multiloop fRG setup based on Eqs.~\eqref{eq:mfRGequationsPhysicalChannels1l}--\eqref{eq:mfRGequationsPhysicalChannelsbeyond2l} and~\eqref{eq:SigmaMdotFlowEquation} consists in neglecting the flow of the 2PI part of the vertex, i.e., $\dot{I}^{\text{2PI}} = 0$ [see Eq.~\eqref{eq:I2PIapprox}]. In the conventional cutoff schemes, where $G_0^\Lambda$ is chosen such that the starting point of the flow coincides with the bare or the noninteracting theory (this applies to all cutoff schemes used in this study; see Sec.~\ref{sec:CutoffSchemes}), this approximation leads to the \emph{parquet approximation}~\cite{Kugler2018a}. An important question we address here is whether one can reproduce the parquet approximation with the multiloop SBE fRG by ignoring the flow of the SBE rest functions $M_{\text{X}}$, i.e., by discarding Eqs.~\eqref{eq:flowequationsMXPhysicalChannels1l},~\eqref{eq:flowequationsMXPhysicalChannels2l}, and~\eqref{eq:flowequationsMXPhysicalChannelsbeyond2l}. We will refer to this scheme as the \emph{SBE approximation} in the following~\footnote{What we are describing here amounts to neglecting the SBE rest functions \emph{after} deriving the flow equations, and is defined as SBEa approximation in Ref.~\cite{Fraboulet2022}. It has to be distinguished from the SBEb approximation obtained by neglecting the SBE rest functions \emph{before} deriving the flow equations, in the sense that $M_r$ is neglected in the self-consistent equations for the SBE objects before introducing the flow parameter $\Lambda$ via $G_0\rightarrow G_0^\Lambda$ to turn them into flow equations. The SBEa and SBEb approximations have been compared in Ref.~\cite{Fraboulet2022}, showing that the SBEa scheme generates an implicit resummation of multiboson exchange contributions not present in the SBEb framework. In contrast, the SBEb results remain cutoff independent by construction. In the present work, the SBE approximation always refers to the SBEa approximation (see also the analysis of Fig.~\ref{fig:chiMDlambdaMDvsQPA} for a related discussion).}. Before analyzing our multiloop SBE fRG results for the 2D Hubbard model, and especially the quality of the SBE approximation in Sec.~\ref{sec:Results}, we will present in more detail how the underlying flow equations are implemented and solved numerically in Sec.~\ref{sec:NumericalImplementation}.

\section{Application to the Hubbard model and numerical implementation}
\label{sec:NumericalImplementation}

\subsection{Model}

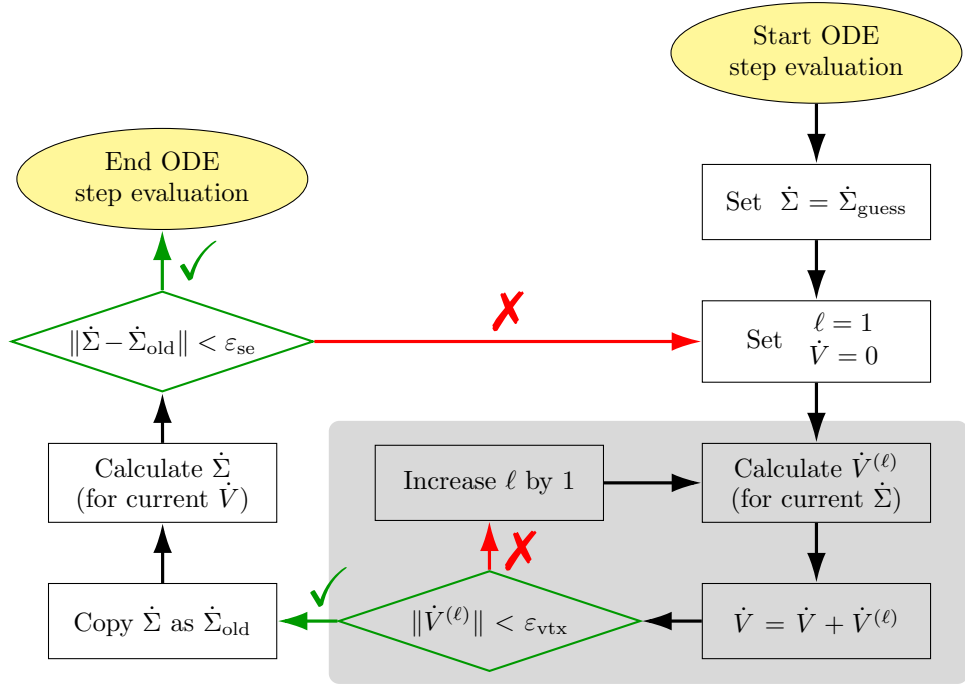
\begin{figure*}
    \centering
    \begin{tikzpicture}[node distance=0.8cm]

\tikzset{
      startend/.style={ellipse, draw, fill=gray!30, text width=2.5cm, align=center, minimum height=1cm},
      start/.style={ellipse, draw, fill=yellow!50, text width=2.5cm, align=center, minimum height=1cm},
      process/.style={rectangle, draw, text width=2.8cm, align=center, minimum height=1cm},
      decision/.style={diamond, aspect=3, draw=green!60!black, thick, text width=2.5cm, align=center, inner sep=1pt},
      box/.style={rectangle, draw=none, fill=gray!30, inner sep=6pt, rounded corners},
      arrow/.style={-{Latex[length=4mm, width=2.5mm]}, line width=1.2pt},
      good/.style={-{Latex[length=4mm, width=2.5mm]}, line width=1.2pt, green!70!black},
      bad/.style={-{Latex[length=4mm, width=2.5mm]}, line width=1.2pt, red},
      markgood/.style={text=green!70!black, font=\bfseries\Large},
      markbad/.style={text=red, font=\bfseries\Large},
    }

    % Start node
    \node[start] (start) {Start ODE \\ step evaluation};

    % First process
    \node[process, below=of start] (setSigma) {Set \ $\dot{\Sigma} = \dot{\Sigma}_{\text{guess}}$};

    % Set ell and Vdot
    \node[process, below=of setSigma] (init) {Set \ $\begin{array}{c}
    \ell = 1 \\
    \dot{V} = 0
    \end{array}$};

    % Gray box for iteration
    \node[box, below right=0.5cm and 0.5cm of init, minimum width=8.5cm, minimum height=3.5cm, anchor=north east] (graybox) {};

    % Inside gray box
    \node[process, below=of init] (calcV) {Calculate $\dot{V}^{(\ell)}$ \\ (for current $\dot{\Sigma}$)};
    \node[process, below=of calcV] (updateV) {$\dot{V} = \dot{V} + \dot{V}^{(\ell)}$};
    \node[decision, left=of updateV] (checkV) {$\|\dot{V}^{(\ell)}\| < \varepsilon_{\text{vtx}}$};
    \node[process, left=1.3cm of calcV] (incell) {Increase $\ell$ by 1};

    % Copy Sigma_old
    \node[process, left=of checkV] (copySigma) {Copy $\dot{\Sigma}$ as $\dot{\Sigma}_{\text{old}}$};

    % Calculate Sigma
    \node[process, above=of copySigma] (calcSigma) {Calculate $\dot{\Sigma}$ \\ (for current $\dot{V}$)};

    % Check convergence Sigma
    \node[decision, left=5.15cm of init] (checkSigma) {$\|\dot{\Sigma} - \dot{\Sigma}_{\text{old}}\| < \varepsilon_{\text{se}}$};

    % End node
    \node[start, above=of checkSigma] (end) {End ODE \\ step evaluation};

    % Connections
    \draw[arrow] (start) -- (setSigma);
    \draw[arrow] (setSigma) -- (init);
    \draw[arrow] (init) -- (calcV);
    \draw[arrow] (calcV) -- (updateV);
    \draw[arrow] (updateV) -- (checkV);
    \draw[arrow] (incell) -- (calcV);
    \draw[arrow] (copySigma) -- (calcSigma);
    \draw[arrow] (calcSigma) -- (checkSigma);
    \draw[arrow,green!60!black] (checkV) -- (copySigma) node[midway, above=2pt, xshift=8pt] {\huge \checkmark};
    \draw[arrow,red] (checkV) -- (incell) node[midway, right=2pt, yshift=-2pt] {\huge \ding{55}};
    \draw[arrow,green!60!black] (checkSigma) -- (end) node[midway, right=1pt] {\huge \checkmark};
    \draw[arrow,red] (checkSigma) -- (init) node[midway, above=0pt] {\huge \ding{55}};
\end{tikzpicture}
    \caption{Flow chart: At each integration step of the multiloop equations, there is a self-consistency cycle between the self-energy derivative $\dot{\Sigma}$ and the vertex derivative $\dot{V}$. The (multiloop) vertex corrections are considered until higher-order corrections are small, indicated by the gray box. The resulting vertex derivative is then used to calculate a new self-energy derivative, which---if different---necessitates recalculating the vertex corrections. Each cycle of updating the self-energy derivative and recalculating the (multiloop) vertex corrections at a given step is referred to as a \emph{self-energy iteration}.}
    \label{fig:self_consistent_problem_every_ODE_step}
\end{figure*}

We consider the Hubbard model on the square lattice defined by the Hamiltonian
\begin{align}
H = \sum_{\bfk,\sigma} (\epsilon(\bfk) - \mu_0)c^\dagger_{\bfk\sigma}c_{\bfk \sigma} + U \sum_{\bfQ, \bfk, \bfk^\prime} c^\dagger_{\bfk+\bfQ \uparrow}c^\dagger_{\bfk^\prime - \bfQ \downarrow} c_{\bfk^\prime \downarrow}c_{\bfk \uparrow},
\end{align}
where the bold symbols $\bfk$ and $\bfQ$ are momentum vectors on a square Brillouin zone $[0,2\pi)\times[0,2\pi)$ and $\sigma \in \{\uparrow,\downarrow\}$ denotes the spin. The integration over momenta is normalized by the Brillouin zone volume. $c^{(\dagger)}_{\bfk\sigma}$ represent annihilation (creation) operators for the fermions with the dispersion given by
\begin{align}
\epsilon(\bfk) = -2t(\cos k_x + \cos k_y) - 4t' \cos k_x \cos k_y.
\end{align}
In real space, $t$ and $t'$ correspond to nearest-neighbor and next-nearest-neighbor hopping amplitudes, respectively.
The chemical potential shift $\mu_0$ is determined by the filling. The bare (repulsive) Hubbard $U>0$ describes a local interaction in real space. By translational invariance and SU(2) spin symmetry of the model~\cite{Rohringer2018}, the bare Matsubara Green’s function further simplifies beyond Eq.~\eqref{eq:FreqMomConservationLawsG0} to
\begin{align}
G_{0;\sigma_{1^\prime}|\sigma_1}(k) = G_0(\bfk,\nu) \delta_{\sigma_{1^\prime},\sigma_1},
\end{align}
with
\begin{align}
G_0(\bfk,\nu) = \frac{1}{i\nu - \epsilon(\bfk) + \mu_0}.
\end{align}
Similarly, the bare interaction vertex simplifies beyond Eq.~\eqref{eq:FreqMomConservationLawsU} to
\begin{align}
U_{\sigma_{1^\prime}\sigma_{2^\prime}|\sigma_1\sigma_2} (Q_r,k_r,k^\prime_{r}) &= -U (1-\delta_{\sigma_1,\sigma_2}) \nonumber\\
&\phantom{=}\times (\delta_{\sigma_{1^\prime}, \sigma_1} \delta_{\sigma_{2^\prime}, \sigma_2} - \delta_{\sigma_{1^\prime}, \sigma_2} \delta_{\sigma_{2^\prime}, \sigma_1}).
\end{align}
In this work, we use $t \equiv 1$ as energy unit.

\subsection{Susceptibilities}

The linear response of a system to an external perturbation coupled to the operator $\hat{O}(\bfQ,\tau)$ is encoded in the susceptibility
\begin{align}
\chi_O(q) = \int_0^\beta \dd{\tau} e^{i\Omega \tau}&\left( \expval{T_\tau \hat{O}(\bfQ, \tau)\hat{O}(\bfQ, 0)}\right. \nonumber \\   & \hspace{-0.16cm}-
\left.\expval{\hat{O}(\bfQ, \tau)}\expval{\hat{O}(\bfQ, 0)} \right),
\end{align}
where $\tau$ is the imaginary time. We here consider for $\hat{O}$ the ($s$-wave) spin in the $z$ direction, ($s$-wave) density, and $s$-wave and $d$-wave pairing operators
\begin{subequations}
\begin{align}
\hat{s}_z(\bfQ,\tau) &= \frac{1}{2}(\hat{n}_{\uparrow}(\bfQ,\tau) - \hat{n}_{\downarrow}(\bfQ,\tau)),\\
\hat{\rho}(\bfQ,\tau) &= \frac{1}{2}(\hat{n}_{\uparrow}(\bfQ,\tau) + \hat{n}_{\downarrow}(\bfQ,\tau)),\\
\hat{\Delta}_{\swave/\dwave}(\bfQ,\tau) &=\frac{1}{2} \sum_{\bfk} f_{\swave/\dwave}(\bfk)(c^\dagger_{\uparrow}(\bfQ - \bfk,\tau)c^{\dagger}_{\downarrow}(\bfk,\tau)\nonumber\\ 
& \hspace{2.55cm} +c_{\uparrow}(\bfQ - \bfk,\tau)c_{\downarrow}(\bfk, \tau)),
\end{align}
\end{subequations}
for the magnetic $\chi_\M$, density $\chi_\D$, and ($s$- and $d$-wave) superconducting $\chi_\SC$ susceptibilities, respectively. The $s$- and $d$-wave form factors are given by
\begin{subequations}
    \begin{align}
        f_{\swave}(\bfk) & = 1, \\
        f_{\dwave}(\bfk) & = \cos(k_x) - \cos(k_y).
    \end{align}
    \label{eq:sdwaveformfactors}
\end{subequations}
The $s$-wave susceptibilities are trivially related to the bosonic propagators by
\begin{align}
\chi_\X(Q) = \frac{U_\X - w_\X(Q)}{U^2_\X}, \label{eq:flowing_swave_susceptibility}
\end{align}
where in an fRG calculation, $w_\X$ is readily available as one of the \emph{flowing} functions $w^{\Lambda = \Lambda_{\text{final}}}_{\X}(Q)$ for $\X = \M, \D, \SC$. An alternative route to obtain the susceptibilities is provided by considering the generalized susceptibility
\begin{align}
\chi_{\text{X}}(Q, k, k^\prime) &= \Pi_{\text{X}}(Q, k)\delta_{k,k^\prime} \nonumber\\
& \phantom{=} + \Pi_{\text{X}}(Q, k)  V_{\text{X}}(Q, k, k^\prime) \Pi_{\text{X}}(Q, k^\prime), \label{eq:generalized_susceptibility}
\end{align}
termed as \emph{postprocessing} susceptibilities.
The $s$-wave susceptibilities are determined by summing over the two fermionic arguments $k$ and $k^\prime$:
\begin{align}
\chi_{\text{X}}(Q) = \sum_{k,k^\prime} \chi_{\text{X}}(Q, k, k^\prime),
\end{align} 
whereas the $d$-wave superconducting susceptibility is obtained by
\begin{align}
\chi_\dSC(Q) = \sum_{k,k^\prime} f_\dwave(\bfk) f_\dwave(\bfk^\prime) \chi_{\SC}(Q, k, k^\prime). 
\end{align}
We note that in the SBE formulation of the parquet equations, the $s$-wave flowing susceptibilities coincide with the postprocessing ones \cite{Krien2021b}.

\begin{figure}[t!]
    \centering
    \adjustbox{max width=0.5\textwidth, scale=0.75}
    {\includegraphics{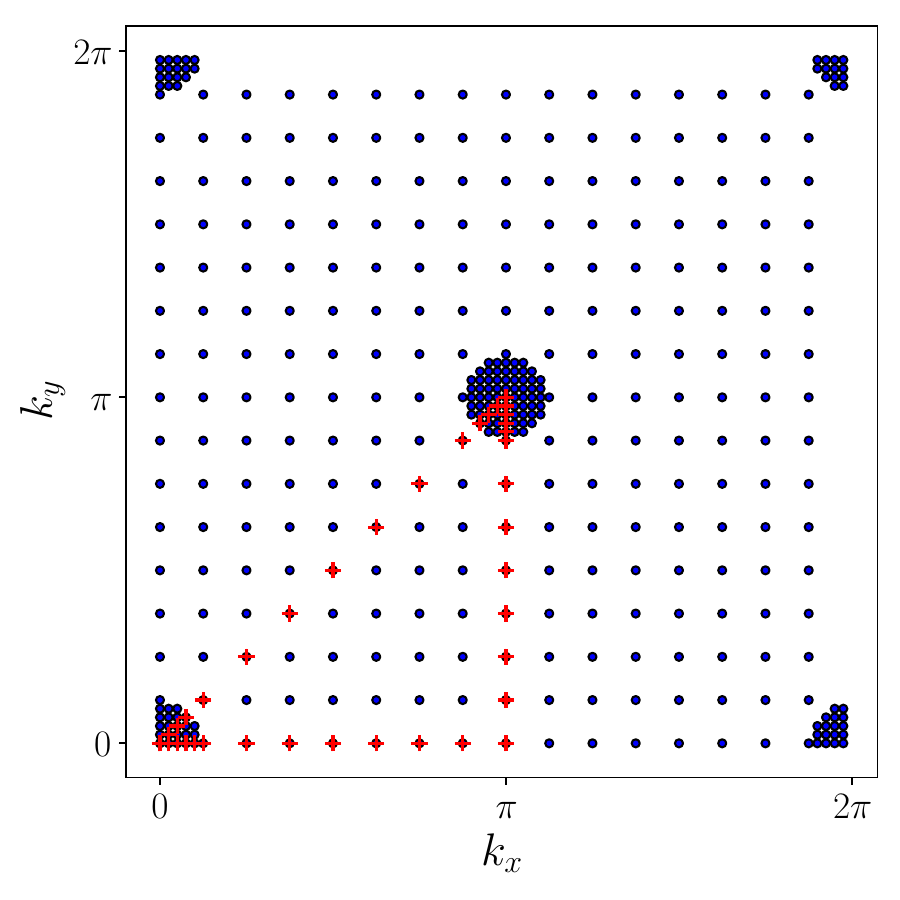}}
    \caption{The coarse grid of $16\times 16$ points is refined for the bosonic momentum dependence of the bosonic propagators $w_\X$,  Yukawa couplings $\lambda_\X$, and the SBE rest functions $M_{\X}$ within a radius of $2\pi/16$ around the $\Gamma$ and $\M$ symmetry points. The red crosses corresponding to the $\Gamma$-$\X$-$\M$-$\Gamma$ symmetry path enclose the reduced Brillouin zone.}
    \label{fig:DiscretizationBZ}
\end{figure}

\subsection{Cutoff schemes}
\label{sec:CutoffSchemes}
In this work, we utilize three different cutoff schemes (also called flow schemes) for the bare Green’s function. The latter features a cutoff-dependent chemical potential $\delta \mu^\Lambda$ that ensures a constant filling throughout the flow, which will be discussed below.  
\begin{enumerate}
\item \textit{$\Omega$-flow.}  
Introduced in Ref.~\cite{Husemann2009}, this flow scheme implements a multiplicative soft frequency cutoff
\begin{equation}
    G_0^{\Lambda}(\bfk,\nu) = \frac{\Theta^\Lambda(\nu)}{G_0^{-1}(\bfk,\nu) + \delta \mu^\Lambda},
    \end{equation}
with
\begin{equation}
\Theta^\Lambda(\nu) = \frac{\nu^2}{\nu^2 + \Lambda^2},
\end{equation}
a low-frequency cutoff function. Here, the flow parameter $\Lambda$ flows from $\Lambda^{\text{init}} = \infty$ to $\Lambda^{\text{final}} = 0$, recovering the full $G_0$.

\item \textit{The $U$-flow.} This flow scheme \cite{Honerkamp2004} is implemented by considering a parameter $\Lambda^2$ multiplying the quartic bare interaction. This parameter is then absorbed in a  redefinition of the fields, leading to
\begin{equation}
    G_0^{\Lambda}(\bfk,\nu) = \frac{\Lambda}{G_0^{-1}(\bfk,\nu) + \delta \mu^\Lambda},
\end{equation}
with $\Lambda$ flowing from $\Lambda^{\text{init}} = 0$ to $\Lambda^{\text{final}} = 1$. The advantage of this flow scheme is that vertex functions at intermediate steps of the flow correspond---after rescaling---to those of a bare interaction $\Lambda^2 U$, allowing to scan over the interaction dependence within a single computation of the flow.

\item \textit{The $T$-flow.} Similarly, the temperature flow \cite{Honerkamp2001c} scans the temperature dependence of the vertex functions along the flow; i.e., its values at intermediate steps of the flow correspond, after a proper rescaling, to the results at the scale $T$. For this, the inverse temperature appearing in the Matsubara summation normalization of the bare interaction is absorbed in a redefinition of the fields. For an instantaneous interaction, this removes all explicit reference to temperature in the quadratic part, which yields
\begin{equation}
G^T_0(\bfk,\nu) = \frac{T^{1/2}}{G_0^{-1}(\bfk,\nu) + \delta \mu^T},
\end{equation}
for the Green’s function. The temperature as flow parameter starts at $T^{\text{init}} = \infty$ and ends at the desired $T^{\text{final}}$. Note that this cutoff is not purely multiplicative since the temperature dependence also enters via the Matsubara frequency $\nu$.
\end{enumerate}

Regardless of the choice of the cutoff scheme, the chemical potential shift $\delta \mu^\Lambda$ compensates any change of the filling induced by the deformation in the Fermi surface due to the self-energy flow~\footnote{For the Hubbard model with a local interaction only, this occurs exclusively at finite doping.}. Following Ref.~\cite{Vilardi2017}, we use the fact that the filling
\begin{equation}
n(\delta \mu^\Lambda) = \sum_{\bfk,\nu} e^{i\nu 0^+} [G^{\Lambda}(\bfk,\nu)](\delta \mu^\Lambda)
\end{equation}
is a monotonic function of the chemical potential shift \cite{Salmhofer1999} and therefore invertible. Thus, it is possible to determine $\delta \mu^\Lambda$ by a root-finding algorithm such that the filling remains fixed at each step $\Lambda$ of the flow. The total scale derivative of $G^\Lambda$ becomes
\begin{align}
\dot{G}^\Lambda(\bfk,\nu) = \left.\frac{\mathrm{d}G^\Lambda}{\mathrm{d}\Lambda}\right|_{\delta \mu^\Lambda = \text{const}} + \left.\frac{\mathrm{d}G^\Lambda}{\mathrm{d}\delta \mu^\Lambda}\right. \frac{\mathrm{d}{\delta\mu}^{\Lambda}}{\mathrm{d}\Lambda} \label{eq:full_G_derivative_with_chemical_potential_shift}
\end{align}
in the multiloop flow equations, where the flow of the chemical potential shift is determined numerically by taking a finite difference in the $j$th integration step~\cite{AlEryani2024,AlEryani2025}
\begin{align}
\frac{\mathrm{d}{\delta\mu}^{\Lambda}}{\mathrm{d}\Lambda} = \frac{\delta \mu^{\Lambda_j} - \delta \mu^{\Lambda_{j-1}}}{\Lambda_j - \Lambda_{j - 1}}.
\end{align}
Note that the first term in Eq.~\eqref{eq:full_G_derivative_with_chemical_potential_shift} already contains the so-called Katanin substitution~\cite{Katanin2004a}.

\subsection{Loop corrections and self-energy iterations}
\label{sec:LoopCorrectionsSigmaIterations}

For the numerical integration of the fRG flow, we first consider the $1\ell$ equations
\begin{subequations}
\begin{align}
\dot{\Sigma} &= \mathbf{F}_{\Sigma}(\Sigma, V),\\
\dot{V} &= \mathbf{F}^{(1)}_V(\Sigma, V),
\end{align}
\end{subequations}
for which a standard solver for ordinary differential equations can be used. For the multiloop extension, the multiloop corrections to the vertex flow equation are added. 
Furthermore, the self-energy flow equation is replaced by the Schwinger-Dyson equation~\cite{Patricolo2025}, leading to
\begin{subequations}
\begin{align}
\dot{\Sigma} &= \mathbf{F}_{\text{SDE}}(\Sigma, V, \dot{\Sigma}, \dot{V}),\\
\dot{V} &= \mathbf{F}_{V}^{\text{multiloop}}(\Sigma, V, \dot{\Sigma}) \nonumber\\ &= \mathbf{F}^{(1)}_V(\Sigma, V, \dot{\Sigma}) + \sum_{\ell = 2}^\infty \mathbf{F}^{(\ell)}_V(\Sigma, V, \dot{\Sigma}).
\label{eq:ml}
\end{align}
\end{subequations}
We note that the Katanin substitution introduces the dependence of $\mathbf{F}_V^{1\ell}$ on $\dot{\Sigma}$. The function $\mathbf{F}^{(\ell)}$ can be constructed from $\mathbf{F}^{(\ell-1)}$ and $\mathbf{F}^{(\ell-2)}$. The most prominent difference with respect to the above conventional $1\ell$ equations lies in the dependence on the derivatives $\dot{\Sigma}$ and $\dot{V}$ on the right-hand side. The associated self-consistent problem has to be solved at every step of the flow. Specifically, one starts with a guess for $\dot{\Sigma}$ from which $\dot{V}$ can be calculated. The resulting $\dot{V}$ is then used to calculate a new $\dot{\Sigma}$. Such a procedure, which we refer to as a \emph{self-energy iteration}, has to be repeated until $\dot{\Sigma}$ (and with it $\dot{V}$) is converged. There are various strategies for its solution. One is shown in Fig.~\ref{fig:self_consistent_problem_every_ODE_step}. We used here $\varepsilon_{\text{se}} = 10^{-3}$  for the tolerance in the convergence of the self-energy iterations and $\varepsilon_{\text{vtx}} = 10^{-4}$ for the convergence of the vertex corrections. A maximum number of self-energy iterations $N_{\Sigma}$ and multiloop vertex corrections $N_{\ell}$, which override the convergence criterion even if the respective tolerances are not achieved, can be specified. In this work, these are used to study the effect of capping the number of self-energy iterations or vertex corrections, respectively. To speed up convergence, we found it useful to use a dynamic $\varepsilon_{\text{vtx}}$ that after the two-loop correction adjusts to $10\%$ of the measured self-energy error in the previous cycle. When the cycle of self-energy iterations achieves the desired tolerance of $\varepsilon_{\text{se}}$, then this will correspond to the vertex also achieving the desired tolerance since $\varepsilon_{\text{vtx}} = \varepsilon_{\text{se}}/10$. Furthermore, to systematically study aspects of multiloop convergence, as will be discussed with the results of Sec.~\ref{sec:Results}, we have opted to fix the guess $\dot{\Sigma}_{\text{guess}} = \mathbf{F}_{\Sigma}(\Sigma, V)$ (i.e., the first guess for the self-energy derivative is the right-hand side of the standard self-energy flow equation).

\begin{table}
    \centering
    \begin{tabular}{c|cccc}
 & \parbox{45pt}{Bos.\\freq.}& \parbox{42pt}{Ferm.\\freq.}& \parbox{45pt}{Bos.\\mom.}& \parbox{40pt}{Ferm.\\mom.}\\
 \\
 \hline
 \\
         $\Sigma$, $\dot{\Sigma}$ &  $\varnothing$&  $20N_w$& $\varnothing$ & $N_k^2$\\
         $\Pi$, $\dot{\Pi}$ &  $128N_w+1$&   $128N_w$& $N_k^2N_\delta^2$ & ff.\\
         $w$, $\dot{w}^{(\ell)}$ &  $128 N_w + 1$&  $\varnothing$& $N_k^2+N_{k,\delta}$ & $\varnothing$\\
         $\lambda$, $\dot{\lambda}^{(\ell)}$ &  $4N_w + 1$&  $4N_w$& $N_k^2+N_{k,\delta}$ & $\text{ff.}$\\
         $M$, $\dot{M}^{(\ell)}$, $\dot{I}^{(\ell)}$ &  $4N_w + 1$&  $2N_w$& $N_k^2+N_{k,\delta}$ & ff. \\
    \end{tabular}
    \caption{Number of frequency and momentum points used for the parametrization, where $N_\delta = 5$ for the bubble (see also text). For the treatment of the momenta within the truncated-unity fRG, the bosonic momenta of the vertices live on a discretized Brillouin zone, whereas the fermionic momentum dependence is restricted to (a few) form factors: an $s$-wave form factor at half filling and an additional $d$-wave one at finite doping. The fermionic momenta of the self-energy live on a coarse uniform $16\times 16$ discretization of the Brillouin zone.
     }
    \label{tab:technical_parameters}
\end{table}

\begin{figure*}[t!]
    \centering
    \adjustbox{max width=\textwidth, scale=1.0}{\includegraphics{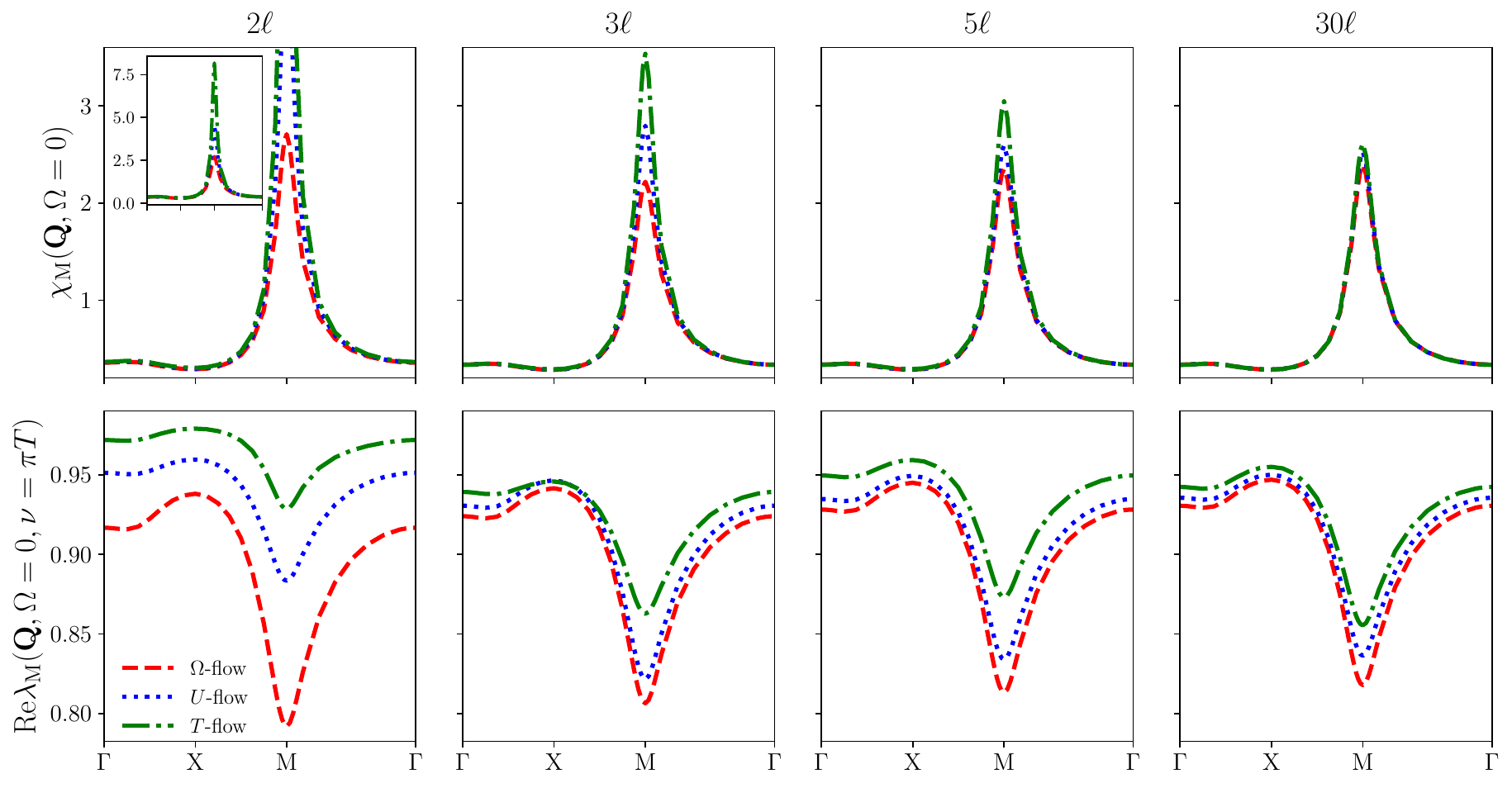}}
    \caption{Static bosonic momentum dependence of the magnetic susceptibility $\chi_{\text{M}}$ and of the real part of the magnetic Yukawa coupling $\lambda_{\text{M}}$ as obtained by the multiloop SBE fRG in the SBE approximation within three different flow schemes ($\Omega$-flow, $U$-flow, and $T$-flow) for different loop orders ($2\ell$, $3\ell$, $5\ell$, and $30\ell$), at $U=2.5$, $t^\prime=0$, $\beta=5$, and half filling ($\mu=0$).}
\label{fig:chiMlambdaMvsQ_SBEaCutoffDependence}
\end{figure*}

\subsection{Technical parameters}
\subsubsection{fRG calculations}

For this study, our fRG calculations were all performed with the \texttt{BosonFlow} codebase, presented in detail in Ref.~\cite{AlEryani2026}. In solving the multiloop flow equations numerically with this setup, we consider different Brillouin zone meshes for different objects. The coarse momentum grid presents a uniform discretization into $N_k \times N_k$ points and is used for the momentum dependence of the self-energy $\Sigma$, where we have used $N_k = 16$ for all our calculations. For the bosonic dependence of the vertex objects $w_\X$, $\lambda_\X$, and $M_\X$, the coarse momentum grid is selectively refined by a factor of $5$, which results in $N_{k,\delta} = 136$ additional points in the vicinity of the high symmetry points $\Gamma = (0, 0)$ and $\M = (\pi, \pi)$ (see Fig.~\ref{fig:DiscretizationBZ}). The bosonic momentum dependence of the bubbles appearing on the right-hand side of the vertex flow equations is calculated on a dense uniform grid of $N_k^2N_\delta^2$ points. On the other hand, in the spirit of the truncated-unity fRG~\cite{Husemann2009,Wang2012,Lichtenstein2017}, the secondary fermionic momentum dependencies of $\lambda_\X$, $M_\X$ and the bubbles are projected into a small set of form factors. Specifically, we keep only the $s$-wave form factor for calculations at half filling and additionally the $d$-wave form factor at finite doping. Moreover, we have neglected the mixed bubbles (we set $\Pi_{\X, nm} = \dot{\Pi}_{\X, nm} = 0$ for unequal form factor indices $n\neq m$), since their contributions vanish at $q = 0$ anyway. Within this approximation, the contributions from the $d$-wave component to the Yukawa couplings vanish identically in the SBE approximation (see Appendix~\ref{sec:vanishing_yukawa} for further explanations). Our numerical implementation thus relies on the multiloop SBE fRG equations~\eqref{eq:mfRGequationsPhysicalChannels1l}--\eqref{eq:mfRGequationsPhysicalChannelsbeyond2l} and~\eqref{eq:SigmaMdotFlowEquation} rewritten in form factor notation (see Appendix~\ref{App:TUmfRG} for more details on this point).

For the frequency dependence, we store the values of the objects at bounded frequency boxes parametrized by $N_w$, where we use $N_w = 6$ for $\beta = 5$, $7.5$ and $N_w = 8$ for $\beta = 10$, $20$. Outside of these boxes, the objects are taken to assume their high-frequency asymptotic values~\cite{Wentzell2016}. The self-energy is taken to decay to zero at large frequencies (outside the box), where the constant part resulting from the Hartree contribution is absorbed in a redefinition of the chemical potential. Table~\ref{tab:technical_parameters} summarizes the information on the frequency and momentum discretization used for the central objects in the code. The dependencies for the derivatives and multiloop vertex corrections are identical to their corresponding objects.

\subsubsection{Parquet calculations}

Our analysis is also based on results obtained from the parquet approximation, determined by solving the Bethe-Salpeter and the Schwinger-Dyson equations fully self-consistently (i.e., without relying on differential equations). The underlying calculations were performed with a different code, the Truncated Unity Parquet Solver~(\texttt{TUPS})~\cite{Eckhardt2020}. For completeness, we describe here in detail how these were performed. In \texttt{TUPS}, the 2PR vertices are not decomposed into single- and multiboson exchange contributions as in Eq.~\eqref{eq:phinablaM}, but only the parquet decomposition [Eq.~\eqref{eq:parquetDecomposition}] is used, together with the Bethe-Salpeter equation~\eqref{eq:ExtendedBSEfirstStep} and the Schwinger-Dyson equation, where the self-energy is directly obtained from the two-particle vertex~\cite{Eckhardt2018}. The fermionic momenta are represented using the truncated-unity approximation. For the results presented in this work, we have used one form factor ($s$-wave) for the fermionic momenta of the 2PR vertices. For the bosonic momenta and the self-energy, we used $N^2_k=40\times 40$, without refinement around the high-symmetry points. An overall frequency grid with $N_{\omega}=160$ was used for all vertices ($N_{\omega}^3$) and the self-energy. Additionally, scan-edge asymptotics was used~\cite{Li2019} for the reducible vertices in all equations. Since in \texttt{TUPS} the Yukawa couplings and bosonic propagators are not part of the computational self-consistent scheme, they were obtained from the converged solution in a postprocessing step, as described in Ref.~\cite{Krien2021b}. The $d$-wave susceptibilities shown in Fig.~\ref{fig:chivsQ_FiniteDoping_U2p5} were obtained by projecting the reconstructed momentum dependence of the 2PR vertices onto the $d$-wave form factor.

For the additional comparison of the SBE fRG to the SBE approximation directly in parquet equations (shown as dotted gray lines in Fig.~\ref{fig:chiMDlambdaMDvsQPA}), another parquet equations solver was used, the \texttt{MBE-parquet} code~\cite{Krien2022}. In this code, the 2PR vertices are, as in the SBE fRG, decomposed into single- and multiboson exchange parts, in accordance with Eq.~\eqref{eq:phinablaM}. The \texttt{MBE-parquet} code does not use the truncated-unity approximation and the full momentum grid is used for all momenta. For the data shown in Fig.~\ref{fig:chiMDlambdaMDvsQPA}, the SBE rest functions, i.e., the multiboson exchange contributions, were set to zero ($M_{\mathrm{X}} =0$ for all channels $\mathrm{X}$). A grid of $N^2_k=16\times 16$ was used for all bosonic and fermionic momenta, with $11\times$ refinement for the one-particle Green’s function. The fermionic momentum dependence of the Yukawa couplings was projected afterward onto the $s$-wave form factor for comparison with the SBE fRG.

\section{Results}
\label{sec:Results}

\subsection{Multiloop convergence and cutoff dependence}

\begin{figure*}[t!]
    \centering
    \adjustbox{max width=\textwidth, scale=1.0}{\includegraphics{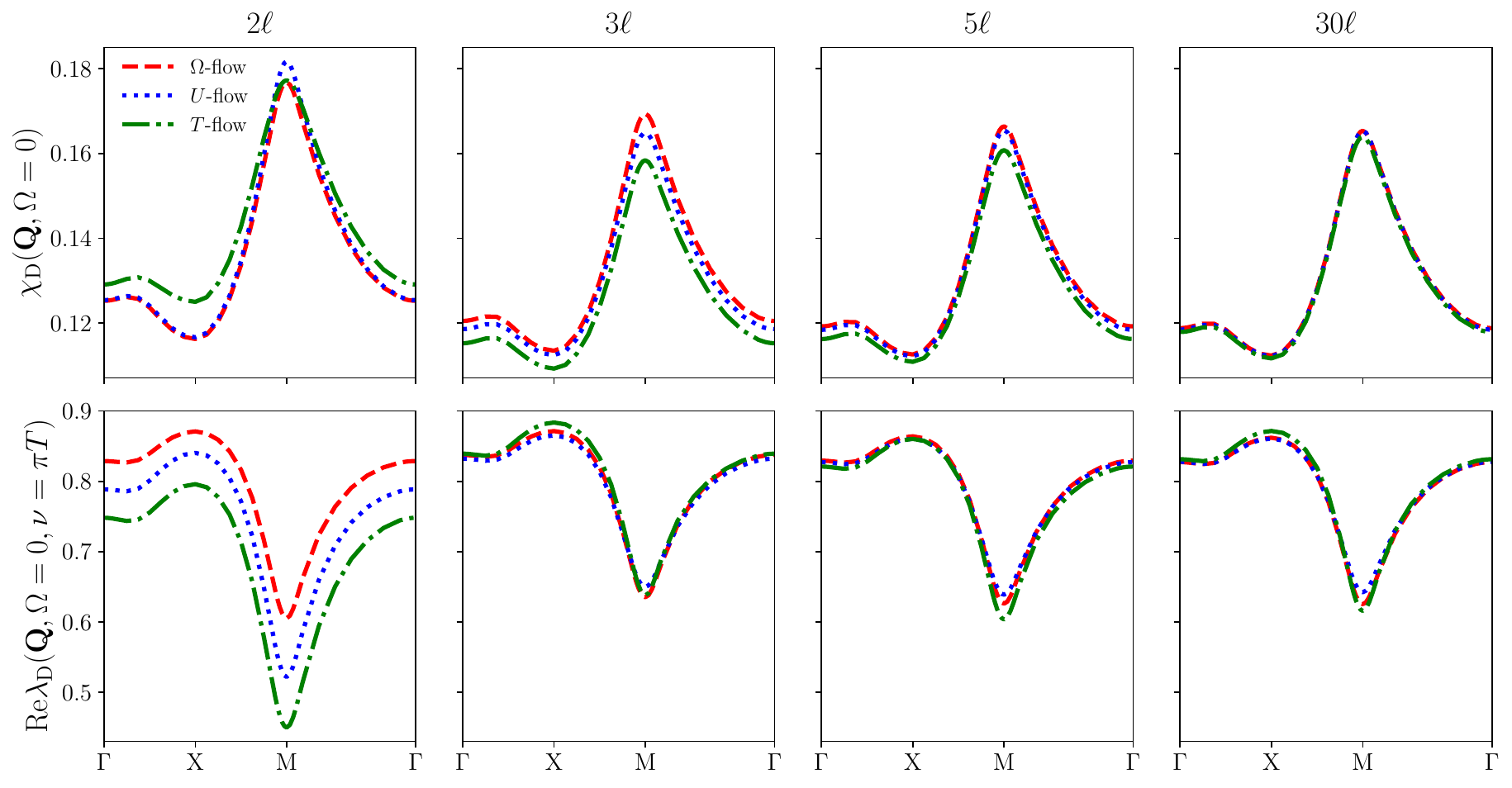}}
    \caption{Same as Fig.~\ref{fig:chiMlambdaMvsQ_SBEaCutoffDependence}, but for the density channel. The corresponding data for the superconducting channel are obtained from the relations $\chi_{\mathrm{SC}}(\mathbf{Q},\Omega)=\chi_{\mathrm{D}}(\mathbf{Q}+(\pi,\pi),\Omega)$ and $\lambda_{\mathrm{SC}}(\mathbf{Q},\Omega,\nu)=\lambda_{\mathrm{D}}(\mathbf{Q}+(\pi,\pi),\Omega,\nu)$.}
    \label{fig:chiDlambdaDvsQ_SBEaCutoffDependence}
\end{figure*}

\begin{figure*}[t!]
    \centering
    \adjustbox{max width=\textwidth, scale=1.0}{\includegraphics{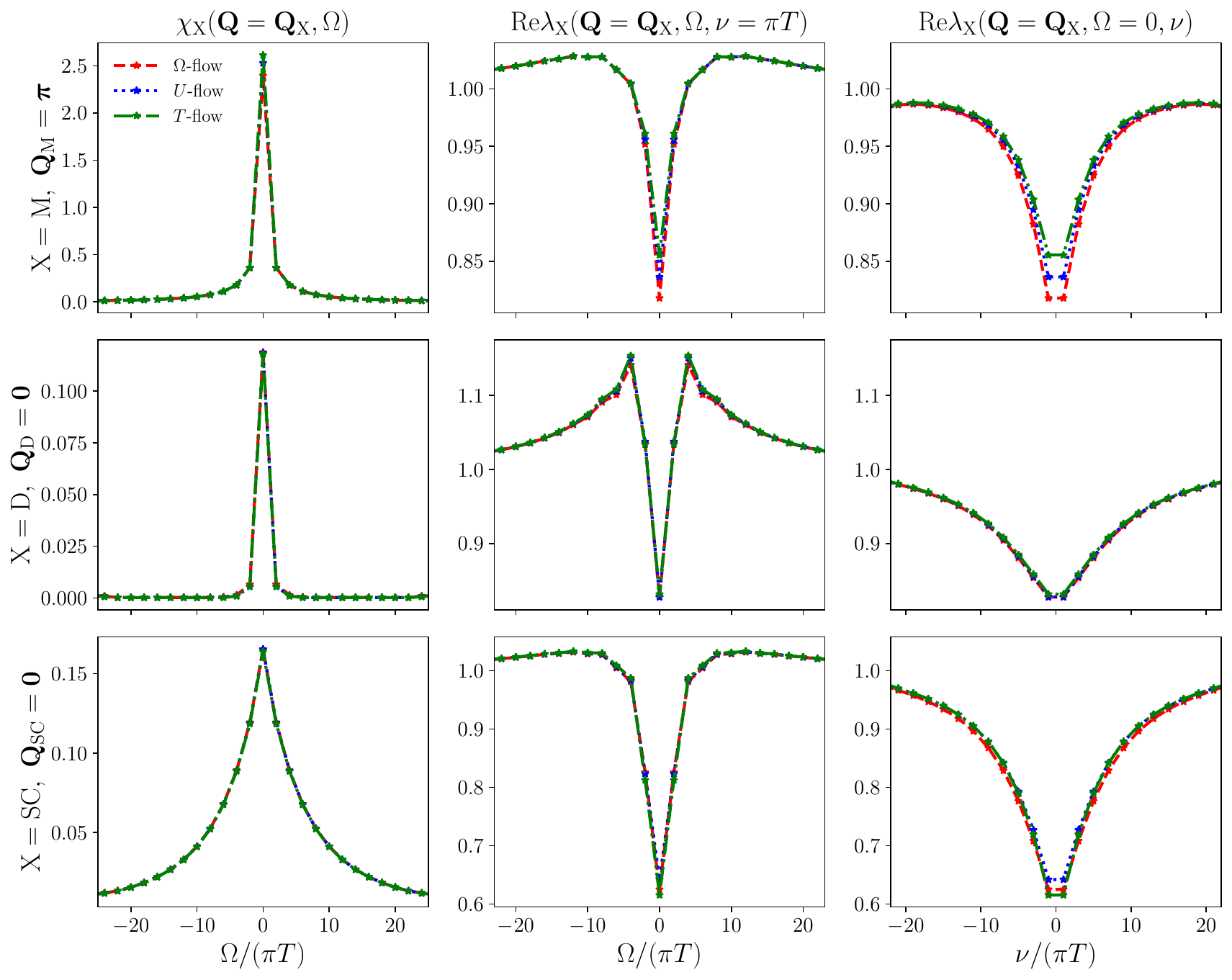}}
    \caption{Frequency dependencies of the susceptibilities and of the real parts of the Yukawa couplings in all three physical channels at bosonic momenta $\mathbf{Q}=\boldsymbol{\pi}=(\pi,\pi)$ (in the $\M$ channel) and $\mathbf{Q}=\mathbf{0}=(0,0)$ (in the $\D$ and $\SC$ channels) as obtained from the multiloop SBE fRG at $30\ell$ in the SBE approximation within three different flow schemes ($\Omega$-flow, $U$-flow, and $T$-flow), at $U=2.5$, $t^\prime=0$, $\beta=5$, and half filling ($\mu=0$).
    }
    \label{fig:chilambdaFreqDep_HalfFilling}
\end{figure*}

\begin{figure*}[t!]
    \centering
    \adjustbox{max width=\textwidth, scale=1.0}{\includegraphics{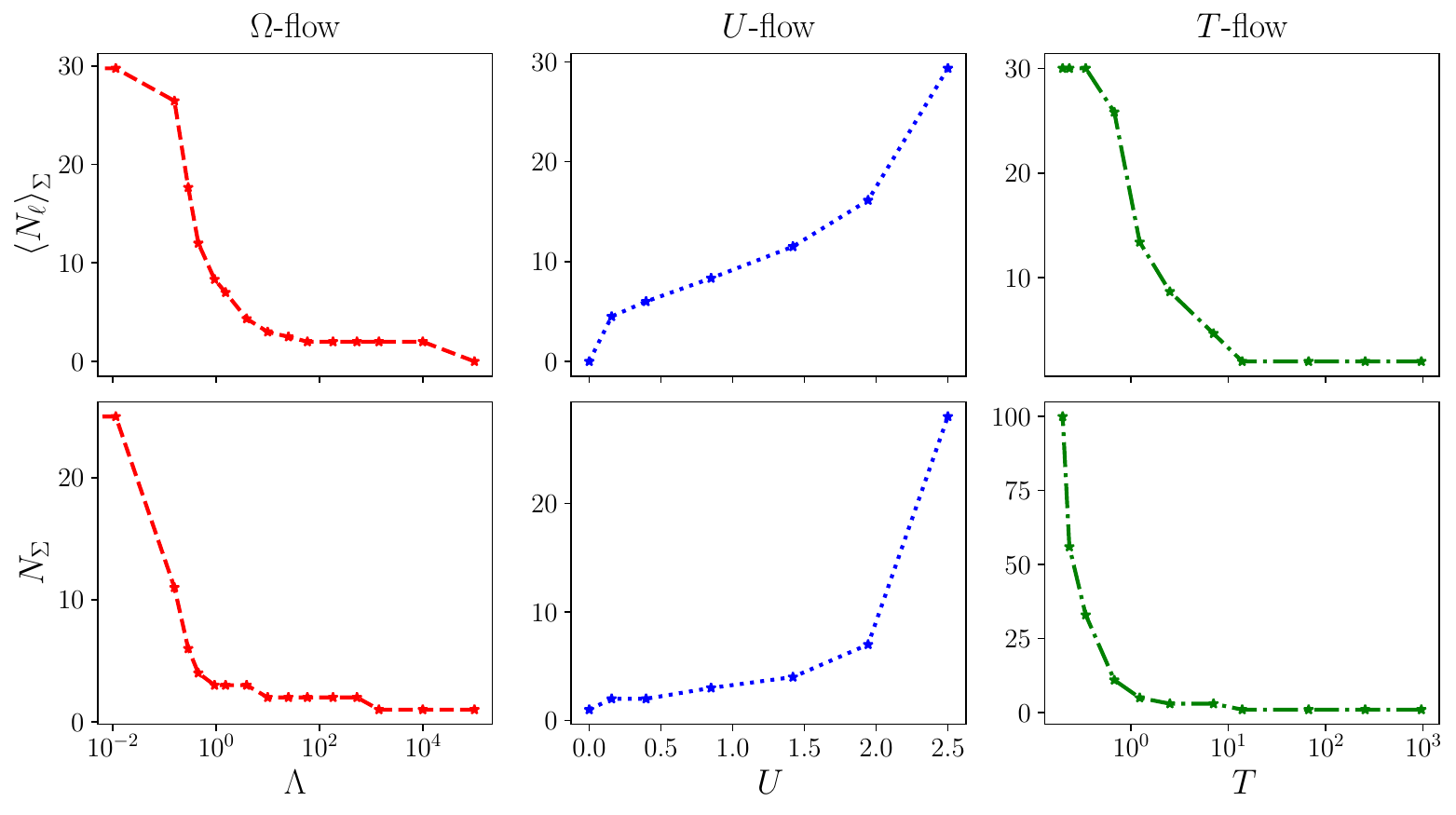}}
    \caption{Convergence parameters for the multiloop SBE fRG at $30\ell$ in the SBE approximation within three different flow schemes ($\Omega$-flow, $U$-flow, and $T$-flow), at $U=2.5$, $t^\prime=0$, $\beta=5$, and half filling ($\mu=0$). $N_{\Sigma}$ and $\left\langle N_\ell \right\rangle_\Sigma$ are the number of self-energy iterations and the average number of loop corrections per self-energy iteration, respectively.
    }
\label{fig:Convergence_CompareCutoffs}
\end{figure*}

In the present section, we analyze the multiloop SBE fRG results for the 2D Hubbard model, relying on the numerical implementation described in Sec.~\ref{sec:NumericalImplementation}. In particular, we use the SBE approximation obtained by neglecting the flow of the SBE rest functions $M_{\mathrm{X}}$, the most complicated objects to calculate.

The first point that we address is the \emph{cutoff dependence} of the multiloop SBE fRG. If the SBE rest functions $M_{\mathrm{X}}$ are included in the flow, the multiloop SBE fRG results are equivalent to those of the multiloop fRG formulated within the parquet decomposition and therefore coincide with the parquet approximation at convergence~\cite{Kugler2018a}. This applies regardless of the chosen cutoff scheme. Put differently, the cutoff dependence of the multiloop SBE fRG results completely vanishes at convergence when the SBE rest functions $M_{\mathrm{X}}$ are included in the flow. However, this is no longer valid within the SBE approximation. Indeed, in that case, the closed system of flow equations to solve is truncated [by discarding Eqs.~\eqref{eq:flowequationsMXPhysicalChannels1l},~\eqref{eq:flowequationsMXPhysicalChannels2l}, and~\eqref{eq:flowequationsMXPhysicalChannelsbeyond2l}], which introduces a dependence with respect to $G_0^\Lambda$ in the results obtained at the end of the fRG flow. We note that the cutoff dependence, a well-known and studied feature of conventional fRG approaches based on the Wetterich equation~\cite{Litim2000,Litim2001a,Litim2001b,Litim2002,Pawlowski2007,Zorbach2024}, is a result of the truncation being implemented after the cutoff function is introduced via $G_0\rightarrow G_0^\Lambda$.

In Figs.~\ref{fig:chiMlambdaMvsQ_SBEaCutoffDependence}--\ref{fig:chilambdaFreqDep_HalfFilling}, we examine the cutoff dependence of the multiloop SBE fRG in the SBE approximation \emph{quantitatively}, by comparing results for the static susceptibilities and the Yukawa couplings of the half-filled 2D Hubbard model obtained within the $\Omega$-flow, the $U$-flow and the $T$-flow, for $U=2.5$ and $\beta=5$. More specifically, Fig.~\ref{fig:chiMlambdaMvsQ_SBEaCutoffDependence} focuses on the magnetic channel and displays the bosonic momentum dependence of $\chi_{\mathrm{M}}(\mathbf{Q},\Omega=0)$ and $\mathrm{Re}\lambda_{\mathrm{M}}(\mathbf{Q},\Omega=0,\nu=\pi T)$ along the $\Gamma$-$\X$-$\M$-$\Gamma$ path in the Brillouin zone defined in Fig.~\ref{fig:DiscretizationBZ}. Comparing the $2\ell$, $3\ell$, $5\ell$, and $30\ell$ results displayed in the different panels (the self-energy is converged according to the criterion in Sec.~\ref{sec:LoopCorrectionsSigmaIterations} for each loop order), we observe that the 
loop corrections from $2\ell$ to $30\ell$ on the peak of $\chi_{\mathrm{M}}(\mathbf{Q}=(\pi,\pi),\Omega=0)$ have a much stronger impact within the $U$-flow and the $T$-flow than within the $\Omega$-flow. The relative difference $\left(\left|\chi_{\mathrm{M}}^{2\ell}-\chi_{\mathrm{M}}^{30\ell}\right|\right)/\chi_{\mathrm{M}}^{30\ell}$ between the $2\ell$ and $30\ell$ results amounts to $13\%$, $80\%$, and $214\%$ at the $\mathrm{M}$ point [$\mathbf{Q}=(\pi,\pi)$], respectively. The $2\ell$ result obtained with the $\Omega$-flow is thus much closer to the converged ones than its $U$-flow and $T$-flow counterparts. In this sense, the $\Omega$-flow can be considered as the optimal flow scheme tested here, consistently with previous multiloop fRG results~\cite{TagliaviniHille2019}. In contrast, the $T$-flow exhibits the slowest convergence for the magnetic susceptibility, with the $2\ell$ peak of $\chi_{\mathrm{M}}(\mathbf{Q}=(\pi,\pi),\Omega=0)$ being more than $3$ times larger than its value at $30\ell$. However, all three flow schemes yield consistent results at $30\ell$. Indeed, the three corresponding curves in the $30\ell$ panel are hardly distinguishable with a few percent of relative difference, with its maximal value reached for $\left(\left|\chi_{\mathrm{M}}^{T\text{-flow}}-\chi_{\mathrm{M}}^{\Omega\text{-flow}}\right|\right)/\chi_{\mathrm{M}}^{\Omega\text{-flow}} \sim 9\%$ at the $\mathrm{M}$ point. The cutoff dependence of our multiloop SBE fRG results for $\chi_{\mathrm{M}}$ is residual at $30\ell$.

The same general conclusions can be drawn for the real part of the magnetic Yukawa coupling $\lambda_{\mathrm{M}}$, shown in the lower panels of Fig.~\ref{fig:chiMlambdaMvsQ_SBEaCutoffDependence}. More precisely, the $T$-flow and $\Omega$-flow results are still the most and the least affected by the loop corrections from $2\ell$ to $30\ell$. At the same time, all three flow schemes yield consistent results at $30\ell$. We note, however, that the deviations between the results for the different flow schemes appear to be larger as compared to the ones for $\chi_{\mathrm{M}}$. Nevertheless, the maximal relative difference between the curves for $\mathrm{Re}\lambda_{\mathrm{M}}$ at $30\ell$ is smaller than $5\%$.

Figure~\ref{fig:chiDlambdaDvsQ_SBEaCutoffDependence} presents analogous results as Fig.~\ref{fig:chiMlambdaMvsQ_SBEaCutoffDependence}, but in the density channel. According to the data for the susceptibility and the Yukawa coupling, we can extend the above conclusions on the convergence properties of the different flow schemes as well as on the residual cutoff dependence at $30\ell$ to the subleading channels. We observe that the discrepancy between the three flow schemes is clearly less pronounced as compared to the magnetic case, but the $\Omega$-flow is still the least affected by the loop corrections from $2\ell$ to $30\ell$. In particular, the relative difference between the $2\ell$ and $30\ell$ susceptibilities, i.e., $\left(\left|\chi_{\mathrm{D}}^{2\ell}-\chi_{\mathrm{D}}^{30\ell}\right|\right)/\chi_{\mathrm{D}}^{30\ell}$, is about $7\%$, $8\%$, and $10\%$ at the $\mathrm{M}$ point for the $\Omega$-flow, the $U$-flow, and the $T$-flow, respectively. This is to be contrasted with the values of $13\%$, $80\%$, and $214\%$ reported previously for the magnetic susceptibility. Moreover, the results for $\chi_{\mathrm{D}}$ at $30\ell$ from all three flow schemes coincide within $1\%$ relative difference for all momenta along the $\Gamma$-$\X$-$\M$-$\Gamma$ path. Similar findings on the loop convergence properties and the reduced cutoff dependence at $30\ell$ also hold for the real part of the Yukawa coupling $\lambda_{\mathrm{D}}$. As a consequence of the symmetry relations $\chi_{\mathrm{SC}}(\mathbf{Q},\Omega)=\chi_{\mathrm{D}}(\mathbf{Q}+(\pi,\pi),\Omega)$ and $\lambda_{\mathrm{SC}}(\mathbf{Q},\Omega,\nu)=\lambda_{\mathrm{D}}(\mathbf{Q}+(\pi,\pi),\Omega,\nu)$ at half filling, these general remarks also apply to $\chi_{\mathrm{SC}}$ and $\lambda_{\mathrm{SC}}$ \cite{Essl2024}.

Figure~\ref{fig:chilambdaFreqDep_HalfFilling} extends this analysis at zero bosonic frequency ($\Omega=0$) and at the lowest fermionic Matsubara frequency ($\nu=\pi T$) by showing the corresponding frequency dependencies. We observe that the discrepancy between the $\Omega$-flow, $U$-flow, and $T$-flow is generally most pronounced at $\Omega=0$ and at $\nu=\pi T$, i.e., for the frequency values considered previously in Figs.~\ref{fig:chiMlambdaMvsQ_SBEaCutoffDependence} and~\ref{fig:chiDlambdaDvsQ_SBEaCutoffDependence}. In fact, the curves corresponding to the three flow schemes are, in most cases, barely distinguishable on the displayed scale. Notable exceptions are the results for
$\mathrm{Re}\lambda_{\mathrm{M}}$, where a clear discrepancy is observed at $\Omega=0$ and $\nu=\pi T$, in accordance with the corresponding results illustrated in  Fig.~\ref{fig:chiMlambdaMvsQ_SBEaCutoffDependence}.
However, the relative difference $\left(\left|\mathrm{Re}\lambda_{\mathrm{M}}^{T\text{-flow}}-\mathrm{Re}\lambda_{\mathrm{M}}^{\Omega\text{-flow}}\right|\right)/\mathrm{Re}\lambda_{\mathrm{M}}^{\Omega\text{-flow}}$ is still smaller than $5\%$ at $\mathbf{Q}=(\pi,\pi)$, $\Omega=0$, and $\nu=\pi T$, as reported previously.

In the considered parameter regime, the cutoff dependence of our multiloop SBE fRG results in the SBE approximation is hence residual for both the frequency- and momentum-dependent susceptibilities $\chi_{\mathrm{X}}(\mathbf{Q},\Omega)$ and the Yukawa couplings $\lambda_{\mathrm{X}}(\mathbf{Q},\Omega,\nu)$, with $\mathrm{X}=\mathrm{M},\mathrm{D},\mathrm{SC}$. This parameter regime of the 2D Hubbard model is still within the reach of the weak-coupling fRG approach used here, but is already fairly nontrivial in the sense that a substantial number of loop corrections is required to reach convergence. We clarify this point in Fig.~\ref{fig:Convergence_CompareCutoffs}, which displays the number of self-energy iterations and the average number of loop corrections per self-energy iteration at each step of the flow for the $30\ell$ results obtained from the $\Omega$-flow, the $U$-flow, and the $T$-flow and shown in Figs.~\ref{fig:chiMlambdaMvsQ_SBEaCutoffDependence}--\ref{fig:chilambdaFreqDep_HalfFilling}. Full convergence is defined by the parameters $\varepsilon_{\text{se}} = 10^{-3}$ and $\varepsilon_{\text{vtx}} = 10^{-4}$ in Sec.~\ref{sec:LoopCorrectionsSigmaIterations}. For the $30\ell$ results of Figs.~\ref{fig:chiMlambdaMvsQ_SBEaCutoffDependence}--\ref{fig:Convergence_CompareCutoffs}, convergence is achieved with respect to self-energy iterations (according to $\varepsilon_{\text{se}} = 10^{-3}$), whereas the number of loop corrections per self-energy iteration is capped at $\ell=30$ if convergence with respect to loop corrections cannot be reached (according to $\varepsilon_{\text{vtx}} = 10^{-4}$) with less than $30$ loops. In fact, none of the three flow schemes are actually fully converged at $U=2.5$, $\beta=5$, and half filling for $\ell=30$. In Fig.~\ref{fig:Convergence_CompareCutoffs}, this implies that the average number of loop corrections per self-energy iteration reaches its maximal value of $30$ (or becomes very close to it) at its last step or before, for each flow scheme. In the plateau observed in the top right panel, this is particularly evident in the $T$-flow. Furthermore, one can see that the $T$-flow requires many more self-energy iterations to achieve convergence at each step of the flow 
as compared to the $\Omega$-flow and the $U$-flow. These findings are consistent with the above discussion of Fig.~\ref{fig:chiMlambdaMvsQ_SBEaCutoffDependence}, assessing the $T$-flow as a less favorable choice for reaching convergence. 

As a next step, we consider the multiloop SBE fRG in the SBE approximation at \emph{full} convergence, as defined in Sec.~\ref{sec:LoopCorrectionsSigmaIterations}. For this, we use the $\Omega$-flow as the optimal flow scheme requiring the smallest number of both loop corrections and self-energy iterations to reach \emph{full} convergence, and compare the obtained results to the parquet approximation.

\subsection{Comparison to the parquet approximation at half filling}

\begin{figure}[t!]
    \centering
    \adjustbox{max width=0.5\textwidth, scale=0.95}{\includegraphics{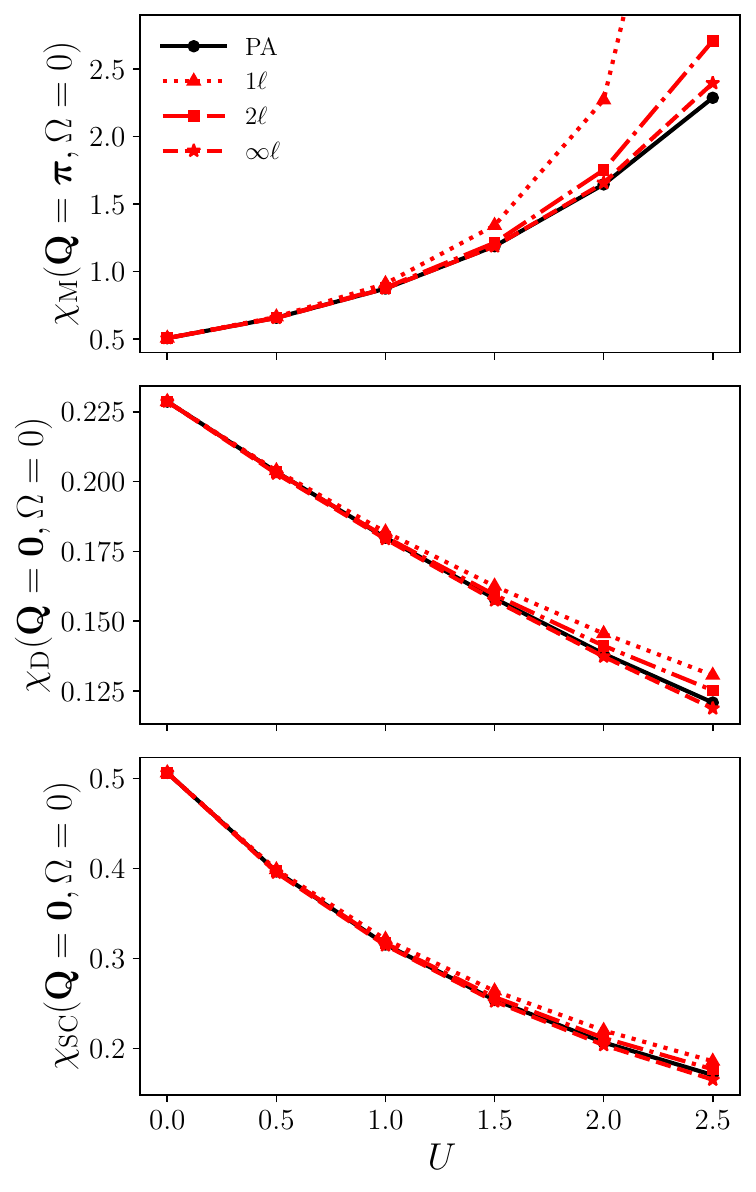}}
    \caption{Static susceptibilities in all three physical channels at bosonic momenta $\mathbf{Q}=\boldsymbol{\pi}=(\pi,\pi)$ (in the $\M$ channel) and $\mathbf{Q}=\mathbf{0}=(0,0)$ (in the $\D$ and $\SC$ channels) as a function of the bare interaction $U$, at $\beta=5$, $t^\prime=0$,  and half filling ($\mu=0$), as obtained from the SBE fRG at $1\ell$, $2\ell$, and at multiloop convergence in the SBE approximation within the $\Omega$-flow, with comparison to the parquet approximation (PA).
    }
    \label{fig:chiMvsU_U0to2p5Beta5}
\end{figure}

\begin{figure}[t!]
    \centering
    \adjustbox{max width=0.49\textwidth, scale=1.0}{\includegraphics{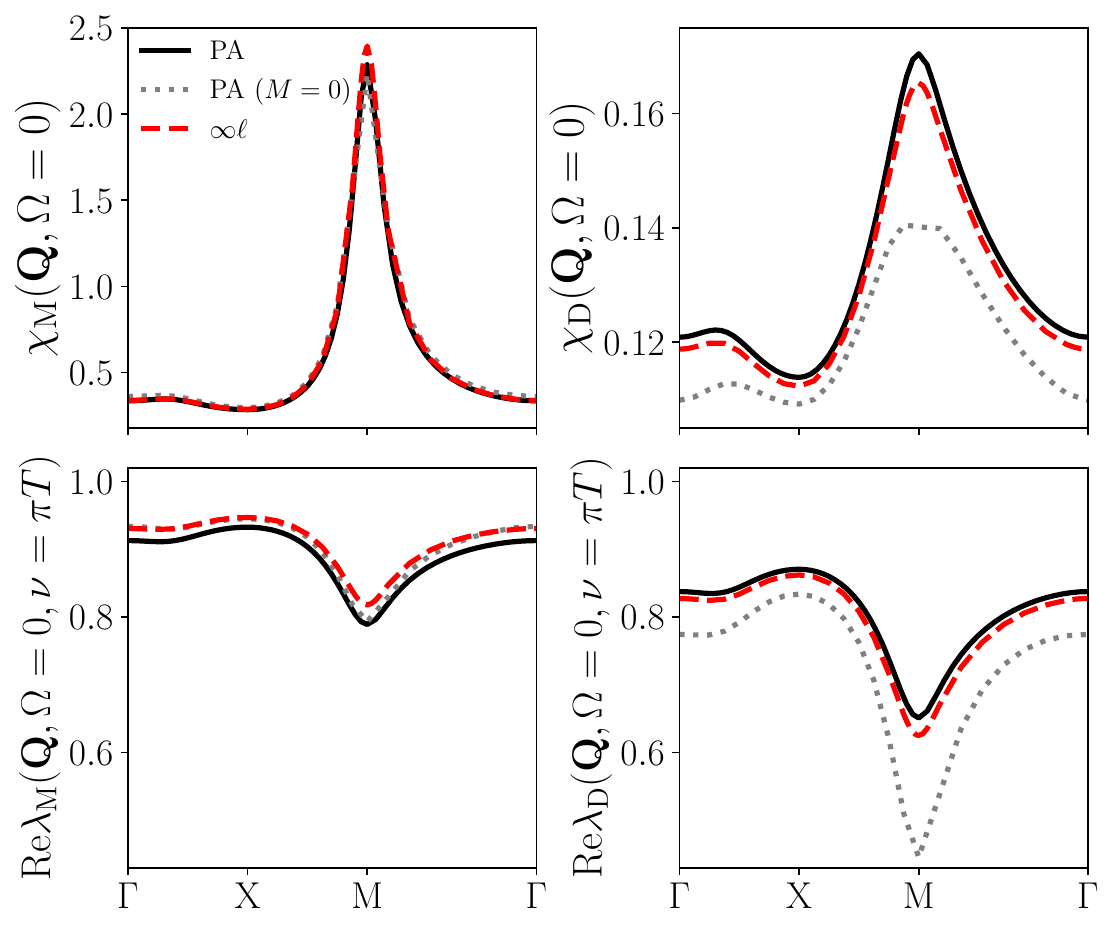}}
    \caption{Static bosonic momentum dependence of the susceptibilities and of the real parts of the Yukawa couplings in the magnetic and density channels at $U=2.5$, $\beta=5$, $t^\prime=0$, and half filling ($\mu=0$), as obtained from the multiloop SBE fRG at convergence in the SBE approximation (within the $\Omega$-flow), with comparison to the parquet approximation (PA). The corresponding data for the superconducting channel are obtained by $\chi_{\mathrm{SC}}(\mathbf{Q},\Omega)=\chi_{\mathrm{D}}(\mathbf{Q}+(\pi,\pi),\Omega)$ and $\lambda_{\mathrm{SC}}(\mathbf{Q},\Omega,\nu)=\lambda_{\mathrm{D}}(\mathbf{Q}+(\pi,\pi),\Omega,\nu)$. The dotted gray lines are parquet results obtained by neglecting the SBE rest functions ($M_{\mathrm{X}}=0$ for all channels $\mathrm{X}$) (see Appendix \ref{sec:SBEapproxPostproc} for its discussion).}
    \label{fig:chiMDlambdaMDvsQPA}
\end{figure}

\begin{figure*}[t!]
    \centering
    \adjustbox{max width=\textwidth, scale=1.0}{\includegraphics{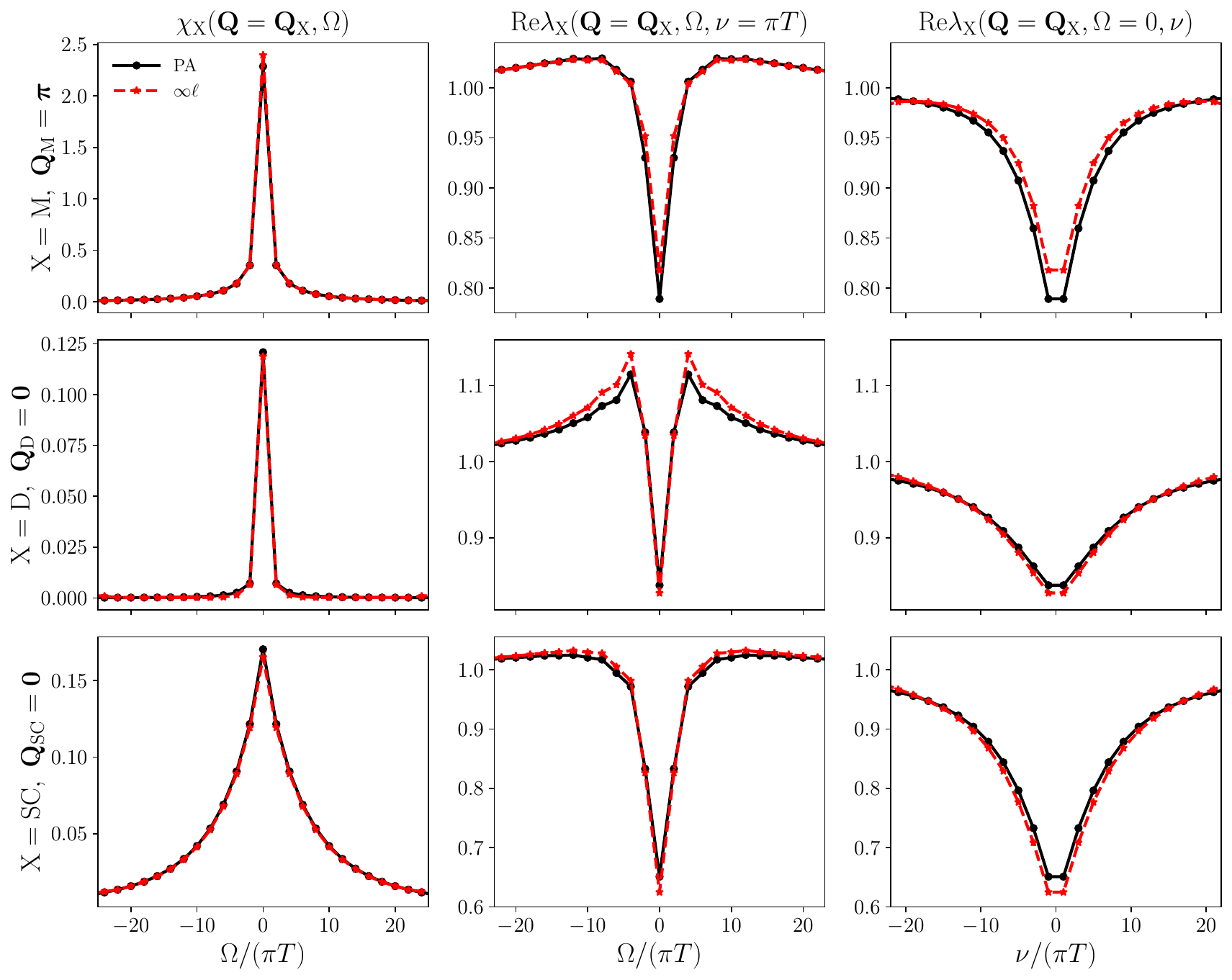}}
    \caption{Frequency dependencies of the susceptibilities and of the real parts of the Yukawa couplings in all three physical channels at bosonic momenta $\mathbf{Q}=\boldsymbol{\pi}=(\pi,\pi)$ (in the $\M$ channel) and $\mathbf{Q}=\mathbf{0}=(0,0)$ (in the $\D$ and $\SC$ channels) at $U=2.5$, $\beta=5$, $t^\prime=0$, and half filling ($\mu=0$), as obtained in the multiloop SBE fRG at convergence in the SBE approximation (within the $\Omega$-flow), with comparison to the parquet approximation (PA).}
    \label{fig:chilambdaFreqDepPA_HalfFilling}
\end{figure*}

The comparison of converged multiloop SBE fRG data of the susceptibilities and the Yukawa couplings (obtained within the $\Omega$-flow, identified as optimal by our previous analysis) to the parquet approximation enables us to determine the importance of the SBE rest functions $M^\text{X}$. When the latter are included, the converged multiloop SBE fRG is equivalent to the parquet approximation, which we also verified numerically. This provides us with another criterion to judge the quality of the SBE approximation within the multiloop SBE fRG approach, alongside the cutoff dependence examined previously. To reach convergence with respect to both self-energy iterations and loop corrections according to the criteria outlined in Sec.~\ref{sec:LoopCorrectionsSigmaIterations} ($\varepsilon_{\text{se}} = 10^{-3}$ and $\varepsilon_{\text{vtx}} = 10^{-4}$), the $\Omega$-flow requires, at its most demanding step, up to $25$ self-energy iterations and an average of $36$ loop corrections per self-energy iteration. The $30\ell$ $\Omega$-flow results shown in Figs.~\ref{fig:chiMlambdaMvsQ_SBEaCutoffDependence}--\ref{fig:Convergence_CompareCutoffs} are therefore almost  converged.

Our converged results for the multiloop SBE fRG in the SBE approximation (obtained within the $\Omega$-flow) are shown in Figs.~\ref{fig:chiMvsU_U0to2p5Beta5}--\ref{fig:chilambdaFreqDepPA_HalfFilling}, together with the corresponding data from the parquet approximation. First, Fig.~\ref{fig:chiMvsU_U0to2p5Beta5} shows the evolution of the susceptibility peaks as a function of the bare interaction $U$ at $\beta=5$ and half filling. It also includes the $1\ell$ and $2\ell$ SBE fRG results obtained with the conventional self-energy flow equation (derived from the Wetterich equation instead of the Schwinger-Dyson equation). These $1\ell$ and $2\ell$ approaches are standard fRG schemes, and the $1\ell$ case was thoroughly investigated in previous work~\cite{Fraboulet2022}. The $1\ell$ SBE fRG overestimates the magnetic peak $\chi_{\mathrm{M}}(\mathbf{Q}=(\pi,\pi),\Omega=0)$ with increasing $U$. Its divergence signaling the onset of antiferromagnetic order is an artifact of the $1\ell$ approximation that violates the Mermin-Wagner theorem~\cite{Mermin1966}. Adding higher-order loop corrections, the parquet approximation, which fulfills the Mermin-Wagner theorem~\cite{Bickers1992,Vilk1997,AlEryani2026c}, is recovered at convergence~\cite{Kugler2018a,TagliaviniHille2019}. In fact, already the $2\ell$ corrections lead to a drastic reduction of the magnetic peak. This behavior is reflected also in the parquet approximation, as can be seen in Fig.~\ref{fig:chiMvsU_U0to2p5Beta5}. In particular, the converged multiloop SBE fRG results in the SBE approximation also agree well with the parquet approximation. The most important discrepancy between the two schemes is found for the magnetic peak at $U=2.5$, with a relative difference $\left(\left|\chi_{\mathrm{M}}^{\infty\ell}-\chi_{\mathrm{M}}^{\mathrm{PA}}\right|\right)/\chi_{\mathrm{M}}^{\mathrm{PA}}$ still smaller than $5\%$.

We now investigate the momentum and frequency dependencies of the susceptibilities and Yukawa couplings at $U=2.5$, $\beta=5$, and half filling, where the most sizable deviations of the converged SBE fRG results and the parquet approximation have been observed (see Figs.~\ref{fig:chiMDlambdaMDvsQPA} and~\ref{fig:chilambdaFreqDepPA_HalfFilling}). Focusing on the momentum dependencies shown in Fig.~\ref{fig:chiMDlambdaMDvsQPA}, we observe a good overall agreement, with a relative difference between the SBE fRG and the parquet approximation on the order of a few percent, reaching a maximum of $5\%$ for $\chi_{\mathrm{M}}(\mathbf{Q},\Omega=0)$ at the $\mathrm{M}$ point. Besides $\chi_{\mathrm{M}}$, the most important discrepancy between fRG and parquet results is found in $\mathrm{Re}\lambda_{\mathrm{D}}(\mathbf{Q},\Omega=0,\nu=\pi T)$ still at the $\mathrm{M}$ point, with a relative difference of $4\%$. The results for the superconducting channel (not displayed in Fig.~\ref{fig:chiMDlambdaMDvsQPA}) can be inferred from the data in the density channel by $\chi_{\mathrm{SC}}(\mathbf{Q},\Omega)=\chi_{\mathrm{D}}(\mathbf{Q}+(\pi,\pi),\Omega)$ and $\lambda_{\mathrm{SC}}(\mathbf{Q},\Omega,\nu)=\lambda_{\mathrm{D}}(\mathbf{Q}+(\pi,\pi),\Omega,\nu)$. Concerning the frequency dependencies presented in Fig.~\ref{fig:chilambdaFreqDepPA_HalfFilling}, we find that the discrepancy between the converged multiloop SBE fRG results and the parquet approximation remains small in all three channels, with a relative difference below $4\%$ for the real parts of all Yukawa couplings. The agreement between fRG and parquet results is of similar quality for the susceptibilities shown in Fig.~\ref{fig:chilambdaFreqDepPA_HalfFilling}. At $\Omega=0$, we recover the susceptibility peaks already shown in Fig.~\ref{fig:chiMvsU_U0to2p5Beta5} for all three channels, with a relative difference between converged multiloop SBE fRG and parquet data always below $5\%$. It should be noted that, at finite $\Omega$ in Fig.~\ref{fig:chilambdaFreqDepPA_HalfFilling}, this relative difference can exceed $5\%$ for the susceptibilities. However, this corresponds to only small differences in absolute values [since we always have $\chi_\mathrm{X}(\mathbf{Q}=\mathbf{Q}_\mathrm{X},\Omega)<0.013$ for all channels $\mathrm{X}$ in that case].

We emphasize that the very good quantitative agreement of the multiloop SBE fRG results in the SBE approximation with the parquet data in the considered parameter regime can be traced back to the effective generation of rest-function-like contributions in the fRG flow, even though these are neglected in the SBE approximation. This property has been already pointed out in Ref.~\cite{Fraboulet2022}, where the underlying diagrammatic resummation is explained in detail (see also Appendix~\ref{sec:SBEapproxPostproc} for a related discussion). This implicit resummation is not accessible within a parquet solution in which the SBE rest functions are not accounted for (see dotted gray lines in Fig.~\ref{fig:chiMDlambdaMDvsQPA} for comparison). This highlights an important advantage of the SBE approximation within the fRG framework that allows one to include contributions of the SBE rest functions at low numerical cost.

For completeness, we also compare the self-energy obtained from the converged multiloop SBE fRG with the parquet approximation. The corresponding frequency- and momentum-dependent results are reported in Appendix~\ref{app:Sigma}, for the same parameter regime considered in Figs.~\ref{fig:chiMlambdaMvsQ_SBEaCutoffDependence}--\ref{fig:Convergence_CompareCutoffs},~\ref{fig:chiMDlambdaMDvsQPA}, and~\ref{fig:chilambdaFreqDepPA_HalfFilling}. The agreement between fRG and parquet data is in that case of similar quality to that observed for the susceptibilities and Yukawa couplings, although with somewhat larger deviations at small frequencies.

\subsection{Comparison to the parquet approximation at finite doping}

\begin{figure}[t!]
    \centering
    \adjustbox{max width=0.49\textwidth, scale=1.0}{\includegraphics{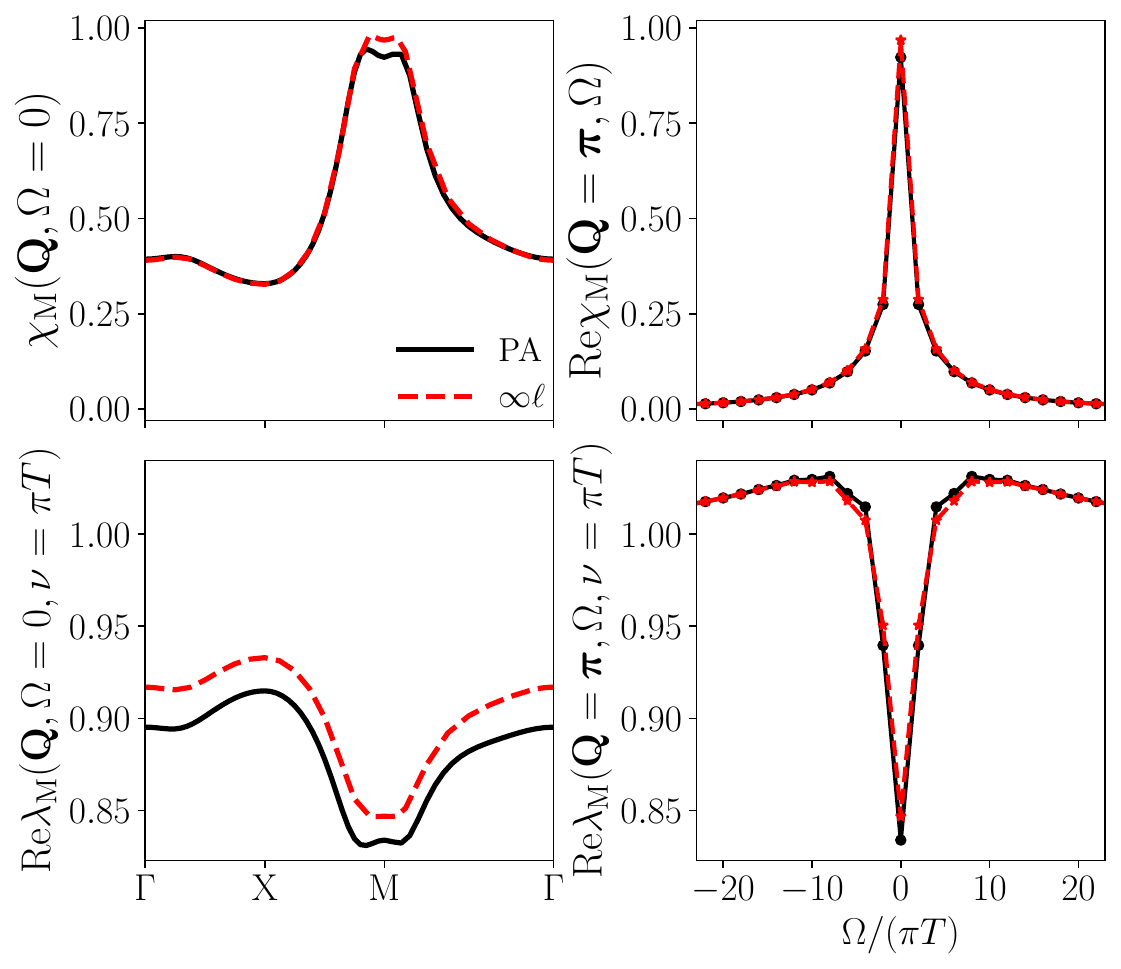}}
    \caption{Bosonic momentum and frequency dependencies of the real parts of the magnetic susceptibility and the magnetic Yukawa coupling at $U=2.5$, $\beta=5$, $t^\prime=-0.2$, and filling $n=0.41$ ($\mu=4 t^\prime$ at the beginning of the flow), as obtained from the multiloop SBE fRG at convergence in the SBE approximation (within the $\Omega$-flow), with comparison to the parquet approximation (PA).}
    \label{fig:chiMlambdaMvsQandOmegaPA}
\end{figure}

We now illustrate that the multiloop SBE fRG in the SBE approximation provides a quantitatively accurate description of the parquet approximation also at finite doping. The results for the bosonic momentum and frequency dependencies of the susceptibility and the Yukawa coupling in the leading channel (which is still the magnetic one) at Van Hove filling for the same bare interaction and temperature ($U=2.5$ and $\beta=5$) are displayed in Fig.~\ref{fig:chiMlambdaMvsQandOmegaPA}. In this figure, the relative difference between the fRG results and the parquet approximation does not exceed $2\%$ for $\mathrm{Re}\lambda_{\mathrm{M}}$, whereas it remains below $5\%$ for $\chi_{\mathrm{M}}(\mathbf{Q},\Omega=0)$ over the whole $\Gamma$-$\X$-$\M$-$\Gamma$ path. Such an agreement between fRG and parquet data also holds for the frequency dependence of the magnetic susceptibility, with the exception of small absolute values of $\mathrm{Re}\chi_{\mathrm{M}}(\mathbf{Q}=(\pi,\pi),\Omega)$ [$\mathrm{Re}\chi_{\mathrm{M}}(\mathbf{Q}=(\pi,\pi),\Omega)<0.012$]. Regarding the analogous results in the density and superconducting channels (not shown), the relative difference is always smaller than $4\%$, except for $\mathrm{Re}\chi_{\D}(\mathbf{Q}=\mathbf{0},\Omega)$ and $\mathrm{Re}\chi_{\SC}(\mathbf{Q}=\mathbf{0},\Omega)$ at small absolute values, i.e., for $\mathrm{Re}\chi_{\D}(\mathbf{Q}=\mathbf{0},\Omega)<0.003$ and $\mathrm{Re}\chi_{\SC}(\mathbf{Q}=\mathbf{0},\Omega)<0.011$.

\begin{figure}[t!]
    \centering
    \adjustbox{max width=0.5\textwidth, scale=0.95}{\includegraphics{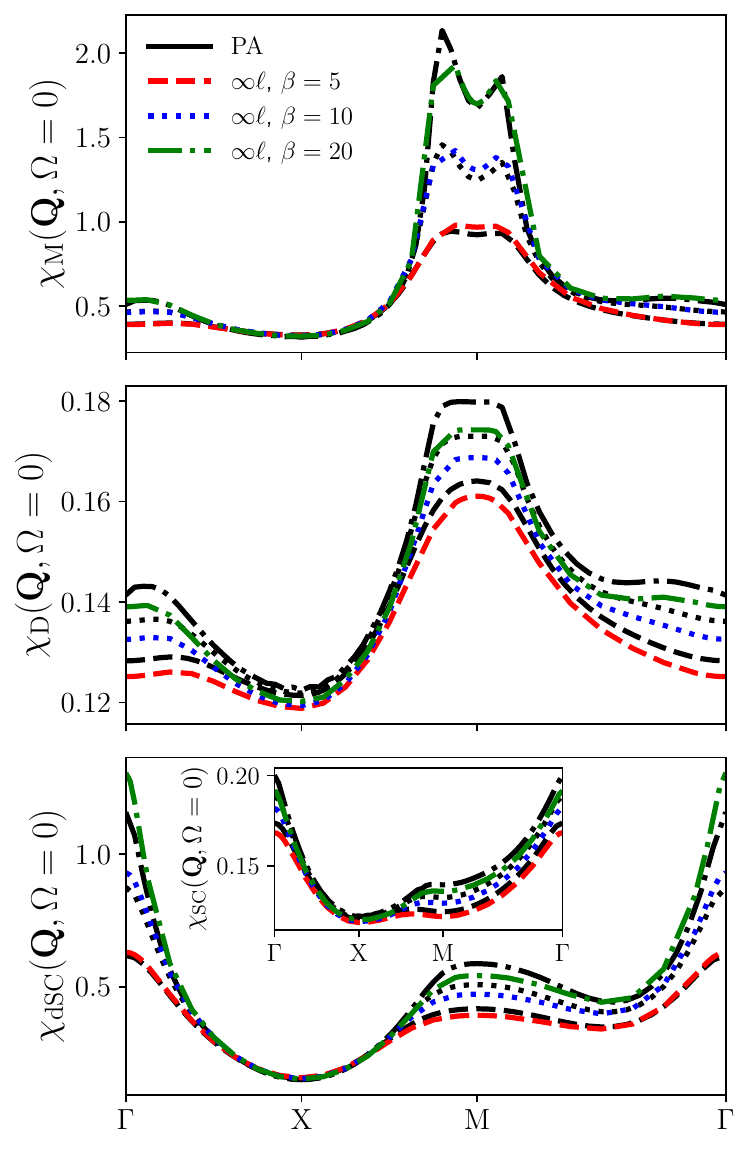}}
    \caption{Bosonic momentum dependence of the static susceptibilities in all three physical channels, as obtained from the multiloop SBE fRG at convergence in the SBE approximation (within the $\Omega$-flow), at $U=2.5$, $t^\prime=-0.2$, and filling $n=0.41$ ($\mu=4 t^\prime$ at the beginning of the flow) and for three different temperatures ($\beta=5,10,20$). Results obtained from the parquet approximation (PA) are also shown for comparison.}
    \label{fig:chivsQ_FiniteDoping_U2p5}
\end{figure}

\begin{figure}[t!]
    \centering
    \adjustbox{max width=0.5\textwidth, scale=0.95}{\includegraphics{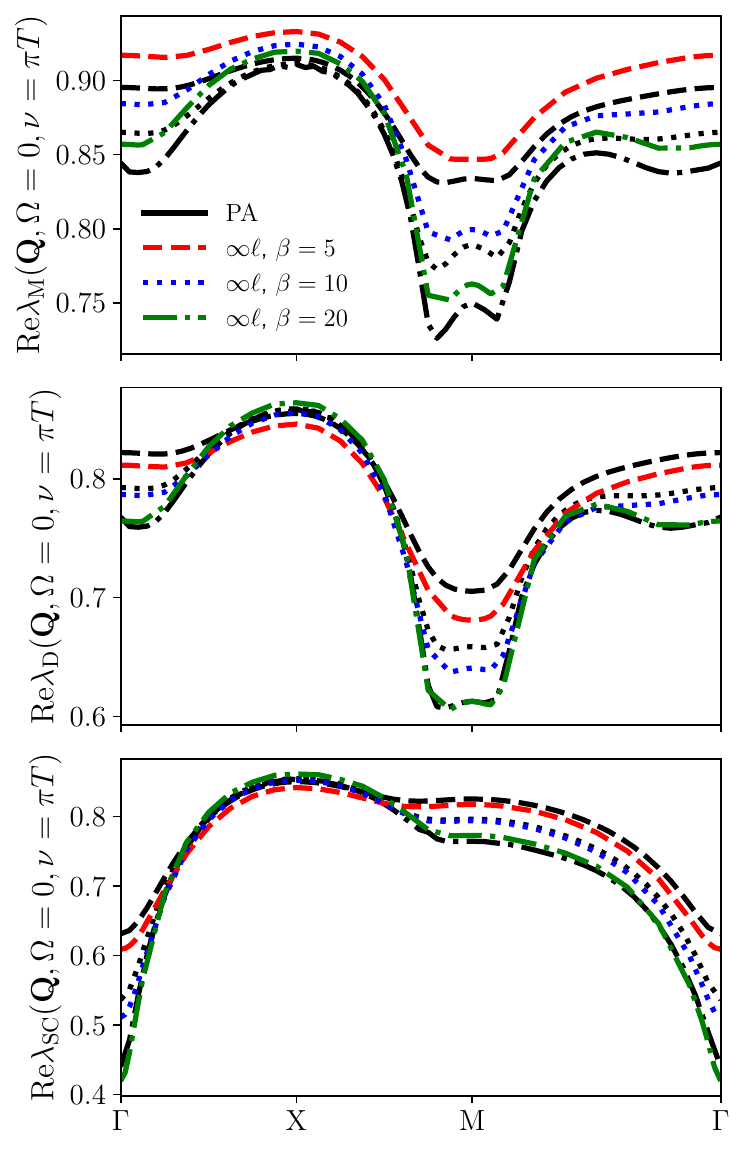}}
    \caption{Bosonic momentum dependence of the real parts of the Yukawa couplings in all three physical channels at zero bosonic frequency ($\Omega=0$) and at the first fermionic Matsubara frequency ($\nu=\pi T$) as obtained from the multiloop SBE fRG at convergence in the SBE approximation (within the $\Omega$-flow), at $U=2.5$, $t^\prime=-0.2$, and filling $n=0.41$ ($\mu=4 t^\prime$ at the beginning of the flow) and for three different temperatures ($\beta=5,10,20$). Results obtained from the parquet approximation (PA) are also shown for comparison.}
    \label{fig:lambdavsQ_FiniteDoping_U2p5}
\end{figure}

\begin{figure*}[t!]
    \centering
    \adjustbox{max width=\textwidth, scale=1.0}{\includegraphics{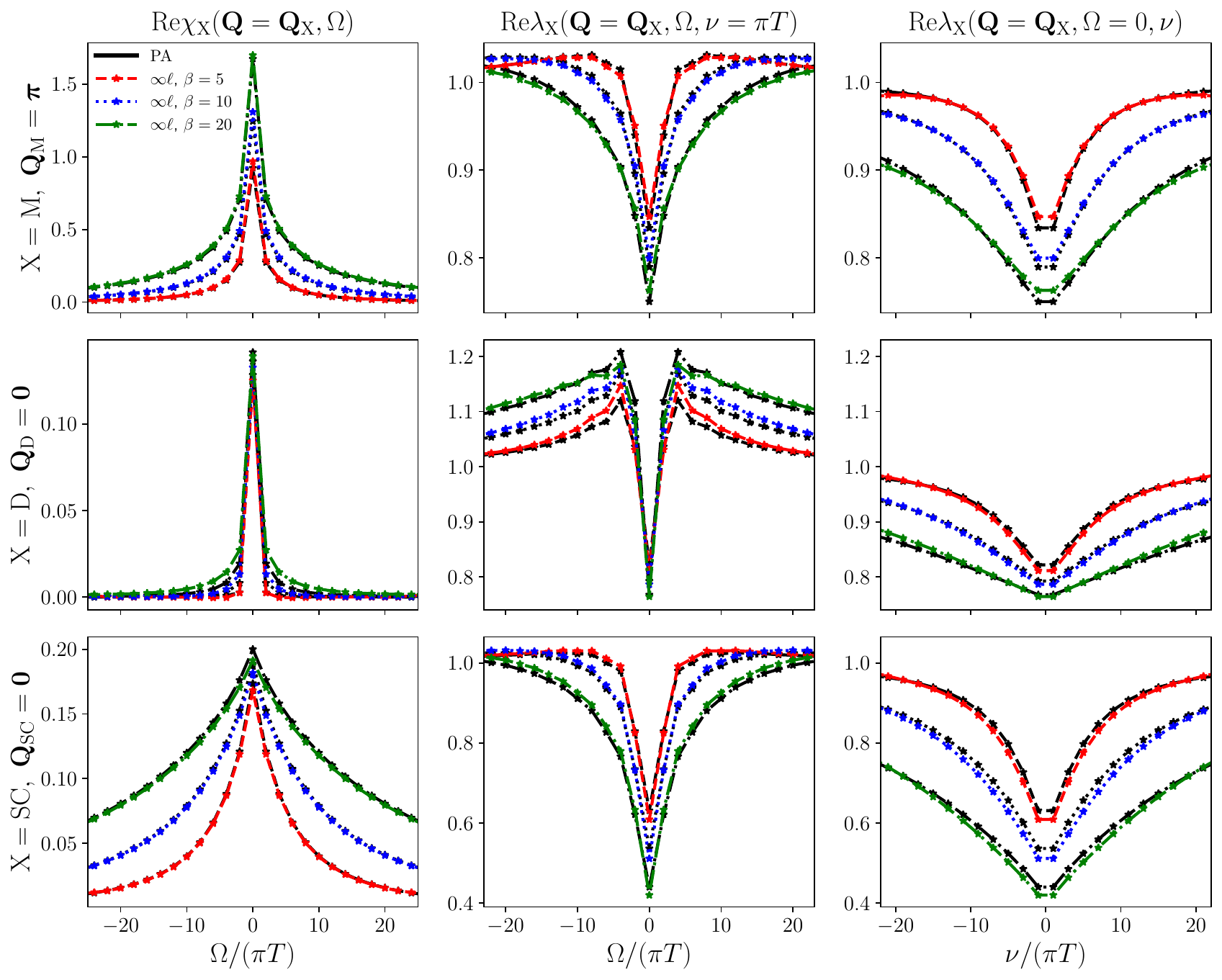}}
    \caption{Frequency dependencies of the real parts of the susceptibilities and of the Yukawa couplings in all three physical channels at bosonic momenta $\mathbf{Q}=\boldsymbol{\pi}=(\pi,\pi)$ (in the $\M$ channel) and $\mathbf{Q}=\mathbf{0}=(0,0)$ (in the $\D$ and $\SC$ channels) as obtained from the multiloop SBE fRG at convergence in the SBE approximation (within the $\Omega$-flow), at $U=2.5$, $t^\prime=-0.2$, and filling $n=0.41$ ($\mu=4 t^\prime$ at the beginning of the flow) and for three different temperatures ($\beta=5,10,20$). Results obtained from the parquet approximation (PA) are also shown for comparison.}
    \label{fig:chilambdaFreqDep_FiniteDoping_U2Beta5}
\end{figure*}

We now analyze the temperature dependence of our converged multiloop SBE fRG results at finite doping, still within the SBE approximation and exploiting the $\Omega$-flow, with convergence achieved according to the parameters $\varepsilon_{\text{se}} = 10^{-3}$ and $\varepsilon_{\text{vtx}} = 10^{-4}$ defined in Sec.~\ref{sec:LoopCorrectionsSigmaIterations} (for completeness, $1\ell$ and $2\ell$ results are reported in Appendix~\ref{app:addres}). In Figs.~\ref{fig:chivsQ_FiniteDoping_U2p5}--\ref{fig:chilambdaFreqDep_FiniteDoping_U2Beta5}, we report the frequency and momentum dependencies of the susceptibilities and the Yukawa couplings in all physical channels at $U=2.5$, $t^\prime=-0.2$, filling $n=0.41$, and three different temperatures, i.e., $\beta=5$, $10$, and $20$ (the results at $\beta=5$ correspond to the data shown in Fig.~\ref{fig:chiMlambdaMvsQandOmegaPA}). Figure~\ref{fig:chivsQ_FiniteDoping_U2p5} illustrates how the competition between magnetic ordering and $d$-wave superconductivity evolves with temperature, as resolved by our multiloop SBE fRG approach, which is an unbiased tool to study competing orders. For all considered values of the temperature, the magnetic channel remains the dominant one (even at $\beta=20$), although one can clearly observe that $d$-wave superconductivity becomes more significant as $\beta$ increases from the panel showing $\chi_{\mathrm{dSC}}$ in Fig.~\ref{fig:chivsQ_FiniteDoping_U2p5}. At finite doping, magnetism occurs at incommensurate wave vectors, characterized by increasingly high and sharp peaks in $\chi_{\mathrm{M}}$. In general, the temperature dependence of the leading magnetic and $d$-wave superconducting static susceptibilities is much more pronounced with respect to the density and $s$-wave superconducting ones: $\chi_{\mathrm{M}}(\mathbf{Q}=(\pi,\pi),\Omega=0)$ and $\chi_{\mathrm{dSC}}(\mathbf{Q}=(0,0),\Omega=0)$ are multiplied by a factor of $2$ from $\beta=5$ to $\beta=20$, while $\chi_{\mathrm{D}}(\mathbf{Q}=(0,0),\Omega=0)$ and $\chi_{\mathrm{SC}}(\mathbf{Q}=(0,0),\Omega=0)$ increase by about $10\%$. The fRG results of Fig.~\ref{fig:chivsQ_FiniteDoping_U2p5} also agree well with the corresponding parquet approximation, with a relative difference of at most $5\%$ for $\chi_{\mathrm{M}}(\mathbf{Q},\Omega=0)$, $\chi_{\mathrm{D}}(\mathbf{Q},\Omega=0)$, and $\chi_{\mathrm{SC}}(\mathbf{Q},\Omega=0)$ over the whole $\Gamma$-$\X$-$\M$-$\Gamma$ path and for $\beta=5$, $10$, and $20$. For the $d$-wave superconducting susceptibility $\chi_{\mathrm{dSC}}(\mathbf{Q},\Omega=0)$, we note that the relative difference between fRG and parquet data reaches slightly higher values, especially at the highest temperature. At the $\Gamma$ point, in particular, it reaches $2\%$, $6\%$, and $13\%$ at $\beta=5$, $10$, and $20$, respectively. This reduction of the quality of the SBE approximation for the $d$-wave superconducting channel is expected from the fact that the contributions of the SBE rest functions $M_{\mathrm{dSC}}$ must be taken into account to describe pairing fluctuations in strongly correlated regimes as already discussed in Refs.~\cite{Bonetti2022,Fraboulet2022}. Improvements of our SBE fRG approach toward the treatment of these fluctuations could be achieved with the help of rebosonization techniques~\cite{Gies2002,Gies2002b,Krahl2007,Pawlowski2007,Floerchinger2009,Bonetti2022} used to recast the information of $M_{\mathrm{dSC}}$ in the form of a bosonic propagator and a Yukawa coupling. Such developments remain, however, beyond the scope of the present study.

We then consider the bosonic momentum dependence of the Yukawa couplings in Fig.~\ref{fig:lambdavsQ_FiniteDoping_U2p5}. For all channels $\mathrm{X}=\mathrm{M},\mathrm{D},\mathrm{SC}$, we observe a strong renormalization of their real part with respect to their initial value $\lambda_{\mathrm{X}}(\mathbf{Q},\Omega,\nu)=1$. This effect can notably be attributed to the Kanamori screening~\cite{Kanamori1963,Katanin2009,Krien2020a,Fraboulet2022} and is substantially enhanced as the temperature decreases. We note that the renormalization of the Yukawa couplings results from the \emph{interplay} between the different channels, absent in the random phase approximation (RPA)~\footnote{The ($1\ell$) fRG scheme reproduces the RPA if only the flow of the bosonic propagators $w_\textrm{X}$ is considered, with $\lambda_{\mathrm{X}}(\mathbf{Q},\Omega,\nu)=1$ for all $\mathrm{X}$ and neglecting the self-energy flow.}. Accounting for their flow is therefore essential for an accurate description of competing orders. Furthermore, the relative difference between the Yukawa couplings obtained from our fRG approach and the parquet approximation shown in Fig.~\ref{fig:lambdavsQ_FiniteDoping_U2p5} remains strictly below $5\%$ for $\mathrm{Re}\lambda_{\mathrm{X}}(\mathbf{Q},\Omega=0,\nu=\pi T)$ over the whole $\Gamma$-$\X$-$\M$-$\Gamma$ path, in all three channels ($\mathrm{X}=\mathrm{M},\mathrm{D},\mathrm{SC}$) and for $\beta=5$, $10$, and $20$. This agreement with the parquet approximation is therefore similar to the one observed for the $s$-wave susceptibilities in Fig.~\ref{fig:chivsQ_FiniteDoping_U2p5}.

The significant renormalization of the Yukawa couplings can also be observed from their frequency dependencies shown in Fig.~\ref{fig:chilambdaFreqDep_FiniteDoping_U2Beta5}. Reducing the temperature, i.e., increasing $\beta$, the significant renormalization of the Yukawa couplings extends to higher bosonic and fermionic frequencies, with a rapid broadening of the affected frequency window. A similar behavior is also observed for the corresponding susceptibilities that exhibit a marked broadening at lower temperatures. These findings demonstrate the merits of our fRG approach, which captures such renormalization effects in the frequency- and momentum-dependent vertices. Finally, concerning the comparison with the parquet approximation, we obtain a particularly small relative difference, always below $3\%$ for all frequencies, between fRG and parquet data of $\mathrm{Re}\lambda_{\mathrm{M}}$ and $\mathrm{Re}\lambda_{\mathrm{D}}$. This relative difference is slightly higher, but remains below $5\%$, for $\mathrm{Re}\lambda_{\mathrm{SC}}$. Finally, the susceptibilities obtained from the fRG agree with the parquet approximation within a relative difference of $5\%$ or less for $\beta=5$, $10$, and $20$ at $\Omega=0$, as found previously in Fig.~\ref{fig:chivsQ_FiniteDoping_U2p5}. For the (real part of the) susceptibilities at finite frequency $\Omega$, the situation at $\beta=5$ was already analyzed previously in Fig.~\ref{fig:chiMlambdaMvsQandOmegaPA}, highlighting an efficient reproduction of the parquet approximation by our fRG results in that case. This efficient reproduction is also observed at $\beta=10$ and $20$ in Fig.~\ref{fig:chilambdaFreqDep_FiniteDoping_U2Beta5}, where the relative difference between fRG and parquet data does not exceed $5\%$ for $\mathrm{Re}\chi_{\mathrm{M}}$ and $\mathrm{Re}\chi_{\mathrm{SC}}$ even at finite $\Omega$ (in the density channel, this relative difference can be larger than $5\%$, but only for $\mathrm{Re}\chi_{\mathrm{D}}<0.028$). Overall, we observe in Figs.~\ref{fig:chivsQ_FiniteDoping_U2p5}--\ref{fig:chilambdaFreqDep_FiniteDoping_U2Beta5} that the quality of the agreement between our converged multiloop SBE fRG results in the SBE approximation and the parquet approximation is not noticeably affected by lowering the temperature (with the exception of $\chi_{\mathrm{dSC}}$ shown in Fig.~\ref{fig:chivsQ_FiniteDoping_U2p5}), at least within the considered temperature range. This highlights a certain robustness of the SBE approximation used within our fRG approach.

\section{Conclusion}
\label{sec:Conclusion}

We have applied the recently introduced SBE formulation of the fRG, which relies on a diagrammatic decomposition in contributions mediated by the exchange of a single boson in the different channels, to the multiloop fRG that resums all diagrams of the parquet approximation. We presented the details of the algorithmic implementation for the 2D Hubbard model and provided a systematic analysis of the fermion-boson Yukawa couplings and of the corresponding physical susceptibilities by studying their evolution with temperature and interaction strength, both at half filling and finite doping. We complemented this analysis with additional results for the self-energy in Appendix~\ref{app:Sigma}. The comparison with the parquet approximation shows that neglecting the SBE rest functions, which describe correlation effects beyond the SBE processes, yields a quantitatively accurate description in the weak-coupling regime, both at half filling and at finite doping in the studied parameter regime. The quality of this approximation should be considered alongside the fact that the SBE rest functions are significantly more demanding to calculate than the vertices describing SBE processes (i.e., bosonic propagators and Yukawa couplings). It should also be stressed that rest-function-like contributions are still generated throughout the fRG flow in the SBE approximation (see also Ref.~\cite{Fraboulet2022}): This resummation has no counterpart in the iterative solution of the parquet equations without the SBE rest functions.

As a result, the SBE formulation of the multiloop fRG provides a promising framework that allows for a substantial reduction of the numerical effort, paving a promising route for future extensions such as temperature flows for electron-phonon systems~\cite{AlEryani2026b} or the combination with DMFT in the so-called DMF$^2$RG~\cite{Taranto2014,Vilardi2019}. The latter could enable access to the nonperturbative regime by means of the multiloop extension of the DMF$^2$RG.

\acknowledgments 
The authors thank M. Gievers, M. Krämer, M. Patricolo, A. Toschi, D. Vilardi, and J. von Delft for valuable discussions. We acknowledge financial support from the Deutsche Forschungsgemeinschaft (DFG, German Research Foundation) within the research unit FOR 5413/1 (Grant No. 465199066). This research was also funded by the Austrian Science Fund (FWF) through 10.55776/I6946, 10.55776/KIN2563725,  10.55776/P36332, and 10.55776/V1018. A.A.-E. acknowledges funding from the DFG under Project No.~277146847 (SFB 1238, project C02). Computational resources provided by the Paderborn Center for Parallel Computing (PC2), specifically the Noctua 2 high-performance computing system, have been used for conducting this research.
In addition, we also used the Austrian Scientific Computing (ASC) infrastructure.

\appendix
\section*{Appendices}

\section{Multiloop SBE fRG equations in form factor notation}
\label{App:TUmfRG}

In the truncated-unity fRG~\cite{Husemann2009,Wang2012,Lichtenstein2017}, the flow equations are rewritten in terms of form factors. For Eqs.~\eqref{eq:mfRGequationsPhysicalChannels1l}--\eqref{eq:mfRGequationsPhysicalChannelsbeyond2l} and~\eqref{eq:SigmaMdotFlowEquation}, this gives
\begin{widetext}
\begin{subequations}
    \begin{align}
        \dot{w}_{\text{X}}^{(1)}(Q) & = \left(w_{\text{X}}(Q)\right)^2 \sum_{\nu,m,m^{\prime}} \lambda_{\text{X},m}(Q,\nu) \dot{\Pi}_{\text{X},m m^{\prime}}(Q,\nu) \lambda_{\text{X},m^{\prime}}(Q,\nu), \\
        \dot{\lambda}_{\text{X},n}^{(1)}(Q,\nu) & = \sum_{\nu^{\prime},m,m^{\prime}} \lambda_{\text{X},m}(Q,\nu^{\prime}) \dot{\Pi}_{\text{X},m m^{\prime}}(Q,\nu^{\prime}) \mathcal{I}_{\text{X},m^{\prime}n}(Q,\nu^{\prime},\nu), \\
        \dot{M}_{\text{X},n n^{\prime}}^{(1)}(Q,\nu,\nu^{\prime}) & = \sum_{\nu^{\prime\prime},m,m^{\prime}} \mathcal{I}_{\text{X},n m}(Q,\nu,\nu^{\prime\prime}) \dot{\Pi}_{\text{X},m m^{\prime}}(Q,\nu^{\prime\prime}) \mathcal{I}_{\text{X},m^{\prime} n^{\prime}}(Q,\nu^{\prime\prime},\nu^{\prime}),
    \end{align}
    \label{eq:TUmfRGequationsPhysicalChannels1l}
\end{subequations}
\begin{subequations}
    \begin{align}
        \dot{w}_{\text{X}}^{(2)}(Q) & = 0, \\
        \dot{\lambda}_{\text{X},n}^{(2)}(Q,\nu) & = \sum_{\nu^{\prime},m,m^{\prime}} \lambda_{\text{X},m}(Q,\nu^{\prime}) \Pi_{\text{X},m m^{\prime}}(Q,\nu^{\prime}) \dot{I}_{\text{X},m^{\prime} n}^{(1)}(Q,\nu^{\prime},\nu), \\
        \dot{M}_{\text{X},n n^{\prime}}^{(2)}(Q,\nu,\nu^{\prime}) & = \sum_{\nu^{\prime\prime},m,m^{\prime}} \dot{I}_{\text{X},n m}^{(1)}(Q,\nu,\nu^{\prime\prime}) \Pi_{\text{X},m m^{\prime}}(Q,\nu^{\prime\prime}) \mathcal{I}_{\text{X},m^{\prime} n^{\prime}}(Q,\nu^{\prime\prime},\nu^{\prime}) \nonumber\\
        & \phantom{=} + \sum_{\nu^{\prime\prime},m,m^{\prime}} \mathcal{I}_{\text{X},n m}(Q,\nu,\nu^{\prime\prime}) \Pi_{\text{X},m m^{\prime}}(Q,\nu^{\prime\prime}) \dot{I}_{\text{X},m^{\prime} n^{\prime}}^{(1)}(Q,\nu^{\prime\prime},\nu^{\prime}),
    \end{align}
    \label{eq:TUmfRGequationsPhysicalChannels2l}
\end{subequations}
and, for $\ell\geq 3$,
\begin{subequations}
    \begin{align}
        \dot{w}_{\text{X}}^{(\ell)}(Q) & = \left(w_{\text{X}}(Q)\right)^2 \sum_{\substack{\nu,m,m^{\prime}, \\ \nu^{\prime},m^{\prime\prime},m^{\prime\prime\prime}}} \lambda_{\text{X},m}(Q,\nu) \Pi_{\text{X},m m^{\prime}}(Q,\nu) \dot{I}^{(\ell-2)}_{\text{X},m^{\prime} m^{\prime\prime}}(Q,\nu,\nu^{\prime}) \Pi_{\text{X},m^{\prime\prime} m^{\prime\prime\prime}}(Q,\nu^{\prime}) \lambda_{\text{X},m^{\prime\prime\prime}}(Q,\nu^{\prime}), \\
        \dot{\lambda}_{\text{X},n}^{(\ell)}(Q,\nu) & = \sum_{\nu^{\prime},m,m^{\prime}} \lambda_{\text{X},m}(Q,\nu^{\prime}) \Pi_{\text{X},m m^{\prime}}(Q,\nu^{\prime}) \dot{I}_{\text{X},m^{\prime} n}^{(\ell-1)}(Q,\nu^{\prime},\nu) \nonumber \\
        & \phantom{=} + \sum_{\substack{\nu^{\prime},m,m^{\prime}, \\ \nu^{\prime\prime},m^{\prime\prime},m^{\prime\prime\prime}}} \lambda_{\text{X},m}(Q,\nu^{\prime}) \Pi_{\text{X},m m^{\prime}}(Q,\nu^{\prime}) \dot{I}^{(\ell-2)}_{\text{X},m^{\prime} m^{\prime\prime}}(Q,\nu^{\prime},\nu^{\prime\prime}) \Pi_{\text{X},m^{\prime\prime} m^{\prime\prime\prime}}(Q,\nu^{\prime\prime}) \mathcal{I}_{\text{X},m^{\prime\prime\prime} n}(Q,\nu^{\prime\prime},\nu), \\
        \dot{M}_{\text{X},n n^{\prime}}^{(\ell)}(Q,\nu,\nu^{\prime}) & = \sum_{\nu^{\prime\prime},m,m^{\prime}} \dot{I}_{\text{X},n m}^{(\ell-1)}(Q,\nu,\nu^{\prime\prime}) \Pi_{\text{X},m m^{\prime}}(Q,\nu^{\prime\prime}) \mathcal{I}_{\text{X},m^{\prime} n^{\prime}}(Q,\nu^{\prime\prime},\nu^{\prime}) \nonumber \\
        & \phantom{=} + \sum_{\nu^{\prime\prime},m,m^{\prime}} \mathcal{I}_{\text{X},n m}(Q,\nu,\nu^{\prime\prime}) \Pi_{\text{X},m m^{\prime}}(Q,\nu^{\prime\prime}) \dot{I}_{\text{X},m^{\prime} n^{\prime}}^{(\ell-1)}(Q,\nu^{\prime\prime},\nu^{\prime}) \nonumber \\
        & \phantom{=} + \sum_{\substack{\nu^{\prime\prime},m,m^{\prime}, \\ \nu^{\prime\prime\prime},m^{\prime\prime},m^{\prime\prime\prime}}} \mathcal{I}_{\text{X},n m}(Q,\nu,\nu^{\prime\prime}) \Pi_{\text{X},m m^{\prime}}(Q,\nu^{\prime\prime}) \dot{I}^{(\ell-2)}_{\text{X},m^{\prime} m^{\prime\prime}}(Q,\nu^{\prime\prime},\nu^{\prime\prime\prime}) \Pi_{\text{X},m^{\prime\prime} m^{\prime\prime\prime}}(Q,\nu^{\prime\prime\prime}) \mathcal{I}_{\text{X},m^{\prime\prime\prime} n^{\prime}}(Q,\nu^{\prime\prime\prime},\nu^{\prime}),
    \end{align}
    \label{eq:TUmfRGequationsPhysicalChannelsbeyond2l}
\end{subequations}
\end{widetext}
and
\begin{align}
    \dot{\Sigma}(k) & = \sum_{Q,m} \dot{w}_{\mathrm{M}}(Q) f_m(\mathbf{k}) \lambda_{\mathrm{M},m}\bigg(Q,\nu + \left\lfloor\frac{\Omega}{2}\right\rfloor\bigg) G(k+Q) \nonumber \\
    & \phantom{=} + \sum_{Q,m} w_{\mathrm{M}}(Q) f_m(\mathbf{k}) \dot{\lambda}_{\mathrm{M},m}\bigg(Q,\nu + \left\lfloor\frac{\Omega}{2}\right\rfloor\bigg) G(k+Q) \nonumber \\
    & \phantom{=} + \sum_{Q,m} w_{\mathrm{M}}(Q) f_m(\mathbf{k}) \lambda_{\mathrm{M},m}\bigg(Q,\nu + \left\lfloor\frac{\Omega}{2}\right\rfloor\bigg) \dot{G}(k+Q), \label{eq:TUmfRGSigmaMdotFlowEquation}
\end{align}
where the indices $n^{(\prime)}$ and $m^{(\prime)}$ label form factors [$\lbrace f_m(\mathbf{k}) \rbrace_{m=0}^\infty$ denotes a complete set of form factors defined on the Brillouin zone]. As mentioned, only $s$-wave and $d$-wave form factors [defined by Eqs.~\eqref{eq:sdwaveformfactors}] are considered for the present study of the 2D Hubbard model. Moreover, $\mathcal{I}_{\text{X},n n^\prime}$ and $\dot{I}_{\text{X},n n^\prime}^{(\ell)}$ also satisfy, respectively, Eqs.~\eqref{eq:ExpressionsMathcalIX} and~\eqref{eq:ExpressionsdotIellX} after replacing each object by the corresponding form factor component, including notably the projection matrices $P^{r\rightarrow r^{\prime}}$ whose expressions in form factor notation can be found in Refs.~\cite{HillePhDThesis,HeinzelmannPhDThesis}.

\section{Vanishing of $d$-wave Yukawa couplings in the SBE approximation}
\label{sec:vanishing_yukawa}

In the main text, we stated 
that neglecting the mixed bubbles leads to $\lambda_{\X,\mathrm{dw}} = 0$, where $\lambda_{\X,\mathrm{dw}}$ is the $d$-wave component of $\lambda_{\X}$. We here demonstrate that if
\begin{enumerate}
    \item the SBE approximation applies ($\dot{M}_\X = 0$),
    \item mixed bubbles and their derivatives are neglected ($\Pi_{\X,nm} = \dot{\Pi}_{\X,nm} = 0$ for $n \neq m$),
    \item only the local $s$-wave form factor is included (no extended $s$-wave form factors, but $d$-wave, etc., form factors may be included),
    \item at the initial fRG scale $\lambda^{\Lambda_{\text{init}}}_{\X, n} = 0$ for $n \neq \mathrm{sw}$,
\end{enumerate}
then $\lambda_{\X,m} = 0$ throughout the flow whenever the form factor $f_m$ changes sign under any of the point group symmetries, i.e.,
\begin{equation}
    f_m(\bfk) = -f_m(R\bfk),
\label{eq:higher_harmonic_ff}
\end{equation}
where $R$ is a matrix representing some lattice symmetry that acts on the momentum vector. In terms of the representations of the point group of the underlying lattice of the model, such a form factor represents a nontrivial (non-$s$-wave) representation. For example, $m = \mathrm{dw}$ satisfies this property on the square lattice for $R$ being a rotation by $\pi/2$.

To show the main claim, one needs to consider the general structure of the flow equation of the Yukawa couplings. At the initial scale $\Lambda_{\text{init}}$, the flow equation takes the form (omitting the frequency dependence that will play no role in our argumentation)
\begin{widetext}
\begin{align}
\dot{\lambda}^{\Lambda_{\text{init}}}_{\X,m}(\mathbf{Q}) &= \sum_{n,n'} \lambda^{\Lambda_{\text{init}}}_{\X,n} \dot{\Pi}^{\Lambda_{\text{init}}}_{\X,nn'}(\mathbf{Q}) \left[\sum_{\bfk,\bfk'} f_{n'}^*(\bfk)f_{m}(\bfk')B_{\X'}^{\Lambda_{\text{init}}}(\mathbf{Q}, \bfk, \bfk')\right]\nonumber\\
&\approx\dot{\Pi}^{\Lambda_{\text{init}}}_{\X,\mathrm{sw} \, \mathrm{sw}}(\mathbf{Q}) \left[\sum_{\bfk'} f_{m}(\bfk')\sum_{\bfk}B_{\X'}^{\Lambda_{\text{init}}}(\mathbf{Q}, \bfk, \bfk')\right].
\label{eq:a2}
\end{align}
\end{widetext}
In going from the first to the second line, we have used that $\lambda_{\X,m}^{\Lambda_{\text{init}}} = \delta_{m, \mathrm{sw}}$, $f_{\mathrm{sw}} = 1$, and the approximation that mixed bubbles are zero (the bubble above may be a nondifferentiated bubble in the multiloop corrections to $\lambda_\X$). If $f_m$ satisfies Eq.~\eqref{eq:higher_harmonic_ff} and 
\begin{align}
\sum_{\bfk}B_{\X}^{\Lambda}(Q, k, k') = \sum_{\bfk}B_\X^{\Lambda}(Q, k, R k'), \label{eq:symmetry_requirement_for_vanishing_of_lambda_dw}
\end{align}
then the integral in the square bracket of Eq. \eqref{eq:a2} vanishes and consequently $\lambda_{\X, m}^{\Lambda_{\text{init}}} = 0$. It follows that this applies to all scales $\Lambda$.

We now show that Eq.~\eqref{eq:symmetry_requirement_for_vanishing_of_lambda_dw} holds in the SBE approximation. Since in the SBE approximation $M_\X^\Lambda \equiv 0$ throughout the flow, $B_\X$ will generally consist of only non-channel-native single-boson contributions, i.e., contributions that involve interchannel projections $P^{r \rightarrow r'}\nabla_r(Q, k, k')$, with $r \neq r'$. For example, in the $1\ell$ equation, $B_\X(Q, k, k') = \mathcal{I}_\X(Q, k, k')$. We consider, as an example, a contribution to $B_{\SC}$ from the magnetic single-boson term given by
\begin{align}
P^{\overline{ph} \rightarrow pp}\nabla_{\M}(\mathbf{Q}, \bfk, \bfk') = \nabla_{\M}(\mathbf{Q} - \bfk - \bfk', \bfk, \bfk'),
\end{align}
where we have performed the translation following Fig.~\ref{fig:FrequencyMomentumParametrization}. From the integral 
\begin{align}
\sum_\bfk P^{\overline{ph} \rightarrow pp}\nabla^\Lambda_{\M}(\mathbf{Q}, \bfk, R\bfk') = \sum_\bfk \nabla^\Lambda_{\M}(\mathbf{Q} - \bfk - R\bfk', \bfk, R\bfk'),
\end{align}
and the explicit expression $\nabla_\X(Q, k, k') = \lambda_\X(Q, k) w_\X(Q) \lambda_\X(Q, k')$, we see that the secondary momentum dependence comes only from the Yukawa couplings. However, at the initial fRG scale, only the $s$-wave components of $\lambda_\X$ are nonzero and hence $\nabla_\X$ does not depend at all on its secondary momentum arguments. We thus have that
\begin{align}
\sum_\bfk P^{\overline{ph} \rightarrow pp}\nabla_\M^{\Lambda_{\text{init}}}(\mathbf{Q}, \bfk, R\bfk') = \sum_\bfk \nabla^{\Lambda_{\text{init}}}_\M(\mathbf{Q} - \bfk - R\bfk') ,\label{eq:simplification_interchannel_nabla}
\end{align}
which is in fact a constant at the initial scale due to the translation invariance of the integration domain (i.e., the periodicity of the Brillouin zone) and Eq.~\eqref{eq:symmetry_requirement_for_vanishing_of_lambda_dw} is trivially satisfied. At a later scale, a nontrivial dependence of the secondary momenta can be generated. However, if Eq.~\eqref{eq:symmetry_requirement_for_vanishing_of_lambda_dw} was satisfied at the previous fRG step, it may only be of the $s$-wave type. If no extended $s$-wave form factors are included, one ends up at the next scale with trivial secondary momentum dependencies as in  Eq.~\eqref{eq:simplification_interchannel_nabla}, and Eq.~\eqref{eq:symmetry_requirement_for_vanishing_of_lambda_dw} holds at every step of the flow. It can be readily checked that these arguments also hold when considering contributions from and to other channels. Therefore, with the assumptions specified at the beginning of this Appendix, we have $\lambda_{\X, \mathrm{dw}} = 0$. More generally, $\lambda_{\X, m} = 0$ for any $m$ for which the corresponding form factor satisfies Eq.~\eqref{eq:higher_harmonic_ff}.

\section{SBE approximation for postprocessing susceptibilities}
\label{sec:SBEapproxPostproc}

\begin{figure}[t!]
    \centering
    \adjustbox{max width=0.49\textwidth, scale=1.0}{\includegraphics{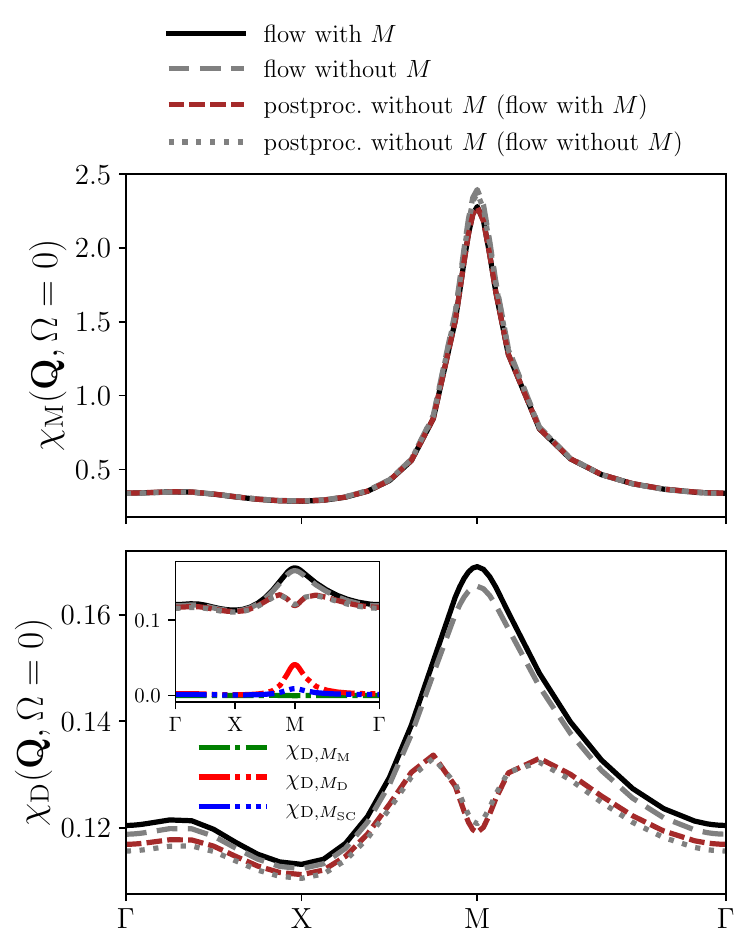}}
    \caption{Static bosonic momentum dependence of the flowing and postprocessing susceptibilities in the magnetic and density channels at $U=2.5$, $\beta=5$, $t^\prime=0$, and half filling ($\mu=0$), as obtained from the multiloop SBE fRG at convergence (within the $\Omega$-flow), by neglecting (or not) the SBE rest functions $M_{\mathrm{X}}$ either at the level of the flow equations [i.e., ignoring Eqs.~\eqref{eq:flowequationsMXPhysicalChannels1l},~\eqref{eq:flowequationsMXPhysicalChannels2l}, and~\eqref{eq:flowequationsMXPhysicalChannelsbeyond2l}] or at the level postprocessing susceptibilities [i.e., setting $\chi_{\mathrm{X},M_{\mathrm{X}^\prime}}=0$ for all $\mathrm{X}$ and $\mathrm{X}^\prime$ in Eq.~\eqref{eq:generalized_susceptibility_appendix}]. The corresponding data for the superconducting channel are obtained by $\chi_{\mathrm{SC}}(\mathbf{Q},\Omega)=\chi_{\mathrm{D}}(\mathbf{Q}+(\pi,\pi),\Omega)$.
    }
    \label{fig:flowVsPostproc}
\end{figure}

\begin{figure}[t!]
    \centering
    \adjustbox{max width=0.49\textwidth, scale=0.95}{\includegraphics{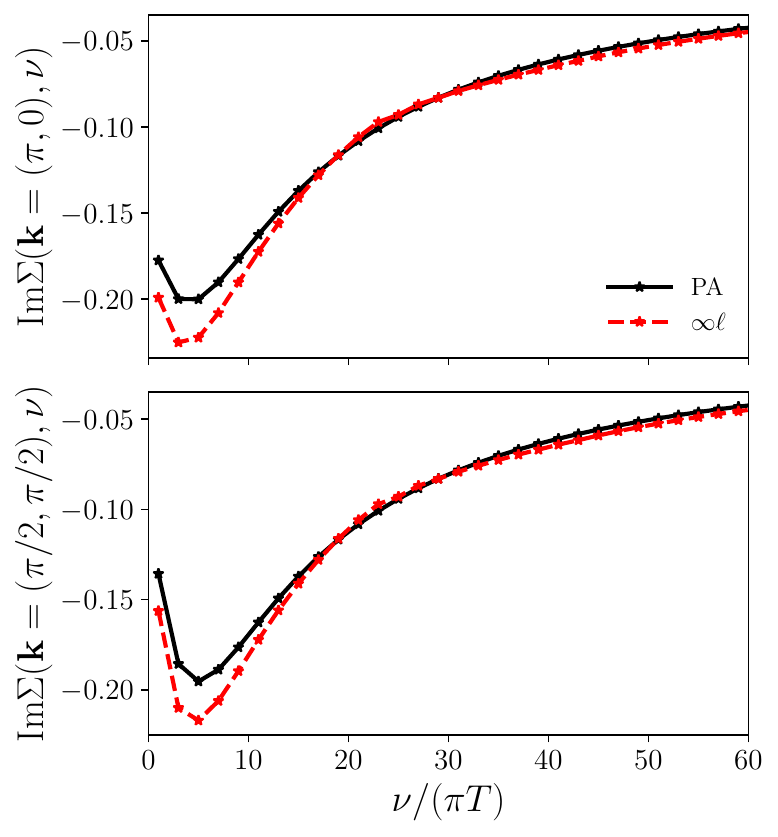}}
    \caption{Frequency dependence of the imaginary part of the self-energy at the node [$\mathbf{k}=(\pi/2,\pi/2)$] and the antinode [$\mathbf{k}=(\pi,0)$], at $U=2.5$, $\beta=5$, $t^\prime=0$, and half filling ($\mu=0$), as obtained from the multiloop SBE fRG at convergence in the SBE approximation (within the $\Omega$-flow), with comparison to the parquet approximation (PA).}
    \label{fig:ImSigvsnu}
\end{figure}

\begin{figure}[t!]
    \centering
    \adjustbox{max width=0.49\textwidth, scale=0.95}{\includegraphics{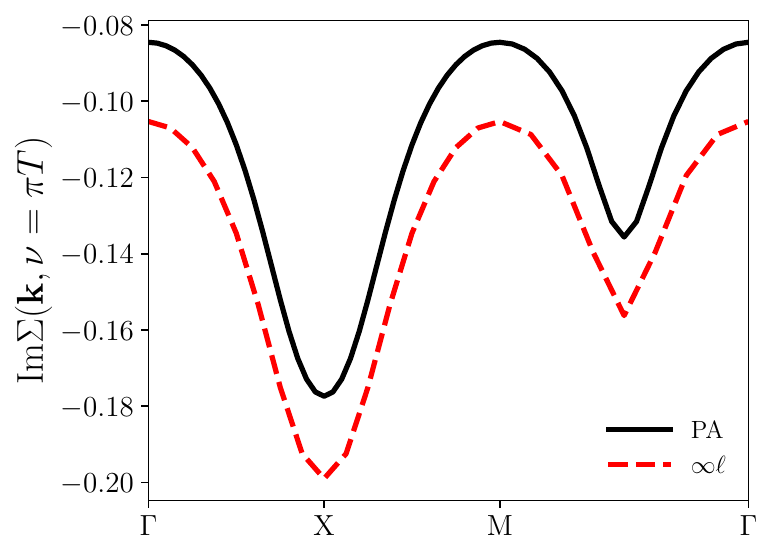}}
    \caption{Momentum dependence of the imaginary part of the self-energy at the first Matsubara frequency ($\nu=\pi T$), at $U=2.5$, $\beta=5$, $t^\prime=0$, and half filling ($\mu=0$), as obtained from the multiloop SBE fRG at convergence in the SBE approximation (within the $\Omega$-flow), with comparison to the parquet approximation (PA).}
    \label{fig:ImSigvsk}
\end{figure}

\begin{figure}[t!]
    \centering
    \adjustbox{max width=0.5\textwidth, scale=0.95}{\includegraphics{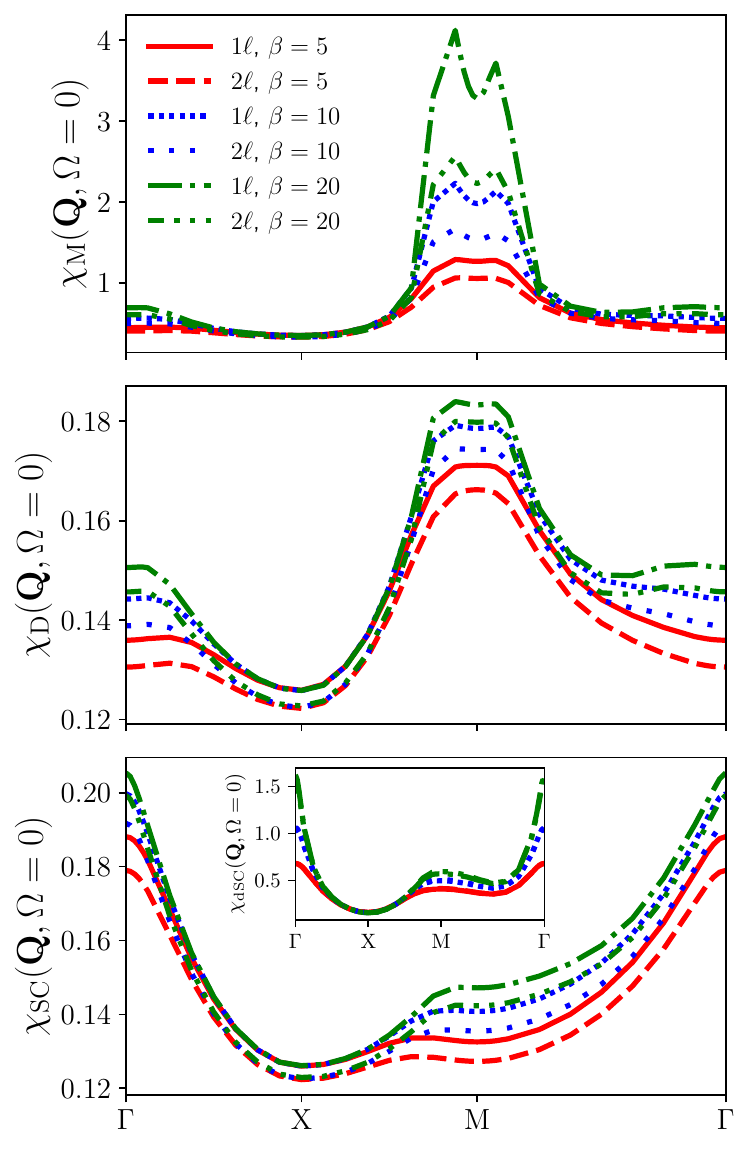}}
    \caption{Same as Fig.~\ref{fig:chivsQ_FiniteDoping_U2p5} but for results obtained from the $1\ell$ and $2\ell$ SBE fRG.
    }
    \label{fig:chivsQ1l2l_FiniteDoping_U2p5}
\end{figure}

\begin{figure}[t!]
    \centering
    \adjustbox{max width=0.5\textwidth, scale=0.95}{\includegraphics{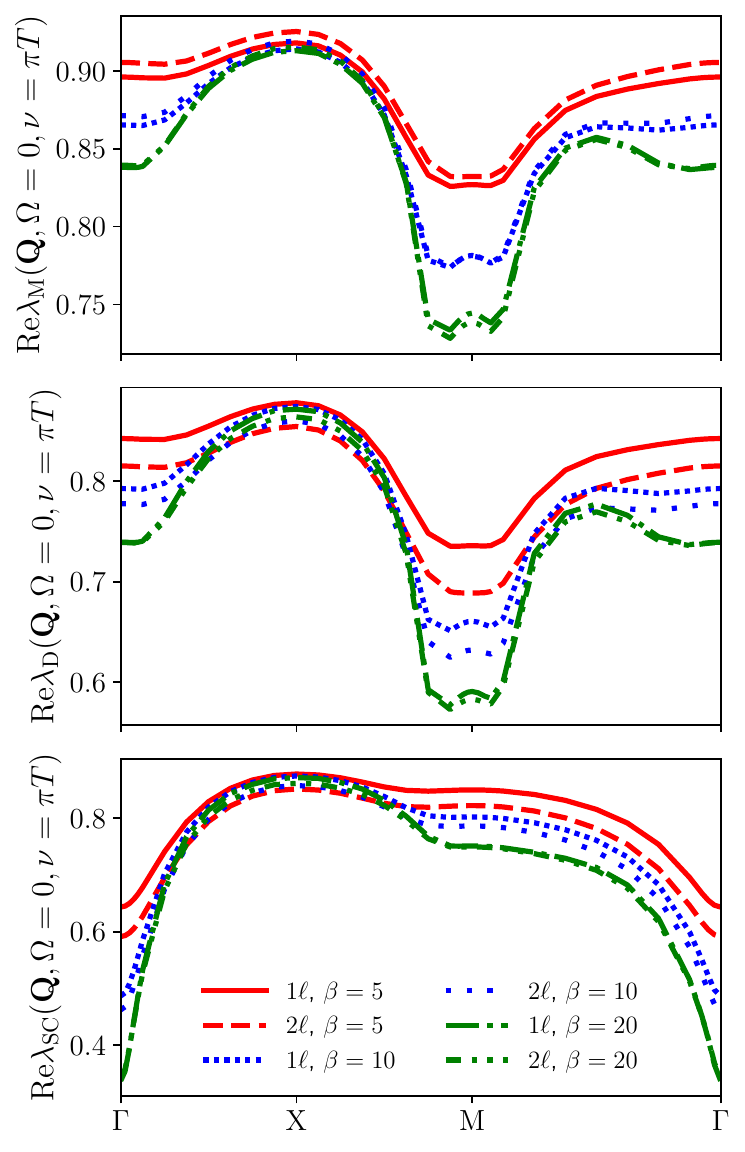}}
    \caption{Same as Fig.~\ref{fig:lambdavsQ_FiniteDoping_U2p5} but for results obtained from the $1\ell$ and $2\ell$ SBE fRG.
    }
    \label{fig:lambdavsQ1l2l_FiniteDoping_U2p5}
\end{figure}

\begin{figure*}[t!]
    \centering
    \adjustbox{max width=\textwidth, scale=1.0}{\includegraphics{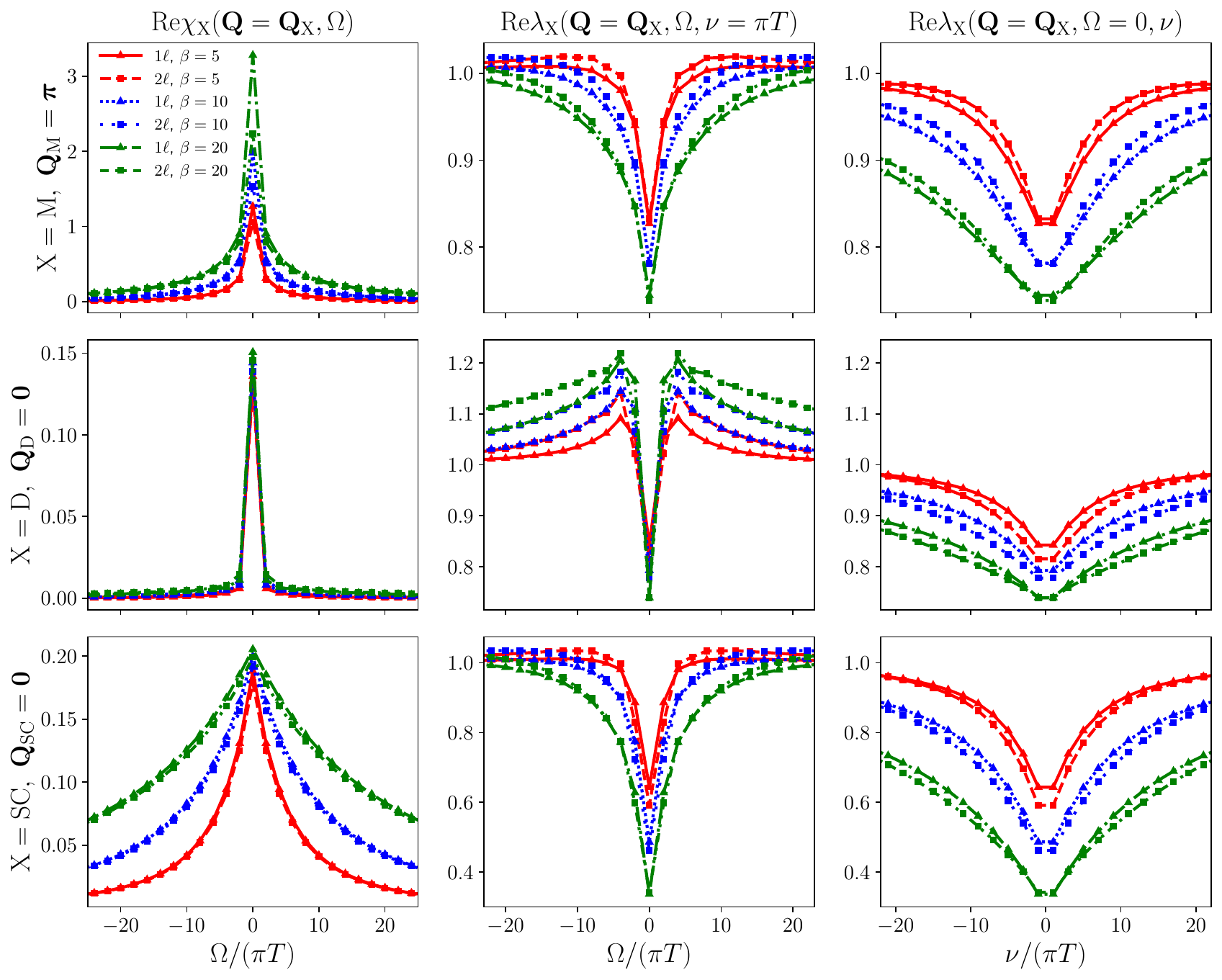}}
    \caption{Same as Fig.~\ref{fig:chilambdaFreqDep_FiniteDoping_U2Beta5} but for results obtained from the $1\ell$ and $2\ell$ SBE fRG.
    }
    \label{fig:chilambdaFreqDep1l2l_FiniteDoping_U2Beta5}
\end{figure*}

In this Appendix, we analyze the SBE approximation at the level of the postprocessing susceptibilities. We show here that it can lead to much more severe errors, compared to the SBE approximation for flowing susceptibilities discussed in the main text. To that end, we consider expression~\eqref{eq:generalized_susceptibility} for the postprocessing susceptibilities, which can be rewritten as
\begin{equation}
\chi_{\mathrm{X}} = \chi_{\mathrm{X},\Pi} + \sum_{\mathrm{X}^\prime}\left(\chi_{\mathrm{X},\overline{\nabla}_{\mathrm{X}^\prime}} + \chi_{\mathrm{X},M_{\mathrm{X}^\prime}}\right) + \chi_{\mathrm{X},U} ,
\label{eq:generalized_susceptibility_appendix}
\end{equation}
where the first term reads
\begin{equation}
\chi_{\mathrm{X},\Pi}(Q, k, k^\prime) = \Pi_{\mathrm{X}}(Q, k)\delta_{k,k^\prime}.
\end{equation}
The remaining terms $\chi_{\mathrm{X},\overline{\nabla}_{\mathrm{X}^\prime}}$, $\chi_{\mathrm{X},M_{\mathrm{X}^\prime}}$, and $\chi_{\mathrm{X},U}$ originate from the vertex contribution $\Pi_{\mathrm{X}}(Q, k)  V_{\mathrm{X}}(Q, k, k^\prime) \Pi_{\mathrm{X}}(Q, k^\prime)$ in Eq.~\eqref{eq:generalized_susceptibility}. They contain the direct contributions from the SBE terms $\overline{\nabla}_{\mathrm{X}^\prime}\equiv\nabla_{\mathrm{X}^\prime}-U_{\mathrm{X}^\prime}$, the SBE rest functions $M_{\mathrm{X}^\prime}$, and the bare interaction $U_{\mathrm{X}}$ in $V_{\mathrm{X}}$, respectively. Their expressions can be found from the relation $V_{\mathrm{X}}=\nabla_{\mathrm{X}}+\mathcal{I}_{\mathrm{X}}=\overline{\nabla}_{\mathrm{X}}+\mathcal{I}_{\mathrm{X}}+U_{\mathrm{X}}$ together with Eqs.~\eqref{eq:ExpressionsMathcalIX} and~\eqref{eq:ExpressionphiX}. For the contribution to $\chi_{\mathrm{M}}$ from the density channel, for example, this gives us
\begin{subequations}
\begin{align}
    \chi_{\mathrm{M},\overline{\nabla}_{\mathrm{D}}}(Q, k, k^\prime) &= -\frac{1}{2}\Pi_{\mathrm{M}}(Q, k) \left[P^{ph\rightarrow \overline{ph}}\overline{\nabla}_{\mathrm{D}}\right](Q, k, k^\prime) \nonumber \\
    & \phantom{=} \times \Pi_{\mathrm{M}}(Q, k^\prime), \\
    \chi_{\mathrm{M},M_{\mathrm{D}}}(Q, k, k^\prime) &= -\frac{1}{2}\Pi_{\mathrm{M}}(Q, k) \left[P^{ph\rightarrow \overline{ph}}M_{\mathrm{D}}\right](Q, k, k^\prime) \nonumber \\
    & \phantom{=} \times \Pi_{\mathrm{M}}(Q, k^\prime),
\end{align}
\end{subequations}
while the contribution from the bare interaction in each channel is given by
\begin{equation}
    \chi_{\mathrm{X},U}(Q, k, k^\prime) = \Pi_{\mathrm{X}}(Q, k) U_{\mathrm{X}}(Q, k, k^\prime) \Pi_{\mathrm{X}}(Q, k^\prime).
\end{equation}
Neglecting the SBE rest functions $M_{\mathrm{X}}$ at the level of the postprocessing susceptibilities amounts to setting $\chi_{\mathrm{X},M_{\mathrm{X}^\prime}}=0$ for all $\mathrm{X}$ and $\mathrm{X}^\prime$. However, the other contributions [$\chi_{\mathrm{X},\Pi}$, $\chi_{\mathrm{X},\overline{\nabla}_{\mathrm{X}^\prime}}$ and $\chi_{\mathrm{X},U}$ in Eq.~\eqref{eq:generalized_susceptibility_appendix}] can still be renormalized by the flow of $M_{\mathrm{X}}$ [through the differential equations~\eqref{eq:flowequationsMXPhysicalChannels1l},~\eqref{eq:flowequationsMXPhysicalChannels2l}, and~\eqref{eq:flowequationsMXPhysicalChannelsbeyond2l}], if the SBE approximation is not implemented at the level of the flow equations.

The different approximations are examined in Fig.~\ref{fig:flowVsPostproc}, for the parameter regime examined previously for the 2D Hubbard model at half filling with $U=2.5$. As discussed earlier, the flowing susceptibilities obtained by including the SBE rest functions in the flow match the parquet approximation and are well reproduced within the SBE approximation (result referred to as ``flow without $M$'' in Fig.~\ref{fig:flowVsPostproc}). Most importantly, Fig.~\ref{fig:flowVsPostproc} shows that, for the studied parameter regime, neglecting the SBE rest functions $M_{\mathrm{X}}$ at the level of the postprocessing susceptibilities drastically affects the peaks of the subleading susceptibilities, whether the flow of $M_{\mathrm{X}}$ is taken into account or not. More precisely, the peak of $\chi_{\mathrm{D}}$ is even spuriously replaced by a dip within this approximation [and the same occurs for $\chi_{\mathrm{SC}}$ due to the symmetry relation $\chi_{\mathrm{SC}}(\mathbf{Q},\Omega)=\chi_{\mathrm{D}}(\mathbf{Q}+(\pi,\pi),\Omega)$]. The inset shows that this spurious dip in $\chi_{\mathrm{D}}$ essentially originates from the missing contribution of $\chi_{\mathrm{D},M_{\mathrm{D}}}$. In contrast, the postprocessing susceptibility $\chi_{\mathrm{M}}$ appears to be well described by the SBE approximation since the rest-function-like contributions $\chi_{\mathrm{M},M_{\mathrm{X}^\prime}}$ are negligible with respect to $\chi_{\mathrm{M}}$ itself.

In conclusion, the SBE approximation implemented at the level of the postprocessing susceptibilities can produce qualitatively wrong results in the studied parameter regime. This is in stark contrast to the quantitatively accurate results obtained with flowing susceptibilities in the SBE approximation as reported in the present study. Implementing the SBE approximation at the level of the differential equations of the fRG thus offers a substantial advantage in that case. This does not apply to the iterative parquet solution referred to in Fig.~\ref{fig:chiMDlambdaMDvsQPA}, where neglecting the SBE rest functions does not include the implicit resummation of multiboson exchange contributions inherent to the fRG framework~\cite{Fraboulet2022}.

\section{Additional results for the self-energy}
\label{app:Sigma}

We present here results for the imaginary part of the self-energy obtained from the converged multiloop SBE fRG within the SBE approximation (and within the $\Omega$-flow) at $U=2.5$, $\beta=5$, and at half filling. These results can be found in Figs.~\ref{fig:ImSigvsnu} and~\ref{fig:ImSigvsk}, where they are also compared with the corresponding parquet approximation. Specifically, Fig.~\ref{fig:ImSigvsnu} shows the frequency dependence of $\mathrm{Im}\Sigma(\mathbf{k},\nu)$ at the node [$\mathbf{k}=(\pi/2,\pi/2)$] and the antinode [$\mathbf{k}=(\pi,0)$]. We observe in this figure a relative difference between converged multiloop fRG and the parquet approximation of $15\%$ ($12\%$) at the node (antinode), and at $\nu=\pi T$. This relative difference is somewhat larger than that reported in the main text for the susceptibilities and the Yukawa couplings in the same parameter regime. However, it decreases rapidly as the frequency $\nu$ increases and remains at or below $6\%$ for $\nu\geq 11 \pi T$, at both the node and the antinode. The converged multiloop SBE results thus adequately reproduce the high-frequency behavior of the self-energy obtained from the parquet approximation, even without including the SBE rest functions in the flow. Finally, the discrepancy observed between fRG and parquet results at the smallest frequencies in Fig.~\ref{fig:ImSigvsnu} appears to be rather constant in the whole Brillouin zone. This is illustrated by Fig.~\ref{fig:ImSigvsk}, where the momentum dependence of $\mathrm{Im}\Sigma(\mathbf{k},\nu)$ is displayed at $\nu=\pi T$ (in the same parameter regime as in Fig.~\ref{fig:ImSigvsnu}). The momentum dependence of the self-energy is thus only weakly affected by the larger discrepancy at small frequencies.

\section{Additional results at finite doping}
\label{app:addres}

For completeness, in Figs.~\ref{fig:chivsQ1l2l_FiniteDoping_U2p5}--\ref{fig:chilambdaFreqDep1l2l_FiniteDoping_U2Beta5} we also report the conventional $1\ell$ and $2\ell$ SBE fRG results for the parameters shown in Figs.~\ref{fig:chivsQ_FiniteDoping_U2p5}--\ref{fig:chilambdaFreqDep_FiniteDoping_U2Beta5}, in analogy to the $1\ell$ and $2\ell$ data of Fig.~\ref{fig:chiMvsU_U0to2p5Beta5} at half filling.

\newpage

\bibliography{main.bbl}
\end{document}